\documentclass[a4paper,fleqn,usenatbib]{mnras}

\usepackage[english]{babel}
\usepackage{graphicx}
\usepackage{amsfonts}
\usepackage{amssymb}
\usepackage{amsmath}
\usepackage{booktabs}
\usepackage{threeparttable}
\usepackage{enumerate}
\usepackage{enumitem}
\usepackage{hyperref}
\usepackage{breakcites}
\usepackage{comment}
\usepackage{bm}
\usepackage{txfonts} 
\allowdisplaybreaks
\usepackage{pifont}

\renewcommand{\S}{Section}
\newcommand{\Ss}{Sections}
\newcommand{\F}{Fig.}

\newcommand{\Eq}{Equation}
\newcommand{\Eqs}{Equations}
\newcommand{\eq}{Equation}
\newcommand{\eqs}{Equations}
\newcommand{\code}{Code fragment}

\newcommand{\ve}[1]{\mathbf{#1}}
\newcommand{\unit}[1]{\hat{\mathbf{#1}}}

\newcommand{\gconst}{\mathcal{G}}
\newcommand{\msun}{\mathrm{M}_\odot}
\newcommand{\rsun}{\mathrm{R}_\odot}
\newcommand{\lsun}{\mathrm{L}_\odot}
\newcommand{\au}{\textsc{au}}
\newcommand{\kms}{\mathrm{km\,s^{-1}}}
\newcommand{\yr}{\mathrm{yr}}
\newcommand{\myr}{\mathrm{Myr}}

\renewcommand{\d}{\mathrm{d}}
\renewcommand{\a}{\mathrm{a}}
\newcommand{\orb}{\mathrm{orb}}

\newcommand{\CE}{\mathrm{CE}}
\newcommand{\TCE}{\mathrm{TCE}}
\newcommand{\spin}{\mathrm{spin}}
\newcommand{\bind}{\mathrm{bind}}
\newcommand{\init}{\mathrm{init}}
\newcommand{\fin}{\mathrm{fin}}
\newcommand{\age}{\mathrm{age}}
\newcommand{\MS}{\mathrm{MS}}
\newcommand{\BGB}{\mathrm{BGB}}

\newcommand{\WA}{\mathrm{WA}}
\newcommand{\comp}{\mathrm{comp}}
\newcommand{\wind}{\mathrm{wind}}
\newcommand{\kick}{\mathrm{kick}}
\newcommand{\BH}{\mathrm{BH}}
\newcommand{\NS}{\mathrm{NS}}
\newcommand{\remnant}{\mathrm{remnant}}
\newcommand{\prog}{\mathrm{prog}}
\newcommand{\CO}{\mathrm{CO}}

\newcommand{\md}{m_\mathrm{d}}
\newcommand{\ma}{m_\mathrm{a}}

\newcommand{\Ad}{{\mathrm{A}_\mathrm{d}}}
\newcommand{\Aa}{{\mathrm{A}_\mathrm{a}}}

\newcommand{\XLz}{X_{\mathrm{L,0}}}

\newcommand{\rAa}{\ve{r}_{\mathrm{A}_\mathrm{a}}}

\newcommand{\betamt}{\beta_\mathrm{MT}}
\newcommand{\taumt}{\tau_\mathrm{MT}}
\newcommand{\mper}{m_\mathrm{per}}
\newcommand{\qper}{r_\mathrm{p,\,per}}
\newcommand{\eper}{e_\mathrm{per}}
\newcommand{\srel}{\sigma_\mathrm{rel}}
\newcommand{\mint}{M_\mathrm{int}}
\newcommand{\eff}{\mathrm{eff}}

\newcommand{\conv}{\mathrm{conv}}
\newcommand{\core}{\mathrm{core}}
\newcommand{\env}{\mathrm{env}}
\newcommand{\mc}{m_{\core,\,i}}
\newcommand{\menv}{m_{\env,\,i}}
\newcommand{\rc}{R_{\core,\,i}}
\newcommand{\renv}{R_{\env,\,i}}
\newcommand{\renc}{R_\mathrm{enc}}
\newcommand{\enc}{\mathrm{enc}}
\newcommand{\per}{\mathrm{per}}

\newcommand{\sse}{\textsc{SSE}}
\newcommand{\bse}{\textsc{BSE}}
\newcommand{\mse}{\textsc{MSE}}

\newcommand{\ssepaper}{HPT00}
\newcommand{\bsepaper}{HTP02}
\newcommand{\emtpaper}{HD19}

\newcommand{\kam}{k_{\mathrm{AM},\,i}}

\newcommand{\deltat}{\Delta t}
\newcommand{\deltatode}{\Delta t_\mathrm{ODE}}
\newcommand{\deltatn}{\Delta t_N}
\newcommand{\deltatnan}{\Delta t_{N,\,\mathrm{an}}}

\newcommand{\rstar}{R_{\star,\,i}}

\usepackage{listings}

\usepackage{color}

\definecolor{dkgreen}{rgb}{0,0.6,0}
\definecolor{gray}{rgb}{0.5,0.5,0.5}
\definecolor{mauve}{rgb}{0.58,0,0.82}

\begin{document}

\title[Multiple-star evolution]{{\it Multiple Stellar Evolution}: a population synthesis algorithm to model the stellar, binary, and dynamical evolution of multiple-star systems}
\author[Hamers, Rantala, Neunteufel, Preece, Vynatheya]{Adrian S. Hamers$^{1}$\thanks{E-mail: hamers@mpa-garching.mpg.de}, Antti Rantala$^{1}$, Patrick Neunteufel$^{1}$, Holly Preece$^{1}$, and \newauthor Pavan Vynatheya$^{1}$ \\
$^{1}$Max-Planck-Institut f\"{u}r Astrophysik, Karl-Schwarzschild-Str. 1, 85741 Garching, Germany}
\date{Accepted 2021 January 29. Received 2021 January 29; in original form 2020 November 9}

\label{firstpage}
\pagerange{\pageref{firstpage}--\pageref{lastpage}}
\maketitle

\begin{abstract}  
In recent years, observations have shown that multiple-star systems such as hierarchical triple and quadruple-star systems are common, especially among massive stars. They are potential sources of interesting astrophysical phenomena such as compact object mergers, leading to supernovae, and gravitational wave events. However, many uncertainties remain in their often complex evolution. Here, we present the population synthesis code {\it Multiple Stellar Evolution} (\mse), designed to rapidly model the stellar, binary, and dynamical evolution of multiple-star systems. \mse~includes a number of new features not present in previous population synthesis codes: (1) an arbitrary number of stars, as long as the initial system is hierarchical, (2) dynamic switching between secular and direct $N$-body integration for efficient computation of the gravitational dynamics, (3) treatment of mass transfer in eccentric orbits, which occurs commonly in multiple-star systems, (4) a simple treatment of tidal, common-envelope, and mass transfer evolution in which the accretor is a binary instead of a single star, (5) taking into account planets within the stellar system, and (6) including gravitational perturbations from passing field stars. \mse, written primarily in the \textsc{C++} language, will be made publicly available and has few prerequisites; a convenient \textsc{Python} interface is provided. We give a detailed description of MSE and illustrate how to use the code in practice. We demonstrate its operation in a number of examples.
\end{abstract}


\section{Introduction}
\label{sect:introduction}
\subsection{The importance of multiple-star systems}
\label{sect:introduction:imp}
Multiple-star systems, stellar systems containing three or more stars, are common. For example, the closest stellar system to the Sun, $\alpha$ Centauri, is a hierarchical triple system \citep{1917CiUO...40..331I}. The Algol system, well known for the Algol paradox (i.e., the less massive star in the binary is more evolved than its higher-mass companion, which can be explained by mass transfer; e.g., \citealt{1998A&AT...15..357P}), is also a triple system \citep{1957ApJ...125..359M}. The binary star Eta Carinae, known for its high mass ($\sim 90\,\msun$ primary and $\sim30\,\msun$ secondary, \citealt{1997NewA....2..107D,1997ARA&A..35....1D}) and its giant mass eruption event in the nineteenth century (e.g., \citealt{1999ASPC..179..216H}), may originally have been a triple system \citep{2016MNRAS.456.3401P,2020MNRAS.491.6000S}. More exotically, through centuries of observations in which more companions have been discovered, Castor, one of the brightest stars in the night sky, is known to harbour six stars in a `(2+2)+2' configuration (e.g., \citealt{1988PASP..100..834H}). A similar sextuple system with a long observational history is Mizar and Alcor (e.g., \citealt{2010AJ....139..919M}). Currently, even two systems with seven stars are known: AR Cassiopeiae, and Nu Scorpii \citep{1997A&AS..124...75T,2018ApJS..235....6T}.

From a more statistical viewpoint, observations of F and G dwarfs within 67 pc of the Sun \citep{2014AJ....147...86T,2014AJ....147...87T} show that $\approx 10\%$ of stellar systems with Solar-like components are triple systems, and $\approx 1\%$ are quadruple systems. In a similar study of Solar-type stars within 25 pc of the Sun \citep{2010ApJS..190....1R}, the triple fraction was found to be $(8\pm 1)\%$, and the fraction of higher-multiplicity systems $(3\pm 1)\%$. For systems with more massive (primary) stars, the multiplicity fraction is significantly higher \citep{2013ARA&A..51..269D}. Further, \citet{2017ApJS..230...15M} find that, among O-type stars, the multiplicity fraction is $(35 \pm 3)\%$ for triple stars, and $(38\pm11)\%$ for quadruple stars, showing that, among massive stellar systems in the field, triples and quadruples significantly outnumber both single and binary stars.

Multiple-star systems are usually arranged in a hierarchical configuration, since they would otherwise be short lived. The simplest hierarchical configuration occurs in triple systems in which two stars are orbited by a more distant, tertiary star. If the inner and outer orbits in such a configuration are initially mutually highly inclined, then the gravitational torque of the outer orbit can induce high-amplitude eccentricity oscillations in the inner binary, known as Lidov-Kozai (LK) or von Zeipel-Lidov-Kozai (ZLK) oscillations (\citealt{1910AN....183..345V,1962P&SS....9..719L,1962AJ.....67..591K}; see \citealt{2016ARA&A..54..441N,2017ASSL..441.....S,2019MEEP....7....1I} for reviews). 

ZLK oscillations have important implications for a large variety of three-body systems, not limited to triple-{\it star} systems. Generally, the high eccentricities that can be reached during these oscillations can drive strong interactions such as efficient tidal dissipation, or even collisions. For example, they have been considered as a possible pathway to produce short-period binaries (e.g., \citealt{1979A&A....77..145M,1998MNRAS.300..292K,2001ApJ...562.1012E,2006Ap&SS.304...75E,2007ApJ...669.1298F,2014ApJ...793..137N}) and hot Jupiters (e.g., \citealt{2003ApJ...589..605W,2007ApJ...669.1298F,2012ApJ...754L..36N,2015ApJ...799...27P,2016MNRAS.456.3671A,2016ApJ...829..132P}), enhancing mergers of compact objects (e.g., \citealt{2002ApJ...578..775B,2011ApJ...741...82T,2013MNRAS.430.2262H,2017ApJ...841...77A,2017ApJ...836...39S,2017ApJ...846L..11L,2018ApJ...863...68L,2018ApJ...865....2H,2018ApJ...856..140H,2018ApJ...853...93R,2018ApJ...864..134R,2018A&A...610A..22T,2019MNRAS.486.4443F}), affecting the evolution of protoplanetary or accretion disks in binaries (e.g., \citealt{2014ApJ...792L..33M,2015ApJ...813..105F,2017MNRAS.467.1957Z,2017MNRAS.469.4292L,2018MNRAS.477.5207Z,2019MNRAS.485..315F,2019MNRAS.489.1797M}), triggering white dwarf pollution by planets (e.g., \citealt{2016MNRAS.462L..84H,2017ApJ...834..116P}), and producing blue straggler stars (e.g., \citealt{2009ApJ...697.1048P,2016ApJ...816...65A,2016MNRAS.460.3494S,2019MNRAS.488..728F}). Also, ZLK cycles can couple with stellar evolution in triples, giving rise to strong interactions during or after the main sequence (MS; e.g., \citealt{2013MNRAS.430.2262H,2013ApJ...766...64S,2014ApJ...794..122M,2016ComAC...3....6T,2016MNRAS.460.3494S,2017ApJ...841...77A,2018A&A...610A..22T,2019ApJ...878...58S,2019ApJ...882...24H,2019ApJ...883...23H}). 

ZLK oscillations can become more complicated when the triple system is marginally hierarchical, such that higher-order expansion terms become important (e.g., \citealt{2011ApJ...742...94L,2011PhRvL.107r1101K,2014ApJ...791...86L,2016MNRAS.459.2827H,2017PhRvD..96b3017W}). Similarly, the dynamics become more complex when more bodies are added to the system (while maintaining a hierarchical configuration with widely separated orbits). In hierarchical quadruples, which occur in either the `2+2' (two binaries orbiting each other's center of mass) or `3+1' (triple orbited by a fourth body) configurations, secular evolution can be more efficient compared to triples \citep{2013MNRAS.435..943P,2015MNRAS.449.4221H,2016MNRAS.461.3964V,2017MNRAS.470.1657H,2018MNRAS.476.4234F,2018MNRAS.474.3547G,2019MNRAS.483.4060L,2019MNRAS.486.4781F}, and this can have implications for, e.g., short-period binaries \citep{2019MNRAS.482.2262H}, and Type Ia Supernovae (SNe Ia; \citealt{2018MNRAS.478..620H,2018MNRAS.476.4234F}). This trend carries over to higher-multiplicity systems (quintuples, sextuples, etc.), in which the likelihood for strong interactions due to secular evolution is even higher \citep{2020MNRAS.494.5298H}.

\subsection{Existing population synthesis codes}
\label{sect:introduction:pop}
Population synthesis codes, which are intended to model the evolution of a large number of systems in order to gain insight into population statistics, have been used extensively during the past several decades to study the evolution of predominantly binary stars (e.g., \citealt{1985MNRAS.214..357W,1987ApJ...321..780D,1987SvA....31..228L,1990ApJ...358..189D,1990ApJ...360...75H,1991ApJ...376..177R,1996A&A...309..179P,1997MNRAS.291..732T,1998AIPC..456...61W,2001A&A...365..491N,2001A&A...368..939N,2014A&A...563A..83C,2009A&A...508.1359I,2012A&A...546A..70T,2013A&A...552A..69V}). In particular, \bse~(\citealt{2002MNRAS.329..897H}, hereafter \bsepaper), based on the rapid evolution algorithm \sse~(\citealt{2000MNRAS.315..543H}, hereafter \ssepaper), has been an industry standard for nearly two decades. Also, \bse, and the \sse~analytic stellar evolution tracks on which it is based, have formed the basis for many other codes such as \textsc{SeBa} \citep{1996A&A...309..179P,2001A&A...365..491N,2001A&A...368..939N,2012A&A...546A..70T,2013A&A...557A..87T}, \textsc{StarTrack} \citep{2002ApJ...572..407B,2002ApJ...571L.147B,2008ApJS..174..223B}, \textsc{binary\_c} \citep{2004MNRAS.350..407I,2006A&A...460..565I,2009A&A...508.1359I,2014A&A...563A..83C,2014ApJ...782....7D,2015A&A...581A..62A}, \textsc{MOBSE} \citep{2018MNRAS.474.2959G,2018MNRAS.480.2011G}, \textsc{COMPAS} \citep{2017NatCo...814906S,2018MNRAS.477.4685B,2018MNRAS.481.4009V,2019MNRAS.490.5228B,2019ApJ...882..121S,2019MNRAS.490.3740N}, and \textsc{COSMIC} \citep{2020ApJ...898...71B}. Other codes based on other methods (i.e., not using analytic fits to stellar evolution tracks) include the Brussels code \textsc{PNS} \citep{2004NewAR..48..861D,2010A&A...515A..89M,2012A&A...543A...4V}, \textsc{BPAS} \citep{2009MNRAS.400.1019E,2012MNRAS.422..794E,2016MNRAS.462.3302E,2017PASA...34...58E,2018MNRAS.479...75S}, \textsc{SEVN} \citep{2015MNRAS.451.4086S,2017MNRAS.470.4739S,2019MNRAS.485..889S}, \textsc{ComBinE} \citep{2018MNRAS.481.1908K}, \textsc{dart\_board} \citep{2018ApJS..237....1A}, and \textsc{METISSE} (a single stellar evolution code based on interpolation; \citealt{2020MNRAS.497.4549A}). The comparative study of \citet{2014A&A...562A..14T} found that in the context of low and intermediate-mass binaries, discrepancies between simulation results of \textsc{binary\_c}, \textsc{PNS}, \textsc{SeBa}, and \textsc{StarTrack} are typically small and can be explained by different physical assumptions.

More recently, population synthesis codes have been developed that can model the evolution of triple stars (as long as the system is hierarchical) taking into account both stellar/binary evolution, and gravitational dynamics\footnote{There exist several publicly available codes that model the secular dynamics of triple systems (e.g., \textsc{kozai}, \citealt{2015MNRAS.452.3610A}, the identically-named \textsc{kozai}, \citealt{2018ApJ...863....7R,2018MNRAS.480L..58A}, and --- not limited to triples --- \textsc{SecularMultiple}, \citealt{2016MNRAS.459.2827H,2018MNRAS.476.4139H,2020MNRAS.494.5492H}). However, these codes do not include stellar/binary evolution.}. The code \textsc{triple\_c} \citep{2013MNRAS.430.2262H} combined binary evolution in \textsc{binary\_c} to model the inner binary system with secular dynamics using the orbit-averaged and expanded equations of motion (e.g., \citealt{1968AJ.....73..190H,2000ApJ...535..385F,2002ApJ...578..775B,2013MNRAS.431.2155N}). It included tidal evolution and common envelope (CE) evolution, both processes which can be enhanced due to high-amplitude ZLK eccentricity oscillations. \textsc{Tres} \citep{2016ComAC...3....6T}, implemented within the \textsc{AMUSE} framework \citep{2013A&A...557A..84P,2009NewA...14..369P}, uses similar routines as \textsc{triple\_c} to model the secular dynamics, whereas \textsc{SeBa} is used to model stellar and binary evolution.

\subsection{Limitations}
\label{sect:introduction:lim}
Current triple population synthesis codes face a number of limitations. In both \textsc{triple\_c} and \textsc{Tres}, mass transfer is included but is limited by the assumption of circular orbits. When mass transfer occurs in isolated binaries, the orbit is usually assumed to be circular (e.g., \bsepaper), since efficient tides in isolated binaries are expected to circularise the orbit prior to the onset of Roche lobe overflow (RLOF). This is often no longer the case for higher-multiplicity systems, since secular evolution in the latter can trigger mass transfer in eccentric orbits (e.g.,  \citealt{2016ComAC...3....6T,2018A&A...610A..22T,2018MNRAS.478..620H,2019MNRAS.482.2262H,2020A&A...640A..16T}). 

The problem of mass loss/mass transfer in binary systems, including eccentric ones, has been studied extensively (e.g., \citealt{1956AJ.....61...49H,1963Icar....2..440H,1964AcA....14..241K,1964AcA....14..251P,1983ApJ...266..776M,1984ApJ...282..522M}), and has received more recent attention in numerical studies (e.g., \citealt{2005MNRAS.358..544R,2009MNRAS.395.1127C,2010ApJ...724..546S,2011ApJ...726...66L,2011ApJ...726...67L,2016MNRAS.455..462V,2017MNRAS.467.3556B}), as well as in (semi)analytical work \citep{2007ApJ...667.1170S,2009ApJ...702.1387S,2010ApJ...724..546S,2011MNRAS.417.2104V,2012MNRAS.422.1648V,2013MNRAS.435.2416V,2014MNRAS.437.1127V,2016ApJ...825...70D,2016ApJ...825...71D}. In particular, \citet{2007ApJ...667.1170S} and \citet{2016ApJ...825...71D} derived equations for the secular (i.e., orbit-averaged) changes of the orbital elements due to mass transfer in eccentric binaries. They assumed that the mass transfer rate is a delta function centered at periapsis, i.e., the donor star transfers its mass in a burst at its closest approach to its companion. This assumption is physically reasonable in the limit of very high eccentricity. However, as pointed out by \citeauthor{2019ApJ...872..119H} (\citeyear{2019ApJ...872..119H}, hereafter \emtpaper), this approximation is no longer appropriate for less eccentric or even circular orbits, and numerical integration of the equations of motion in the latter case can yield unphysical results such as negative eccentricities. In \emtpaper, an alternative model was considered in which the mass transfer rate function smoothly transitions between the high-eccentricity regime (in which the mass transfer rate behaves like a delta function), and the circular regime (in which the mass transfer rate is constant during the orbit). 

Furthermore, current population synthesis codes are limited to triple systems\footnote{\citet{2018MNRAS.478..620H} included stellar and dynamical evolution in quadruples, but CE evolution and mass transfer were not taken into account.}, whereas, as mentioned above, quadruple systems are common among massive systems. Recent studies of the dynamical evolution of quadruples have shown that the latter could play an important role for black hole (BH) and neutron star (NS) mergers with implications for gravitational wave (GW) observations (e.g., \citealt{2019MNRAS.483.4060L,2019MNRAS.486.4781F,2020ApJ...888L...3S,2020ApJ...898...99H,2020ApJ...895L..15F}). 

Another limitation of current triple population synthesis codes is that they model the long-term evolution using the secular equations of motion. However, there exist situations in which the secular approximation breaks down. For example, the dynamical stability of the system can be affected by stellar evolution-induced mass loss (e.g., \citealt{2012ApJ...760...99P,2013MNRAS.430.2262H}), mass loss and kicks from SNe, fly-bys (e.g., \citealt{2019ApJ...882...24H}), or secular evolution in high-multiplicity systems (e.g., \citealt{2017MNRAS.466.4107H,2018MNRAS.478..620H,2020MNRAS.494.5298H}). Furthermore, the secular approximation can break down when the timescale for angular-momentum changes due to secular evolution becomes shorter than the orbital timescale \citep{2012ApJ...757...27A,2014ApJ...781...45A,2016MNRAS.458.3060L,2018MNRAS.481.4907G,2018MNRAS.481.4602L,2019MNRAS.490.4756L}. 

In addition, triple systems can give rise to more complicated processes compared to binary systems, such as mass transfer or CE evolution of the tertiary with respect to the inner binary (e.g., \citealt{2014MNRAS.438.1909D,2020MNRAS.498.2957C,2021MNRAS.500.1921G}). Such processes are not taken into account in previous population synthesis codes. 

Lastly, with the advent of exoplanetary astronomy (e.g., \citealt{2015ARA&A..53..409W}), it has become clear that exoplanets not only occur in single-star systems, but also in higher-order systems. Most notably, many planets have been found in binary systems (see, e.g., \citealt{2020Galax...8...16B}, for a recent review), including the {\it Kepler} transiting circumbinary planets (e.g., \citealt{2011Sci...333.1602D,2013ApJ...770...52K,2016ApJ...827...86K}). Planets have also been found in triples (e.g., \citealt{2007A&A...469..755M}), and even quadruples (e.g., \citealt{2013ApJ...768..127S}). In particular in systems such as triples and quadruples, the long-term evolution of planets can be complex and chaotic (e.g., \citealt{2015PNAS..112.9264M,2015MNRAS.453.3554M,2016MNRAS.455.3180H,2017ApJ...835L..24H,2017MNRAS.466.4107H}). Also, stellar evolution can have important implications for planetary dynamics, for example in evolving binary systems (e.g., \citealt{2016MNRAS.462L..84H,2017ApJ...834..116P}). These processes are not included in most existing population synthesis codes.

\subsection{A new population synthesis code}
\label{sect:introduction:code}
In this work, we present a new population synthesis code, \textsc{Multiple Stellar Evolution} (\mse), aimed at modelling the stellar evolution, binary (such as mass transfer and CE) evolution, and gravitational dynamics of multiple-star systems. The core components of \mse~are the stellar evolution fits of \ssepaper, several aspects of binary evolution adopted from \bsepaper, the model for eccentric mass transfer of \emtpaper, secular dynamical evolution with \textsc{SecularMultiple} \citep{2016MNRAS.459.2827H,2018MNRAS.476.4139H,2020MNRAS.494.5492H}, and dynamical evolution for non-secular systems with a direct $N$-body code \citep{2020MNRAS.492.4131R}. 

The main new features of \mse~which differentiate it from previous codes are:
\begin{enumerate}[leftmargin=0.5cm]
\item an arbitrary number of stars, as long as the initial system is hierarchical;
\item hybrid integration techniques (secular and direct $N$-body) to efficiently model the gravitational dynamics;
\item including the effects of mass transfer in eccentric binary subsystems;
\item incorporating simple treatments for `triple' interactions such as CE and mass transfer evolution in which one of the components is a binary instead of a single star (single-binary-star interactions);
\item including planets within the system (taking into account their Newtonian gravitational and tidal evolution);
\item taking into account gravitational perturbations from passing field stars.
\end{enumerate}

\mse~is written primarily in the \textsc{C++} programming language. It includes some linking to existing \textsc{Fortran} routines from \sse~(\ssepaper)~and \bse~(\bsepaper), as well as to the \textsc{C}-code \textsc{MSTAR} \citep{2020MNRAS.492.4131R}. Although \mse~is accessible directly via \textsc{C++}, a \textsc{Python} interface is included which makes it easy to use the code in combination with the large existing library of plotting and analysis tools within \textsc{Python}. Only a \textsc{C++} and \textsc{Fortran} compiler, and a \textsc{Python} installation with \textsc{Numpy} are required for installation.

At the time of writing, \mse~is part of a private repository on \texttt{GitHub}\footnote{\href{https://github.com/hamers/mse}{https://github.com/hamers/mse}.}. Access to this repository can be requested by contacting the authors. In the future, the repository will be made publicly available. 

\begin{table*}
\begin{tabular}{lp{7.0cm}p{7.0cm}}
\toprule
Symbol & Description & Notes \\
\midrule
{\it Physical constants} \\
$\gconst$ 					& Gravitational constant. \\
$c$						& Speed of light. \\
\midrule
{\it Stars} \\
$k_i$					& Stellar type of star $i$. 											& Default initial value 1 (ZAMS). See Table~\ref{table:st}.			\\
$m_i$					& Mass of star $i$. 												&										\\
$\mc$					& Core mass of star $i$.											& Determined by \sse.						\\
$\menv$					& Convective envelope mass of star $i$.								& Determined by \sse.						\\
$L_i$					& Luminosity of star $i$.											& Determined by \sse.						\\
$Z_i$					& Metallicity of star $i$.											& Default value $0.02$.									\\
$\rstar$					& Radius of star $i$.												& Determined by \sse.						\\
$\rc$						& Core radius of star $i$.											& Determined by \sse.						\\
$\renv$						& Convective envelope radius of star $i$.								& Determined by \sse.					\\
$\ve{\Omega}_i$			& Spin angular frequency vector of star $i$. 							& Initial default magnitude set according to \eq~(\ref{eq:init_spin}). 	\\
$t_{\mathrm{V},\,i}$			& Viscous timescale of star $i$. 									& Computed from the stellar properties using a prescription (\S~\ref{sect:dyn:sec:tide}). 	\\
$k_{\mathrm{AM},\,i}$			& Apsidal motion constant of star $i$.								& Computed from fits to stellar models, not part of \sse~(see \S~\ref{sect:stellar:apsmot}).\\
$r_{\mathrm{g},\,i}$			& Gyration radius of star $i$.										& Given by \sse.								\\
$t_{\mathrm{dyn},\,i}$ 		& Dynamical timescale of star $i$.									& $t_{\mathrm{dyn},\,i} \equiv \sqrt{\rstar^3/(\gconst m_i)}$	\\
$t_{\mathrm{KH},\,i}$ 		& Kelvin-Helmholtz timescale of star $i$.								& See \eq~(\ref{eq:t_kh}).							\\
$\ve{R}_i$					& Position vector of star $i$.										& Computed from system orbital properties.					\\
$\ve{V}_i$					& Velocity vector of star $i$.										& Computed from system orbital properties.					\\
\midrule
{\it Orbits} \\
$a_k$					& Semimajor axis of orbit $k$. 										&											\\
$e_k$					& Eccentricity of orbit $k$. 										&											\\
$i_k$						& Inclination of orbit $k$. 										&											\\
$i_{kl}$					& Inclination of orbit $k$ relative to orbit $l$ (mutual inclination).							&											\\
$\omega_k$				& Argument of periapsis of orbit $k$. 									&										\\
$\Omega_k$				& Longitude of the ascending node of orbit $k$. 							&										\\
$\ve{e}_k$					& Eccentricity vector of orbit $k$. 									&										\\
$\ve{\jmath}_k$				& Dimensionless angular-momentum vector of orbit $k$. 					& $\jmath_k=\sqrt{1-e_k^2}$						\\
$M_{k.\mathrm{C}l}$			& Mass of all bodies contained within child $l$ of orbit $k$ ($l$ can be either 1 or 2).  & 										\\
$M_k$					& Mass of all bodies contained within orbit $k$. 							& $M_k \equiv M_{k.\mathrm{C}1}+M_{k.\mathrm{C}2}$ 	\\
$P_{\orb,\,k}$					& Period of orbit $k$.		 										& $P_{\orb,\,k}= 2\pi \sqrt{a_k^3/(\gconst M_k)}$ \\
$\ve{h}_k$					& Angular-momentum vector of orbit $k$.								& $\ve{h}_k = (M_{k.\mathrm{C}1} M_{k.\mathrm{C}2}/M_k) \, \sqrt{\gconst M_k a_k} \, \ve{\jmath}_k$ \\
$\mu_{ij}$					& Reduced mass for objects $i$ and $j$ (bodies and/or orbits).				& $\mu_{ij} \equiv m_i m_j/(m_i+m_j)$				\\
$n_k$					& Mean motion of orbit $k$.		 										& $n_k \equiv 2\pi/P_{\orb,\,k}$				\\
$\unit{q}_k$				& Dimensionless vector perpendicular to $\unit{h}_k$ and $\unit{e}_k$.			& $\unit{q}_k \equiv \unit{h}_k \times \unit{e}_k$ \\
\bottomrule
\end{tabular}
\caption{Description of various quantities used in this paper.} 
\label{table:not}
\end{table*}

\begin{table}
\begin{center}
\begin{tabular}{cl}
\toprule 
$k_i$ & Description \\
\midrule
0 & Main sequence ($m_i\lesssim0.7 \, \msun$) \\
1 & Main sequence ($m_i\gtrsim0.7 \, \msun$) \\
2 & Hertzsprung gap (HG) \\
3 & Red giant branch (RGB) \\
4 & Core helium burning (CHeB) \\
5 & Early asymptotic giant branch (EAGB) \\
6 & Thermally pulsing AGB (TPAGB) \\
7 & Naked helium star MS (He MS) \\
8 & Naked helium star Hertzsprung gap (He HG) \\
9 & Naked helium star giant branch (He GB) \\
10 & Helium white dwarf (He WD) \\
11 & Carbon-oxygen white dwarf (CO WD) \\
12 & Oxygen-neon white dwarf (ONe WD) \\
13 & Neutron star (NS) \\
14 & Black hole (BH) \\
15 & Massless remnant \\
\bottomrule
\end{tabular}
\end{center}
\caption{\small Description of the different stellar types used in \mse, reproduced from \ssepaper.}
\label{table:st}
\end{table}

\subsection{Notation and contents}
\label{sect:introduction:struc}
In table~\ref{table:not}, we summarise the notation of the most important quantities referenced to in this paper. \sse~uses integer numbers, `stellar types' $k_i$, which denote the type of star. Definitions of the stellar types and a table showing the stellar types are given in \ssepaper; for easy reference, we include a similar table here (Table~\ref{table:st}).

The structure of this paper is as follows. In \Ss~\ref{sect:dyn} through \ref{sect:alg}, we discuss the evolution algorithm, focusing sequentially on gravitational dynamics, stellar evolution, binary evolution, triple evolution, fly-bys, and the main evolution algorithm. These sections contain detailed information and are mostly tailored to a more specialised audience. We discuss more practical information on how to use the code in \S~\ref{sect:usage}. A number of example systems are presented in \S~\ref{sect:ex}. We discuss current limitations and future directions in \S~\ref{sect:discussion}, and conclude in \S~\ref{sect:conclusions}.

\section{Gravitational dynamics}
\label{sect:dyn}
The long-term gravitational dynamics of hierarchical multiple systems are often complex. Since the orbital timescales are typically much shorter than the timescales on which orbits evolve, integrating the long-term dynamical evolution is challenging from a computational point of view. In \mse, we take a hybrid approach in which the gravitational dynamics of the system are modelled using two complementary methods. The code switches between these two methods dynamically during runtime. 

\subsection{Secular integration}
\label{sect:dyn:sec}
Initially, the system is assumed to be dynamically stable (if not, the code will immediately switch to direct $N$-body integration; see \S~\ref{sect:dyn:Nbody}), and it is integrated using the secular approximation. The latter is based on an expansion of the Hamiltonian $H$ of the system in terms of ratios of adjacent orbital separations, $x_i$, and an averaging of the expanded $H$ over the orbits in the system. The secular approximation is well justified for highly hierarchical systems (e.g., \citealt{2016MNRAS.459.2827H}). It has the advantage of being computationally much faster than direct $N$-body integration, since the orbital phases are not resolved. The core implementation for the secular evolution in \mse~is adopted from the (freely-available\footnote{\href{https://github.com/hamers/secularmultiple}{https://github.com/hamers/secularmultiple}.}) code \textsc{SecularMultiple} \citep{2016MNRAS.459.2827H,2018MNRAS.476.4139H,2020MNRAS.494.5492H}. The secular equations of motion are formulated in a set of ordinary differential equations (ODEs). The latter is evolved using the \textsc{C}-code library \textsc{CVODE} \citep{1996ComPh..10..138C}, which is suited for both stiff and non-stiff ODEs. By default, \mse~takes into account Newtonian terms in the expansion of $H$ up to and including fifth order (dotriacontupole) in $x_i$ for pairwise binary interactions, and up to including third order (octupole order) for interactions involving three binaries simultaneously. Schematically, the secular evolution for an orbit $k$ is described by
\begin{subequations}
\label{eq:eom_sec}
\begin{align}
\left ( \frac{\d \ve{e}_k}{\d t} \right )_\mathrm{sec} &= \ve{f}_\ve{e} (\ve{h}_l, \ve{e}_l); \\
\left ( \frac{\d \ve{h}_k}{\d t} \right )_\mathrm{sec} &= \ve{f}_\ve{h} (\ve{h}_l, \ve{e}_l),
\end{align}
\end{subequations}
where $ \ve{f}_\ve{e}$ and $ \ve{f}_\ve{h}$ are functions of the angular-momentum and eccentricity vectors of some or all orbits in the system (including orbit $k$). For more information on the secular method, we refer to \citet{2016MNRAS.459.2827H,2018MNRAS.476.4139H,2020MNRAS.494.5492H}. 

There are situations, however, in which the secular approximation breaks down. This can occur, for example, as a result of orbital changes due to stellar evolution-induced mass loss (e.g., \citealt{2012ApJ...760...99P,2013MNRAS.430.2262H}), mass loss and kicks from SNe, fly-bys (e.g., \citealt{2019ApJ...882...24H}), or secular evolution itself in high-multiplicity systems (e.g., \citealt{2017MNRAS.466.4107H,2018MNRAS.478..620H,2020MNRAS.494.5298H}). Therefore, also incorporated into \mse~is a direct $N$-body code, to which it will switch if the secular approximation breaks down. This is discussed in more detail in \S~\ref{sect:dyn:Nbody}.

\subsubsection{Tidal evolution}
\label{sect:dyn:sec:tide}
In addition to the point-mass Newtonian gravitational dynamics, we take into account tidal evolution when integrating the secular equations of motion. Tidal evolution is modelled according to the equilibrium tide model \citep{1879RSPT..170....1D,1973Ap&SS..23..459A,1981A&A....99..126H,1998ApJ...499..853E}. In this model, which strictly only applies to a single star with a point-mass companion, the subject star is assumed to have two symmetric bulges that are in quasi-hydrostatic equilibrium. These bulges can be modelled as two additional point masses at the stellar surface on opposite sides of the star. The presence of these bulges generally gives rise to orbital apsidal motion. Furthermore, if the bulges are misaligned with respect to the relative orbital separation, this leads to dissipation of orbital energy while conserving total angular momentum \citep{1973ApJ...180..307C,1980A&A....92..167H}. 

The equilibrium tide model is implemented in \mse~for orbits containing a star and any companion. The companion can be a single star, but also a multi-body subsystem (possibly containing more than two stars). We take an {\it ad hoc} approach and apply the equilibrium tides in each of these cases, ignoring possible interaction terms between different orbits. A more self-consistent treatment of equilibrium tides in \mse~for triple and higher-order (sub)systems (e.g., \citealt{2018MNRAS.479.3604G,2020MNRAS.491..264G}) is left for future work. Due to the high sensitivity of the strength of tidal interactions to the separation between the subject star and its companion, we expect that tides in orbits not containing two stars (which, by necessity, are wide in order to guarantee dynamical stability) are typically unimportant.

\paragraph{Equations of motion} Specifically, \mse~implements the equations for dissipative equilibrium tides of \citet{2009MNRAS.395.2268B}. For completeness, they are repeated here. Dissipation in a body $i$ (the effects are added for each of the bodies in each orbit) gives rise to changes in its parent orbit $k$ according to
\begin{subequations}
\begin{align}
\nonumber \left ( \frac{\d \ve{e}_k}{\d t} \right )_\mathrm{tides,\,diss} &= - \frac{1}{t_{\mathrm{f},\,i}} \Biggl [ \frac{\ve{\Omega}_i \cdot \ve{e}_k}{2 n_k} f_2(e_k) \unit{h}_k + 9 f_1(e_k) \ve{e}_k \\
&\qquad - \frac{11}{2} \frac{\ve{\Omega}_i \cdot \unit{h}_k}{n_k} f_2(e_k) \ve{e}_k \Biggl ]; \\
\nonumber \left ( \frac{\d \ve{h}_k}{\d t} \right )_\mathrm{tides,\,diss} &= - \frac{1}{t_{\mathrm{f},\,i}} \Biggl [ \frac{\ve{\Omega}_i \cdot \ve{e}_k}{2 n_k} f_5(e_k) h_k \ve{e}_k - \frac{\ve{\Omega}_i}{2 n_k} f_3(e_k) h_k + f_4(e_k) \ve{h}_k \\
&\qquad- \frac{\ve{\Omega}_i \cdot \ve{h}_k}{2n_k} f_2(e_k) \unit{h}_k \Biggl ].
\end{align}
\end{subequations}
Here, hats denote units vectors, and
\begin{align}
t_{\mathrm{f},\,i} = \frac{t_{\mathrm{V},\,i}}{9}  \left ( \frac{a_k}{\rstar} \right )^8 \frac{m_i^2}{m_{\comp,\,i} M_k} \left (1 + 2 \kam  \right )^{-2},
\end{align}
where $t_{\mathrm{V},\,i}$ is the viscous timescale (see below). The eccentricity functions are given by
\begin{subequations}
\begin{align}
f_1(e) &= \left(1-e^2 \right)^{-13/2} \left (1 + \frac{15}{4} e^2 + \frac{15}{8} e^4 + \frac{5}{64} e^6 \right ); \\
f_2(e) &= \left(1-e^2 \right)^{-5} \left (1 + \frac{3}{2} e^2 + \frac{1}{8} e^4 \right ); \\
f_3(e) &= \left(1-e^2 \right)^{-5} \left (1 + \frac{9}{2} e^2 + \frac{5}{8} e^4 \right ); \\
f_4(e) &= \left(1-e^2 \right)^{-13/2} \left (1 + \frac{15}{2} e^2 + \frac{45}{8} e^4 + \frac{5}{16} e^6 \right ); \\
f_5(e) &= \left(1-e^2 \right)^{-5} \left (3 + \frac{1}{2} e^2 \right ); \\
f_6(e) &= \left(1-e^2 \right)^{-8} \left (1 + \frac{31}{2} e^2 + \frac{255}{8} e^4 + \frac{185}{16} e^6 + \frac{25}{64} e^8 \right ).\end{align}
\end{subequations}
The non-dissipative parts of the equilibrium tides give rise to changes in the directions of $\ve{e}_k$ and $\ve{h}_k$ described by \citep{1998ApJ...499..853E}
\begin{subequations}
\begin{align}
\left ( \frac{\d \ve{e}_k}{\d t} \right )_\mathrm{tides,\,non-diss} &= e_k \left (Z_{k,\,i} \, \unit{q}_k - Y_{k,\,i} \, \unit{h}_k \right ); \\
\left ( \frac{\d \ve{h}_k}{\d t} \right )_\mathrm{tides,\,non-diss} &= h_k \left (- X_{k,\,i} \, \unit{q}_k + Y_{k,\,i} \, \unit{e}_k \right ),
\end{align}
\end{subequations}
where
\begin{subequations}
\begin{align}
X_{k,\,i} &= - C_{k,\,i} \left (1-e_k^2 \right )^{-2} \left (\ve{\Omega}_i \cdot \unit{h}_k \right ) \left (\ve{\Omega}_i \cdot \unit{e}_k \right ); \\
Y_{k,\,i} &= - C_{k,\,i} \left (1-e_k^2 \right )^{-2} \left (\ve{\Omega}_i \cdot \unit{h}_k \right ) \left (\ve{\Omega}_i \cdot \unit{q}_k \right ); \\
Z_{k,\,i} &= \frac{1}{2} C_{k,\,i} \left (1-e_k^2 \right )^{-2} \left [ 2 \left (\ve{\Omega}_i \cdot \unit{h}_k \right )^2 - \left (\ve{\Omega}_i \cdot \unit{q}_k \right )^2 - \left (\ve{\Omega}_i \cdot \unit{e}_k \right )^2 \right ],
\end{align}
\end{subequations}
and with
\begin{align}
C_{k,\,i} \equiv \frac{\kam}{n_k} \frac{M_k}{m_{i}} \left ( \frac{\rstar}{a} \right )^5.
\end{align}
The response of the spin frequency of star $i$ is computed assuming conservation of orbital and spin angular momentum due to tidal evolution only, i.e.,
\begin{align}
\left ( \frac{\d \ve{\Omega}_i}{\d t} \right )_\mathrm{tides} &= - \frac{1}{I_i} \left ( \frac{\d \ve{h}_k}{\d t} \right )_\mathrm{tides},
\end{align}
where $I_i$ is the moment of inertia of star $i$ (see \S~\ref{sect:bin:mt:stable:spin}). See \S~\ref{sect:stellar:sse} for how the spins are initialised by default. 

\paragraph{Prescriptions for $\kam/T_i$} We quantify the tidal dissipation strength by the viscous timescale $t_{\mathrm{V},\,i}$. The latter is computed as a function of the stellar properties, in particular, the \sse~stellar type (see Table~\ref{table:st}), mass, radius, convective envelope mass and radius, and stellar spin period, using the prescription of \bsepaper. The prescription gives the combination of the apsidal motion constant $\kam$ divided by a tidal dissipation timescale $T_i$, i.e., $\kam/T_i$. The latter is related to $t_{\mathrm{V},\,i}$ according to
\begin{align}
t_{\mathrm{V},\,i} = 3 \left (1 + 2 \kam \right )^2 \left (\frac{\kam}{T_i} \right )^{-1}.
\end{align}
However, tidal terms associated with no dissipation contain $\kam$ separately, and they give rise to apsidal motion. In isolated binaries, such motion is immaterial. However, this is no longer the case in multiple systems with more than two stars. The apsidal motion constant is not provided separately by \sse~(\S~\ref{sect:stellar}). Therefore, in \mse, we implemented a stand-alone calculation of $\kam$, which is discussed in \S~\ref{sect:stellar:apsmot}. 

For completeness, we briefly describe the prescription of \bsepaper~for $\kam/T_i$. It makes a distinction between cases when tidal damping is dominated by radiative, convective, or degenerate regions.  
Radiative damping is assumed if $k_i=1$ and $m_i > 1.2\,\msun$, or if $k_i \in \{4,7\}$ (see Table~\ref{table:st} for the stellar types). Otherwise, convective damping is assumed if $k_i < 10$. If neither radiative or convective damping applies, we assume degenerate damping. 

For radiative damping, the model of \citet{1975A&A....41..329Z,1977A&A....57..383Z} for dynamical tides is adopted, i.e.,
\begin{align}
\left (\frac{\kam}{T_i} \right )_{\mathrm{rad}} = E_2 \left (1 + \frac{m_{\comp,\,i}}{m_i} \right )^{5/6} \rstar \sqrt{\frac{\gconst m_i}{a_k^5}},
\end{align}
where $E_2$ is given by
\begin{align}
E_2 = 1.592 \times 10^{-9} \left ( \frac{m_i}{\msun} \right )^{2.84}.
\end{align}

Convective damping is described by adopting a modified model of \citet{1996ApJ...470.1187R}, i.e.,
\begin{align}
\left (\frac{\kam}{T_i} \right )_\conv = \frac{2}{21} \frac{f_\conv}{\tau_\conv} \frac{\menv}{m_i}.
\end{align}
Here, the convective eddy turnover timescale is given by
\begin{align}
\tau_\conv \equiv \left [ \frac{\menv \, \renv \,(\rstar - \frac{1}{2} \renv)}{3 L_i} \right ]^{1/3},
\end{align}
and
\begin{align}
f_\conv = \mathrm{min} \left [1,  \left (\frac{P_{\mathrm{tid}}}{2 \tau_\conv} \right )^2 \right ],
\end{align}
with the tidal forcing frequency given by
\begin{align}
\frac{1}{P_{\mathrm{tid}}} = \left | \frac{1}{P_{\orb,\,k}} - \frac{\Omega_i}{2\pi} \right |.
\end{align} 

Lastly, for degenerate stars, $\kam/T_i$ is estimated based on the calculations of \citet{1984MNRAS.207..433C}, i.e.,
\begin{align}
\left (\frac{\kam}{T_i} \right )_{\mathrm{deg}} = 2.564 \times 10^{-8} \, r^2_{\mathrm{g},i} \, \left ( \frac{L_i}{\lsun} \frac{\msun}{m_i} \right )^{5/7} \, \yr^{-1}.
\end{align}

The details of tidal evolution are still highly uncertain (see, e.g., \citealt{2014ARA&A..52..171O} for a review). Here, we adopt the equilibrium tide model, whereas this could break down at high eccentricities (see, e.g., \citealt{2018ApJ...854...44M,2020MNRAS.496.3767V}). Furthermore, the prescription of \bsepaper~gives only a rough estimate of the efficiency of tidal dissipation in convective envelopes; it can be incorrect by at least one to several orders of magnitude \citep{2017ApJ...835..209N}. However, the prescription does provide an estimate for all stages of stellar evolution so is better than the alternative of naively assuming a fixed viscous timescale. For example, the efficiency of tidal dissipation is much larger during the giant stages, and this effect is taken into account with the prescription of \bsepaper. 

\subsubsection{Post-Newtonian terms}
\label{sect:dyn:sec:PN}
When in the secular integration mode, post-Newtonian (PN) terms are included to the 1PN and 2.5PN orders in all orbits in the system (i.e., including PN expansion terms proportional to $1/c^2$ and $1/c^5$, respectively). We ignore PN `interaction' terms that can arise between different orbits (e.g., \citealt{2013ApJ...773..187N,2014CQGra..31x4001W,2020PhRvD.102f4033L}). 

The 1PN terms give rise to apsidal motion, described for an orbit $k$ according to \citep{1972gcpa.book.....W}
\begin{align}
\label{eq:1PN}
\left (\frac{\mathrm{d} \ve{e}_k}{\mathrm{d} t} \right )_{1\mathrm{PN}} = 3 \left ( \frac{\gconst M_k}{a_k^3} \right )^{1/2} \frac{\gconst M_k}{c^2 a_k} \left (1-e_k^2 \right )^{-1} \unit{h}_k \times \ve{e}_k.
\end{align}
The 2.5PN describe orbital shrinkage due to GW emission; for an orbit $k$ \citep{1964PhRv..136.1224P},
\begin{subequations}
\label{eq:25PN}
\begin{align}
\left (\frac{\mathrm{d} \ve{e}_k}{\mathrm{d} t} \right )_{2.5\mathrm{PN}} &= -\frac{304}{15} e_k \frac{\gconst^3 M_{k.\mathrm{C}1} M_{k.\mathrm{C}2} M_k}{c^5 a_k^4 \left(1-e_k^2 \right )^{5/2}} \left (1 + \frac{121}{304} e_k^2 \right ) \, \unit{e}_k; \\
\left (\frac{\mathrm{d} \ve{h}_k}{\mathrm{d} t} \right )_{2.5\mathrm{PN}} &= -\frac{32}{5} \frac{\gconst^{7/2} M^2_{k.\mathrm{C}1} M^2_{k.\mathrm{C}2}}{c^5 a_k^{7/2} \left(1-e_k^2 \right )^{2}} \left (1 + \frac{7}{8} e_k^2 \right )\, \unit{h}_k.
\end{align}
\end{subequations}

We also include the lowest-order spin-orbit coupling terms describing precession of the spins around the orbit. Specifically, for a body $i$ in orbit $k$ \citep{1975PhRvD..12..329B},
\begin{align}
\label{eq:so}
    \frac{\mathrm{d}\unit{\Omega}_i}{\mathrm{d} t } = \frac{2 \gconst }{c^2 a_k^3 \left(1-e_k^2\right)^{3/2}} \left ( 1 + \frac{3}{4} \frac{m_{\comp,\,i}}{m_i} \right ) \, \ve{h}_k \times \unit{\Omega}_i.
\end{align}
Note that the magnitude of the spin, $\Omega_i$, is unaffected by the 1PN spin-orbit terms. Due to PN spin-orbit coupling, the orbit $\ve{h}_k$ also precesses around the spins; however, the latter effect is negligible if $S_i \ll h_k$, where $S_i$ is the spin angular momentum of body $i$. The latter is satisfied in roughly equal mass-ratio systems. We also ignore general relativistic spin-spin coupling during secular integration.

\subsection{Direct $N$-body integration}
\label{sect:dyn:Nbody}
\subsubsection{Direct $N$-body integrator}
\label{sect:dyn:Nbody:mstar}
As mentioned in \S~\ref{sect:dyn:sec}, the secular approximation can break down due to evolutionary processes. To handle such cases, \mse~also implements direct integration of the Newtonian equations of motion,
\begin{align}
\frac{\d^2 \ve{R}_i}{\d t^2} =  -\gconst \sum_{\substack{j=1 \\ j\neq i}}^{N} m_j \frac{ \ve{R}_i - \ve{R}_j}{\left |\left | \ve{R}_i - \ve{R}_j \right |\right |^3},
\end{align}
by linking it to the \textsc{MSTAR} code \citep{2020MNRAS.492.4131R}. \textsc{MSTAR} uses algorithmic chain regularisation \citep{2006MNRAS.372..219M,2008AJ....135.2398M,2010CeMDA.106..143H} with minimum spanning tree coordinates, allowing for highly accurate integration for arbitrary mass ratios. The minimum spanning tree, treated as a branching chain of inter-particle vectors, is constructed by finding the shortest inter-particle coordinate vector, and successively finding nearest neighbours until all particles are included in the minimum spanning tree. For $N\leq 3$ particles, the minimum spanning tree in \textsc{MSTAR} is identical to a (non-branching) chain used by the \textsc{AR-CHAIN} code \citep{2006MNRAS.372..219M,2008AJ....135.2398M,2010CeMDA.106..143H}. The Logarithmic Hamiltonian method \citep{1999MNRAS.310..745M,1999AJ....118.2532P} is used to transform time to a the fictitious time, $s$.

The equations of motion for the chained particle coordinates read
\begin{subequations}
\begin{align}
\frac{\d t}{\d s} &= \frac{1}{T + B}; \\
\frac{\d \tilde{\ve{R}}_k}{\d s} &= \frac{1}{T + B} \tilde{\ve{V}}_k,
\end{align}
\end{subequations}
and the velocity equations are given by
\begin{subequations}
\begin{align}
\frac{\d \tilde{\ve{V}}_k}{\d s} &= \frac{1}{U} \left ( \tilde{\ve{A}}_k + \tilde{\ve{G}}_k \right ); \\
\frac{\d B}{\d s} &= -\frac{1}{U} \sum_{i=1}^{N} m_i \ve{V}_i \cdot \ve{g}_i.
\end{align}
\end{subequations}
Here, tildes indicate chained coordinates (for example, when $k$ indicates a particle's index within the chain, $\tilde{\ve{R}}_k = \ve{R}_{k_{j}} - \ve{R}_{k_{i}}$). $T$ is the (Newtonian) kinetic energy, $B=-H$, where $H$ is the (Newtonian) Hamiltonian, and $U$ is the negative of the (Newtonian) potential energy. The vector $\tilde{\ve{A}}_k$ denotes the chained Newtonian acceleration vector. Lastly, the vector $\ve{g}_i$ denotes additional velocity-dependent perturbations on particle $i$, and $\tilde{\ve{G}}_k$ denotes its chained version. See Appendices 1 and 2 of \citet{2017ApJ...840...53R} for a more general derivation of algorithmic regularisation and detailed implementation instructions of the chained Logarithmic Hamiltonian method.

The \textsc{MSTAR} integrator uses the Gragg–Bulirsch–Stoer (GBS) extrapolation method \citep{1965SJNA....2..384G,bulirsch_stoer_1966} to determine the substeps within the global $N$-body timestep, in order to achieve high numerical accuracy. The default GBS parameter in \mse~(which is user adjustable) is $10^{-10}$. The integrator will not finish before the required tolerance is achieved, where the default time tolerance parameter in \mse~(also user adjustable) is $10^{-6}$. In \mse, the serial version of \textsc{MSTAR} is implemented, since the number of bodies is typically too low for parallelisation to be beneficial (e.g., \citealt{2007NewA...12..641P}).

The \textsc{MSTAR} code includes pairwise PN terms to the 1PN, 2PN, 2.5PN, 3PN, and 3.5PN orders (e.g., \citealt{2004PhRvD..69j4021M,2006LRR.....9....3W}), as well as spin-orbit, spin-spin, and quadrupole terms \citep{1975PhRvD..12..329B,1995PhRvD..52..821K}. Tidal evolution is, however, not currently implemented in the direct $N$-body code. The version of \textsc{MSTAR} used in \mse~also includes collision detection. For more details on \textsc{MSTAR}, we refer to \citet{2020MNRAS.492.4131R}. 

\subsubsection{Switch to $N$-body}
\label{sect:dyn:Nbody:to}
In \mse, we invoke direct integration, i.e., from secular to direct, in the following cases.
\begin{enumerate}[leftmargin=0.5cm]
\item The system becomes dynamically unstable according to the analytic stability criterion of \citet{2001MNRAS.321..398M}, i.e., 
\begin{align}
\label{eq:dynstab}
\frac{a_\mathrm{out}(1-e_\mathrm{out})}{a_\mathrm{in}} > 2.8 \, \left [ (1+q_\mathrm{out}) \frac{1+e_\mathrm{out}}{\sqrt{1-e_\mathrm{out}}} \right ]^{2/5} \, \left (1-0.3\, \frac{\Phi}{\pi} \right ).
\end{align}
This criterion is applied to any pair of orbits in the system; the subscripts `in' and `out' refer to the inner and outer orbits for such a pair. The mass ratio $q_\mathrm{out}$ is defined here as $q_\mathrm{out}\equiv (M_\mathrm{out}-M_\mathrm{in})/M_\mathrm{in}$, where $M_\mathrm{in}$ is the mass of all bodies contained within the inner orbit, and $M_\mathrm{out}$ is the mass of all bodies contained within the outer orbit (including those in the inner orbit). The angle $\Phi$ is the mutual inclination between the pair of orbits (expressed in radians). 

The criterion of \citet{2001MNRAS.321..398M} applies strictly only to hierarchical triple systems, with masses that are not too unequal (see, e.g., \citealt{2015ApJ...808..120P} for an investigation of the criterion in a situation with disparate masses). Our use of the criterion for any hierarchical system is an extrapolation, and may not be accurate in all cases. However, to our knowledge, no generalised stability criterion exists for an arbitrary number of bodies. Moreover, the criterion of \citet{2001MNRAS.321..398M} tends to be conservative (e.g., \citealt{2018MNRAS.474...20H}), and once the criterion is met, we switch to direct $N$-body integration. 
\item The system enters the `semisecular regime', defined according to 
\begin{align}
t_{\jmath_k} \equiv \left | \frac{1}{\jmath_k} \left ( \frac{\mathrm{d} \jmath_k}{\mathrm{d} t} \right )_\mathrm{sec} \right |^{-1} < P_{\orb,\,k},
\end{align}
for any orbit $k$. Here, $(\mathrm{d} \jmath_k/\mathrm{d} t)_\mathrm{sec}$ is the time derivative of $\jmath_k$ according to the secular equations of motion (cf. \eq~\ref{eq:eom_sec}). In other words, the semisecular regime is entered when the timescale for angular-momentum change due to secular evolution is shorter than the orbital timescale; in the latter case, the orbit-averaging approximation likely breaks down \citep{2012ApJ...757...27A,2014ApJ...781...45A,2016MNRAS.458.3060L,2018MNRAS.481.4907G,2018MNRAS.481.4602L,2019MNRAS.490.4756L}. 

\item One or more of the orbits in the system becomes unbound following mass loss and/or kicks in SNe explosions, or recoil velocities in the case of mergers. 

\item Directly after CE evolution or direct collisions. In this case, the effect of mass loss on the orbits of other bodies in the system is taken into account during the $N$-body integration (see \S~\ref{sect:bin:mlorbit}).
\end{enumerate}

To switch to direct $N$-body integration, the position and velocity vectors of all bodies in the system (relative to an arbitrary inertial reference frame) are computed based on the current orbital configuration. The orbital phases are fundamentally not modelled in the secular approach; here, we assume that the mean anomalies of all orbits during secular integration evolve linearly with time. The masses and radii are assumed to be constant during the $N$-body evolution, which we justify by limiting the $N$-body timestep by the stellar evolution timestep; the latter is set such that the masses and radii do not change significantly. After the $N$-body integration, we update the stellar masses and radii according to \sse~(see \S~\ref{sect:stellar} below). 

The timestep of the $N$-body integration is initially determined by the prior evolution of the system (see \S~\ref{sect:alg}). When remaining in direct $N$-body integration mode in future steps (see \S~\ref{sect:dyn:Nbody:from}), the new timestep is determined in part by $N$-body evolution. In particular, the $N$-body timestep is given by
\begin{align}
\label{eq:deltatn}
\deltatn = \alpha_{N,\,\Delta t} \, \mathrm{max}_k \left ( P_{\orb,\,k} \right ),
\end{align}
where the maximum is taken over the orbital periods determined from the new system after $N$-body evolution (see \S~\ref{sect:dyn:Nbody:from}), and $\alpha_{N,\,\Delta t}$ is a tuning parameter. After dynamical instability, we set $\alpha_{N,\,\Delta t}=1.5$. After having entered the semisecular regime, we set $\alpha_{N,\,\Delta t}=10^2$. After an SNe event, we set $\alpha_{N,\,\Delta t}=1.5$. In all other cases, $\alpha_{N,\,\Delta t}=1.5$.

We check for physical collisions during the $N$-body integration (implemented assuming parabolic trajectory interpolation). When collisions occur, we handle them as described in \S~\ref{sect:bin:col}.

\subsubsection{Switch (back) to secular}
\label{sect:dyn:Nbody:from}
Once direct $N$-body integration has been invoked, the system might experience strong interactions such as collisions (\S~\ref{sect:bin:col}). In less extreme cases, for example after SNe explosions, some stars might become unbound from the multiple system, whereas the remaining stars could remain bound. Generally, after a strong interaction, one or more stable orbits could persist in the system,  possibly in a hierarchical configuration. In the latter case, it is computationally advantageous to switch back to secular integration. 

Therefore, at the end of each invocation of the $N$-body code, we evaluate the stability of the system, and switch back to secular integration if the system is deemed stable. We adopt the following strategy to analyse the state of the system after $N$-body integration (i.e., based on the positions and velocities), and to evaluate stability. First, from the $N$-body positions $\ve{R}_i$ and velocities $\ve{V}_i$, we compute the orbital elements for all pairs of bodies $(i,j)$ in the system, i.e.,
\begin{subequations}
\begin{align}
\ve{h}_{ij} &= \mu_{ij} \left (\ve{R}_i - \ve{R}_j \right ) \times \left ( \ve{V}_i - \ve{V}_j \right ); \\
\ve{e}_{ij} &= \frac{1}{\gconst(m_i+m_j)} \left ( \ve{V}_i - \ve{V}_j \right ) \times \frac{\ve{h}_{ij}}{\mu_{ij}} - \frac{\ve{R}_i - \ve{R}_j}{\left | \left | \ve{R}_i - \ve{R}_j \right | \right |},
\end{align}
\end{subequations}
from which
\begin{subequations}
\begin{align}
e_{ij} &= || \ve{e}_{ij}||; \\
a_{ij} &= \frac{|| \ve{h}_{ij}||^2 (m_i+m_j)}{\gconst m_i^2 m_j^2 (1-e_{ij}^2)}.
\end{align}
\end{subequations}
After computing the elements for all pairs of bodies, we also iteratively compute orbital elements between previously found orbits and bodies that have not been yet assigned an orbit (including orbit-orbit pairs), until no new pairs are found. For the resulting set of orbital elements, if the orbital elements correspond to a bound orbit ($a_{ij}>0$, and $0\leq e_{ij} < 1$), we consider the pair as a potential orbit. However, in a given hierarchical system, even if it is dynamically stable, not all pairs of bodies always correspond to physical orbits. Instead, some orbits identified in this way may be spurious. 

If the system is truly dynamically stable, then spurious orbits are temporary. Therefore, to evaluate which potential orbits identified above are physical orbits, we integrate the system for an additional short duration given by
\begin{align}
\deltatnan = \alpha_{N,\,\Delta t,\,\mathrm{an}} \, \deltat,
\end{align}
where $\alpha_{N,\,\Delta t,\,\mathrm{an}}$ is a dimensionless tuning parameter; we set $ \alpha_{N,\,\Delta t,\,\mathrm{an}}=0.05$ by default. After this second $N$-body integration, we re-evaluate the orbital elements for all pairs identified before. If the number of bound pairs found does not match the previous number, this suggests dynamical instability, and/or the presence of spurious orbits. We then continue future integration in direct $N$-body mode. If the number of bound pairs remains the same, then we compare the new semimajor axes ($a'_{ij}$) to the old ones ($a_{ij}$). Specifically, we consider a pair (of two bodies, one body and an orbit, or two orbits) to be stable if
\begin{align}
\frac{\left | a_{ij} - a'_{ij} \right | }{a_{ij}} < \alpha_{N,\,a},
\end{align}
where $\alpha_{N,\,a}$ is taken to be $\alpha_{N,\,a}=0.01$ by default. In other words, for stability, we require that the fractional change in the semimajor axis should not be larger than 1\%. We do not consider eccentricity changes in this evaluation during the stability check integration (i.e., the second $N$-body integration with a timestep of $\deltatnan$), since the eccentricity could change due to secular evolution in dynamically stable systems.  

If the system is deemed stable according to the above criterion for all orbits, {\it and} if the entire system is also stable according to the criterion of \citet{2001MNRAS.321..398M} (cf. \S~\ref{sect:dyn:Nbody:to}), then future evolution will be carried out using the secular integration method. A switch back to direct $N$-body integration at later times is always allowed.

\subsection{Unbound bodies}
\label{sect:dyn:unbound}
It is possible that, due to various processes (e.g., SNe kicks), bodies become unbound from the parent system. In direct integration mode, the positions and velocities for unbound bodies are taken into account self-consistently (see also \S~\ref{sect:stellar:sse}). In secular integration mode, which only takes account the evolution of bound orbits, we update the positions and velocities of all unbound bodies, as described in the following.

In secular integration mode, we assume that the velocities of unbound bodies are affected by wind mass loss only. Linear momentum conservation implies an acceleration on body $i$ given by
\begin{align}
\label{eq:unbound_acc}
\dot{\ve{V}}_i = - \frac{\dot{m}_i}{m_i} \ve{V}_i,
\end{align}
where $\dot{m}_i$ is the time derivative due to wind mass loss only ($\dot{m}_i\leq0$). Assuming a constant $\dot{m}_i$ during a secular timestep $\deltatode$, 
\begin{align}
\label{eq:unbound_m}
m_i = m_{i,\,0} + \dot{m}_i \, \deltatode,
\end{align}
where $m_{i,\,0}$ is the mass at the beginning of the timestep. \Eqs~(\ref{eq:unbound_acc}) and (\ref{eq:unbound_m}) then imply that the velocity after a timestep $\deltatode$ is given by
\begin{align}
\label{eq:unbound_V}
\ve{V}_i = \ve{V}_{i,\,0} \left (1 + \frac{\dot{m}_i}{m_{i,\,0}} \deltatode \right )^{-1},
\end{align}
where $\ve{V}_{i,\,0}$ is the velocity at the beginning of the timestep. The position after $\deltatode$ is therefore given by
\begin{align}
\label{eq:unbound_R}
\ve{R}_i = \ve{R}_{i,\,0} + \ve{V}_{i,\,0} \frac{m_{i,\,0}}{\dot{m}_i} \ln \left (1 + \frac{\dot{m}_i}{m_{i,\,0}} \deltatode \right ),
\end{align}
where $\ve{R}_{i,\,0}$ is the position vector at the beginning of the timestep. In the limit that $\dot{m}_i/m_{i,\,0} \rightarrow 0$ (no wind mass loss), these expressions reduce to
\begin{subequations}
\begin{align}
\ve{V}_i &= \ve{V}_{i,\,0}; \\
\ve{R}_i &= \ve{R}_{i,\,0} + \ve{V}_{i,\,0} \, \deltatode.
\end{align}
\end{subequations}

\section{Stellar evolution}
\label{sect:stellar}

\subsection{\sse}
\label{sect:stellar:sse}
The evolution of single stars in \mse~is modelled by linking the code to \sse~(\ssepaper). The latter is a \textsc{Fortran} code that implements comprehensive analytic fit formulae for the radii, luminosities,  core masses and core radii, spin frequencies, and gyration radii, all as a function of mass, age, and metallicity, from zero-age MS (ZAMS) up to and including remnant stages. The fits are based on the detailed stellar models of \citet{1998MNRAS.298..525P}, which consisted of a grid of tracks for masses between 0.5 and 50 $\msun$, and seven different metallicities between $10^{-4}$ and 0.03. 

\sse~includes mass loss from stellar winds using a number of prescriptions for the mass loss rate. Also included is spin-down due to magnetic breaking in stars with appreciable convective envelopes. In \mse, by default we initialise the spins to be parallel with the orbit (i.e., the initial obliquity $\theta_{\mathrm{s},\,i}=0^\circ$). Furthermore, by default, the initial spin frequency $\Omega_i$ of all stars is taken to be consistent with \sse, i.e., $\Omega_i$ is determined from \ssepaper's fit to data of the equatorial speed of MS stars of \citet{1992adps.book.....L}, given explicitly by
\begin{align}
\label{eq:init_spin}
v_{\mathrm{rot},\,i} = 330 \, \kms \, \left (\frac{m_i}{\msun} \right )^{3.3} \left [15 + \left (\frac{m_i}{\msun} \right )^{3.45} \right ]^{-1}.
\end{align}
The spin frequency $\Omega_i$ is then given by $\Omega_i = v_{\mathrm{rot},\,i}/R_i$. 

In \mse, each star is evolved with \sse~for a timestep that is at least as short as the shortest stellar evolution timestep for any star ($\Delta t_{\sse,\,i}$), where $\Delta t_{\sse,\,i}$ is obtained from the timestep function of \sse~(cf. \ssepaper, \S~8). The latter gives the timescale on which the mass loss due to stellar winds is less than 1\%, and the change in radius is less than 10\%. 

After evolving each star for a short timestep imposed by \mse, most stellar properties are updated immediately. The dynamical evolution is handled after stellar evolution (see \S~\ref{sect:alg}), and these parameters are assumed to be constant during the dynamical evolution (with either the secular or direct methods). However, there are some evolutionary phases in which the orbital dynamics are very sensitive to some stellar parameters. In particular, this can occur when stars evolve to become RGB or AGB stars --- the radii then change rapidly and the stars develop deep convective envelopes, strongly increasing the efficiency of tidal dissipation. This can lead to difficulties in the numerical integration of the secular ODEs. 

Testing has shown that these numerical difficulties can be overcome by assuming that the masses and radii vary linearly with time, rather than being constant. We therefore treat $m_i$, $\rstar$, and the spin frequency $\Omega_i$ in the secular integration as ODE variables, with $\dot{m}_i$, $\dot{R}_{\star,\,i}$, and $\dot{\Omega}_i$ constant and determined in part by \sse. Note that the masses and spins can also be affected by mass transfer and/or wind accretion. Specifically, during the ODE integration, we set
\begin{subequations}
\label{eq:rmdot}
\begin{align}
\dot{R}_{\star,\,i} &= \dot{R}_{\star,\,i,\,\sse}; \\
\dot{m}_i &= \dot{m}_{i,\,\sse} + \dot{m}_{i,\,\mathrm{MT}} + \dot{m}_{i,\,\mathrm{WA}}; \\
\dot{\Omega}_i &= \dot{\Omega}_{i,\,\sse} + \dot{\Omega}_{i,\,\mathrm{MT}}. 
\end{align}
\end{subequations}
Here, $\dot{R}_{\star,\,i,\,\sse}$, $\dot{m}_{i,\,\sse}$, and $\dot{\Omega}_{i,\,\sse}$ are the radius, mass, and spin frequency time derivatives from \sse, respectively, computed from the differences in radii, masses, or spin frequencies between the new and old steps and the timestep. Note that $\dot{m}_{i,\,\sse}$ takes into account wind mass loss only. The other mass derivatives, $\dot{m}_{i,\,\mathrm{MT}}$ and $\dot{m}_{i,\,\mathrm{WA}}$, denote mass changes due to mass transfer (MT) and wind accretion (WA) respectively. The latter are discussed in more detail in \Ss~\ref{sect:bin:mt} and \ref{sect:bin:wa}, respectively. The spin frequency rate of change from \sse, $\dot{\Omega}_{i,\,\sse}$, takes into account stellar spin slowdown due to wind mass loss, and magnetic breaking. Changes in the spins due to mass loss/accretion are encapsulated in the term $\dot{\Omega}_{i,\,\mathrm{MT}}$, and are discussed in \S~\ref{sect:bin:mt:stable:spin}.

Tides are currently not implemented in the $N$-body code. The masses, radii, and spins are assumed to be constant in direct $N$-body mode (an exception to this, where the masses do vary, occurs when using direct $N$-body integration to evaluate the effects of mass loss on external orbits, see \S~\ref{sect:bin:mlorbit}). They are updated after the integration according to \eq~(\ref{eq:rmdot}) but without the MT and WA terms, since we do not take into account these binary interactions while in $N$-body integration mode (see \S~\ref{sect:bin}). 

Another exception to \eq~(\ref{eq:rmdot}) occurs when stars evolve to become an NS or BH. In this case, the mass and radius changes are usually large, and tidal evolution is not important. In this instance, the masses and radii are updated immediately after stellar evolution. Also, we take into account the effects of the fast mass loss and possible SNe kicks on the system (see \S~\ref{sect:stellar:sne}). 

\begin{table*}
\begin{center}
\begin{tabular}{llp{3cm}p{6cm}l}
\toprule 
Source of mass change(s) & Interaction type & Specification & Effect on orbit(s) & \S(s) \\
\midrule
Wind mass loss & Single star & & Adiabatic for all parent orbits. & \ref{sect:stellar:wind} \\
Wind mass accretion & Star--star & &Adiabatic for all parent orbits. & \ref{sect:bin:wa} \\
Wind mass accretion & Star--binary & &Not affected. & \ref{sect:bin:wa} \\
SNe event & Single star & & Instantaneous. & \ref{sect:stellar:sne} \\
Mass loss in collision product & Star--star & Parent orbits & Instantaneous, including possible SNe kicks. & \ref{sect:bin:col} \\
RLOF & Star--star & Transferred mass & \textsc{emt} model. & \ref{sect:bin:mt:stable:orb} \\
RLOF & Star--star & Mass not transferred to companion & Adiabatic wind from accretor. & \ref{sect:bin:mt:stable:orb} \\
RLOF (dynamical) & Star--star & Inner orbits & According to prescription. & \ref{sect:bin:mt:dynlow} \& \ref{sect:bin:mt:dynwd} \\
RLOF (dynamical) & Star--star & Parent orbits & According to short-term direct $N$-body integration with mass loss. & \ref{sect:bin:mlorbit} \\
RLOF & Star--binary & Outer orbit & Non-conservative mass transfer in circular orbit. & \ref{sect:triple:mt:out} \\
RLOF & Star--binary & Inner orbit & CE-like behaviour. & \ref{sect:triple:mt:in} \\
RLOF & Star--binary & Orbits exterior to triple (sub)system & Adiabatic for all parent orbits. & \ref{sect:triple:mt:out} \\
CE & Star--star & Subject orbit & According to CE prescription. & \ref{sect:bin:ce:energy} \\
CE & Star--star & Parent orbits & According to short-term direct $N$-body integration with mass loss. & \ref{sect:bin:mlorbit} \\
CE & Star--binary & Inner orbit & Not affected. & \ref{sect:triple:ce} \\
CE & Star--binary & Outer orbit & According to modified CE prescription. & \ref{sect:triple:ce} \\
CE & Star--binary & Orbits exterior to triple (sub)system & According to short-term direct $N$-body integration with mass loss. & \ref{sect:bin:mlorbit} \\
\bottomrule
\end{tabular}
\end{center}
\caption{\small Overview of the different mass loss mechanisms in \mse, and the associated assumptions on how the mass loss affects the orbits in the system. }
\label{table:mass_loss}
\end{table*}

\subsection{Orbital response to wind mass loss}
\label{sect:stellar:wind}
We assume that mass lost in a stellar wind from any star in the system is lost adiabatically, i.e., the wind mass loss timescale is always much longer than the orbital timescale. This implies that the semimajor axis of an orbit $k$ changes according to 
\begin{align}
\label{eq:adotwind}
\dot{a}_k = - a_k \,\frac{\dot{M}_{k,\,\wind}}{M_i},
\end{align}
where $\dot{M}_{k,\,\wind}$ is the combined wind mass loss rate of all stars contained within orbit $i$ including the amount accreted due to wind accretion (see \S~\ref{sect:bin:wa}), and $M_i$ is the total mass contained within orbit $i$. Note that, in the absence of wind accretion, $\dot{M}_{k,\,\wind}<0$, i.e., orbits always expand in respond to wind mass loss only, within our approximation. Conversely, accretion of wind from companions could lead to a net positive $\dot{m}_{i,\,\wind}$ for an individual star. Wind accretion is discussed further in \S~\ref{sect:bin:wa}.

Other sources of mass loss or mass accretion in the system (for example, mass loss in SNe events, or mass transfer) can affect the orbits in the system in \mse~in various ways. These other processes are described in the sections below, and an overview of them is given in Table~\ref{table:mass_loss}.

\subsection{Supernovae}
\label{sect:stellar:sne}
\subsubsection{General}
\label{sect:stellar:sne:gen}
When a star evolves to become an NS or BH, the remnant mass and radius, as given by \sse, are immediately updated in \mse. We take into account the effects of mass loss and possible kicks on the orbits in the system by using the routines described in \citet{2018MNRAS.476.4139H}, which are part of \textsc{SecularMultiple}. In short, a realization of the positions and velocities of all bodies in the system is made based on the pre-SNe masses. The mass of the star undergoing the SNe event is then updated, and a kick velocity is added, if applicable (see \S~\ref{sect:stellar:sne:kick}). The new orbits are then determined, assuming that the structure of the system has not changed. With this approach, we assume that the mass from the SNe is lost from the system effectively instantaneously, and that it does not interact with any other bodies. 

If the new system contains one or more unbound orbits ($e_i<0$ or $e_i\geq1$ for any orbit $i$), we switch to direct $N$-body integration (cf. \S~\ref{sect:dyn:Nbody:to}). In higher-order multiplicity systems, mass loss and/or kicks could trigger interesting dynamical interactions and possibly lead to collisions (e.g., \citealt{2012ApJ...760...99P,2020ApJ...895L..15F}).

\subsubsection{Kicks}
\label{sect:stellar:sne:kick}
Based on the typically high space velocities of young pulsars, NSs are thought to receive natal kicks at their birth (e.g., \citealt{2005MNRAS.360..974H}). BHs may also receive natal kicks, although the latter are poorly constrained observationally (e.g., \citealt{2015MNRAS.453.3341R,2016MNRAS.456..578M}). There exists some observational evidence for small natal kicks of WDs based on wide WD+MS and WD+WD binaries in the field \citep{2018MNRAS.480.4884E}. Theoretically, NS and BH kicks are highly uncertain (see, e.g., \citealt{2012ARNPS..62..407J} for a review). WD kicks, if existent, are likely very small (e.g., \citealt{2003ApJ...595L..53F,2012ApJ...761L..23J,2017Sci...357..680V}).

Given the large uncertainties in the natal kick velocities of BHs and NSs, we adopt in \mse~different models for their distributions (currently, we do not take into account natal WD kicks). In all cases, we assign a random direction of the kick on the unit sphere; the magnitude of the kick velocity, $V_\kick$, is sampled from an assumed distribution. The models for $V_\kick$ that are currently implemented in \mse~(more can be added in the future) are described below. Here, we use $m_\prog$ to denote the mass of the progenitor star (just before becoming an NS or BH), $m_\CO$ the mass of the CO core of the progenitor, and $m_\remnant$ the remnant mass. These quantities are extracted from the \sse~code.

\paragraph{Kick distribution model 1}
\label{sect:stellar:sne:kick:m1}
In model 1, we adopt Maxwellian distributions for both NS and BH kicks, i.e.,
\begin{align}
\label{eq:kickmax}
\frac{\mathrm{d} N}{\mathrm{d} V_\kick} = \sqrt{ \frac{2}{\pi}} \frac{V_\kick^2}{\sigma_\kick^3} \exp \left ( - \frac{V_\kick^2}{2\sigma_\kick^2} \right ).
\end{align}
Here, $\sigma_\kick$ is user-adjustable separately for NSs (default value $\sigma_{\kick,\,\NS} = 265\,\kms$, \citealt{2005MNRAS.360..974H}) and BHs (default value $\sigma_{\kick,\,\BH} = 50\,\kms$). 

\paragraph{Kick distribution model 2}
\label{sect:stellar:sne:kick:m2}
In model 2, the kick distribution for NSs is the same as in model 1 (Maxwellian with $\sigma_\kick = \sigma_{\kick,\,\NS}$). For BHs, the kick is sampled from the Maxwellian NS kick distribution, but scaled back such that the linear momentum of the BH would be the same as for an NS, i.e.,
\begin{align}
V_{\kick,\,\BH} = V_{\kick,\,\NS} \frac{ m_\NS}{m_\BH},
\end{align}
where $m_\NS$ is an adjustable parameter (by default, $m_\NS = 1.4\,\msun$), and $m_\BH = m_\remnant$ is the remnant mass of the BH. 

\paragraph{Kick distribution model 3}
\label{sect:stellar:sne:kick:m3}
In model 3, we adopt one of the prescriptions of \citet{2012ApJ...749...91F} in which a kick speed is sampled from \Eq~(\ref{eq:kickmax}), and scaled down according to the amount of mass expected to fall back onto the compact object. In particular,
\begin{align}
V_{\kick} = V_{\kick,\,\mathrm{Max}} (1 - f_{\mathrm{fb}}),
\end{align}
where $V_{\kick,\,\mathrm{Max}}$ is a speed sampled from the distribution in \Eq~(\ref{eq:kickmax}), and the fallback fraction $f_{\mathrm{fb}}$ depends on the progenitor's CO core mass ($m_\CO$):
\begin{align}
f_{\mathrm{fb}} = \left \{ 
\begin{array}{cc}
0, & m_\CO < 5 \, \msun; \\
0.378 \, (m_\CO/\msun) - 1.889, & 5 \,\msun < m_\CO < 7.6 \,\msun; \\
1, & m_\CO \geq 7.6 \, \msun.
\end{array} \right.
\end{align}

\paragraph{Kick distribution model 4}
\label{sect:stellar:sne:kick:m4}
In model 4, we adopt one of the prescriptions of \citet{2020ApJ...891..141G}, in which a kick speed is sampled from \Eq~(\ref{eq:kickmax}) and scaled down according to 
\begin{align}
V_{\kick} = V_{\kick,\,\mathrm{Max}} \frac{m_\prog - m_\remnant}{m_\remnant} \frac{\langle m_\NS \rangle}{\langle m_{\mathrm{ej}} \rangle},
\end{align}
where $\langle m_\NS \rangle = 1.2\,\msun$, and $\langle m_{\mathrm{ej}} \rangle = 9\,\msun$ (by default). 

\paragraph{Kick distribution model 5}
\label{sect:stellar:sne:kick:m5}
In model 5, we adopt the prescription of \citet{2020MNRAS.499.3214M} in which the kick speed is sampled from a normal distribution centred at $\mu_\kick$,
\begin{align}
\frac{\mathrm{d} N}{\mathrm{d} V_\kick} = \sqrt{\frac{2}{\pi}} \frac{1}{\sigma_\kick} \exp \left [ - \frac{ \left (V_\kick - \mu_\kick\right)^2}{2\sigma_\kick^2} \right ].
\end{align}
Here, $\mu_\kick = v_\NS (m_\CO-m_\remnant)/m_\remnant$ for NS, and $\mu_\kick = v_\BH (m_\CO-m_\remnant)/m_\remnant$ for BHs, and with $v_\NS = 400\,\kms$ and $v_\BH = 200\,\kms$ by default. The width of the normal distribution is given by $\sigma_\kick = \sigma_\kick' \mu_\kick$, where $\sigma_\kick'=0.3$ by default. 

\begin{figure*}
\center
\includegraphics[width=1\textwidth,trim=10mm 15mm 10mm 0mm]{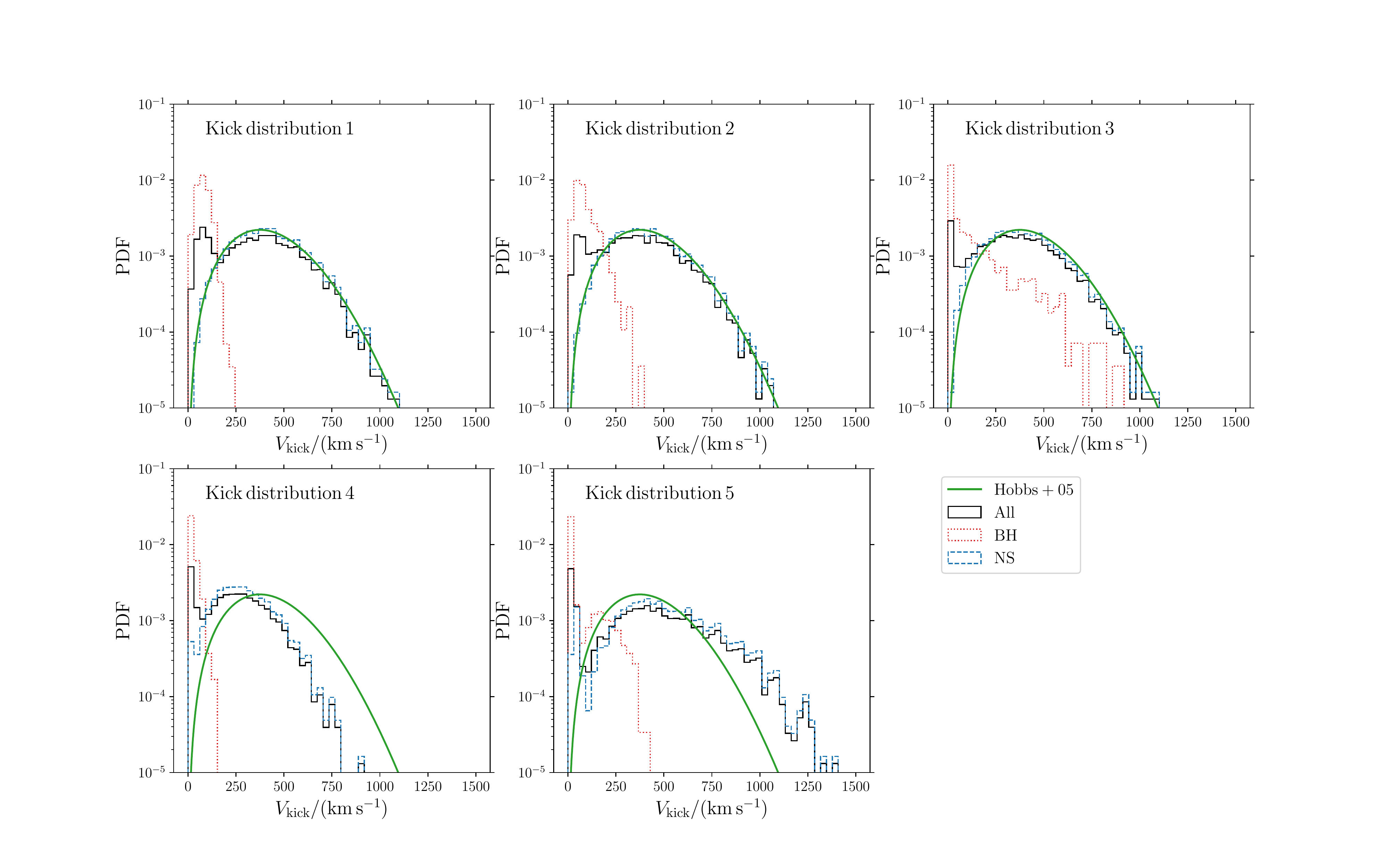}
\caption{Distributions of the SNe kick speed according to the five prescriptions described in \S~\ref{sect:stellar:sne:kick}. These distributions are generated by sampling initial masses from a \citet{1993MNRAS.262..545K} distribution, and evolving the stars until becoming a compact object (assuming Solar metallicity). Probability density functions (PDFs) are shown for all compact objects (solid black lines), and separately for BHs (red dotted lines), and NSs (blue dashed lines). The solid green line in each panel shows a Maxwellian distribution (cf. \eq~\ref{eq:kickmax}) with $\sigma_\kick=265\,\kms$ \citep{2005MNRAS.360..974H}. }
\label{fig:kick}
\end{figure*}

In \F~\ref{fig:kick}, we compare the different kick distributions described above. The distributions are generated by sampling an initial stellar mass, $m_{i,\,\init}$, between 8 and 100 $\msun$ from a Kroupa mass distribution ($\d N/\d m_{i,\,\init} \propto m_{i,\,\init}^{-2.7}$ in this mass range; \citealt{1993MNRAS.262..545K}). A star with the sampled initial mass is then evolved (assuming a metallicity $Z_i=0.02$) until becoming a compact object and a kick velocity is sampled. In \F~\ref{fig:kick}, we show the kick distributions for all compact objects (solid black lines), and separately for BHs (red dotted lines), and NSs (blue dashed lines). 

Kick distribution 1 shows a peak around 50 $\kms$ as a result of the assumed distribution for BHs; NS kicks cause a wider Maxwellian tail extending to $\sim 1000\,\kms$. Overall, the kick distributions for NSs are not very much different between the various models (with model 4 having somewhat smaller kicks, and model 5 somewhat larger ones). The largest differences arise in the BH kick distributions. In particular, model 3 has significantly larger BH kicks compared to other prescriptions. The smallest BH kicks occur in model 4.

\subsection{Apsidal motion constant}
\label{sect:stellar:apsmot}
As described in \S~\ref{sect:dyn:sec:tide}, the tidal dissipation prescription of \bsepaper~does not give the apsidal motion constant $\kam$ separately (only the combination $\kam/T_i$), whereas $\kam$ is needed to describe tidal evolution in systems with more than two stars. In addition, \sse~does not include fits to $\kam$. In \mse, we therefore implement a separate calculation of $\kam$ as a function of mass, stellar type, and age. 

Specifically, for low-mass MS stars, we approximate the star as being fully convective, such that an $n=3/2$ polytrope is appropriate, and $\kam \simeq 0.1433$ \citep{1955MNRAS.115..101B}. For higher-mass MS stars and other stars up to and including giant stars, $1\leq k_i \leq 6$, we compute $\kam$ as a function of mass, stellar type, and age using linear fits obtained from the detailed stellar evolution models of \citet{2004A&A...424..919C}. We note that the latter models assumed $Z_i=0.02$, whereas \mse~allows for a larger range of metallicities (i.e., the same range as \sse). Here, we neglect the dependence of $\kam$ on metallicity, and assume $Z_i=0.02$ when determining $\kam$. For stripped He stars, $7 \leq k_i \leq 9$, and WDs, $10 \leq k_i \leq 12$, we use analytic fit functions to the data of \citet{1977ApJ...213..464V} based on mass alone. For NSs, $k_i=13$, we assume an $n=1$ polytrope model such that $\kam \simeq 0.2560$ \citep{1955MNRAS.115..101B}. For BHs, we do not take into account tides, but $\kam$ is nevertheless set to $\kam=0$.

\section{Binary evolution}
\label{sect:bin}
In this section, we describe various binary evolutionary processes (i.e., involving star-star interactions) included in \mse. Most of these processes are modelled only in the secular integration mode, when the system is assumed to be dynamically stable. Exceptions to this are collisions in direct $N$-body mode, which could trigger CE evolution (\S~\ref{sect:bin:col}). Many of the assumptions and prescriptions made regarding binary evolution are adopted directly from \bse~(a notable exception lies in the treatment of stable mass transfer in eccentric orbits). However, in order to provide here a more self-contained description, we repeat some material from \bsepaper. 

\subsection{Mass transfer}
\label{sect:bin:mt}
When integrating secularly, we check (as a stopping condition in the ODE integration) for RLOF of one star $i$ in an orbit $k$ at periapsis to a companion (the latter can be a single star, or a binary). The criterion used is
\begin{align}
\label{eq:mt_crit}
\rstar \geq R_{\mathrm{L},\,k} (r_{\mathrm{p},\,k}) = R_{\mathrm{L},\,k} (a_k[1-e_k]),
\end{align}
with the instantaneous Roche lobe radius
\begin{align}
\label{eq:rlof}
R_{\mathrm{L},\,k}(r_k) = r_k \frac{0.49 \, q_i^{2/3}}{0.6 \, q_i^{2/3} + \ln\left (1 + q_i^{1/3} \right )}.
\end{align}
Here, $r_k$ is relative orbital separation of star $i$'s orbit, $a_k$ and $e_k$ refer to the orbital semimajor axis and eccentricity, respectively, and the mass ratio $q_i$ is defined as $q_i = m_i/m_{\comp,\,i}$, where $m_{\comp,\,i}$ is the mass of the companion to star $i$ (the companion can be a star or a binary, and the latter could contain multiple components). The Roche lobe radius used in \eq~(\ref{eq:rlof}) is adopted from \citet{1983ApJ...268..368E}, replacing the semimajor axis with the instantaneous orbital separation. This is an approximation in multiple ways. First, \citet{2007ApJ...660.1624S} provided fits for corrections of the Roche Lobe radius in the binary-star case to the {\it ad hoc} \eq~(\ref{eq:rlof}) as a function of the mass ratio, spin frequency, eccentricity, and orbital phase. However, in modelling the long-term evolution due to stable mass transfer (\S~\ref{sect:bin:mt:stable}), we adopt the model of \emtpaper~in which, for practical reasons, \eq~(\ref{eq:rlof}) was used for the Roche lobe radius. Therefore, in \mse, we adopt \eq~(\ref{eq:rlof}) instead of using the correction fits of \citet{2007ApJ...660.1624S}. Second, we neglect modifications of the Roche lobe radius from the fact that the companion is not necessarily a single star, but could be a multiple subsystem (the case of a binary companion is studied in detail by \citealt{2020MNRAS.491..495D}). 

If the donor satisfies \eq~(\ref{eq:mt_crit}), we handle mass transfer differently depending on the properties of the companion. If the companion is itself a star (star-star RLOF), we adopt a scheme similar to that of \bsepaper, and which is outlined below. If the companion is itself a binary consisting of two stars (star-binary RLOF, or `triple RLOF'), we adopt a different scheme, which is described in \S~\ref{sect:triple:mt}. Currently, we do not model mass transfer from a star to a subsystem which itself consists of more than two stars. We expect that the latter case (e.g., transfer from a star to an inner triple) is marginal, since the companion subsystem would need to be very compact in order for the donor star to fill its Roche lobe around it, strongly limiting the possibilities for $N>2$ hierarchies within the companion subsystem that are dynamically stable. 

For star-star RLOF, we classify the type of mass transfer according to the following cases. These cases, which are discussed in more detail below, are adopted largely from the prescriptions of \bsepaper.
\begin{enumerate}[leftmargin=0.5cm]
\item Dynamical mass transfer from a low-mass MS donor (\S~\ref{sect:bin:mt:dynlow}) if $k_i=0$ and $q_i > 0.695$ \citep{1997MNRAS.291..732T}.
\item CE evolution (\S~\ref{sect:bin:ce}) if $k_i \in \{2,3,4,5,6,8,9\}$, i.e., giant-like envelopes. Also required is $t_\mathrm{MT} < t_{\mathrm{dyn},\,i}$ or $q_i>q_\mathrm{crit,\,CE}$, or $t_\mathrm{MT} < P_{\orb,\,k}$. 
\item Dynamical mass transfer from a WD donor (\S~\ref{sect:bin:mt:dynwd}) if $10 \leq k_i \leq 12$ and $q_i > 0.628$ \citep{1997MNRAS.291..732T}.
\item Stable mass transfer (\S~\ref{sect:bin:mt:stable}) in all other cases. 
\end{enumerate} 
Here, the general mass transfer timescale $t_\mathrm{MT}$ is computed to be consistent with the orbit-averaged rate in the model of \emtpaper, i.e., 
\begin{align}
t_\mathrm{MT} \equiv \frac{P_{\orb,\,k}}{\left | f_{\dot{m}} \right |},
\end{align}
where $f_{\dot{m}}$ is a dimensionless quantity defined through \eq~(35) of \emtpaper. The latter relates the instantaneous mass transfer rate to the orbit-averaged mass transfer rate, assuming that the mass transfer rate is proportional to the radius excess to the third power (see \S~\ref{sect:bin:mt:stable:orb} below). 

If both stars in an orbit fill their Roche lobes around each other, then contact evolution is assumed to ensue. In this case, we invoke CE evolution (\S~\ref{sect:bin:ce}) if both the donor and accretor are giant-like stars ($k_i \in 2, 3, 4, 5, 6, 8, 9\}$); otherwise, we let the two stars merge (\S~\ref{sect:bin:col}). 

\subsubsection{Dynamical mass transfer from a low-mass MS donor}
\label{sect:bin:mt:dynlow}
In this case of mass transfer with $q_i > 0.695$ (cf. \S~2.6.4 of \bsepaper), the donor star has a deep convective envelope, and mass transfer is expected to proceed on a fast, short timescale, somewhat similar to CE evolution. Following \bsepaper, the donor star is assumed to be completely disrupted after the mass transfer process. We treat this case in \mse~as a fast process and do not include its evolution as part of the ODE integration, but in separate routines that immediately update the system. 

The absolute value of the amount of mass lost by the donor in this case is assumed to be $\Delta \md = \md$, where $\md$ is the donor mass. The accreted amount by the companion, $\Delta \ma$, depends on the type of the companion star, as described below. 

If the accretor is an MS star ($k_\a \in \{0,1\}$), we limit accretion to the thermal timescale of the accretor, i.e.,
\begin{align}
\Delta \ma = \frac{\tau_\d}{t_\mathrm{KH,\,\a}},
\end{align}
where
\begin{align}
\tau_\d \equiv \sqrt{ t_\mathrm{KH,\,\d} \, t_{\mathrm{dyn},\,\d}}
\end{align}
is the geometric mean of the donor's Kelvin-Helmholtz and dynamical timescales. The Kelvin-Helmholtz timescale $t_{\mathrm{KH},\,i}$ of a star $i$ is generally computed according to
\begin{align}
\label{eq:t_kh}
t_{\mathrm{KH},\,i} &= \frac{\gconst m_i}{2 \rstar L_i} \times \left \{
\begin{array}{lc}
m_i, & k_i  \in \{0,1,7,10,11,12,13,14\}; \\
(m_i - \mc), & k_i \in \{2,3,4,5,6,8,9\}. \\
\end{array}
\right.
\end{align}
The accretor is rejuvenated, i.e., its age in \sse~is updated by comparing the MS timescales before and after accretion of the material (determined from the \sse~routines). Specifically, if the accretor has no convective core ($0.35 \,\msun < \ma < 1.25 \,\msun$), then its new age is determined according to
\begin{align}
t'_{\mathrm{age},\,\a} = t_{\mathrm{age},\,\a} \times \frac{t'_{\mathrm{MS},\,\a}}{t_{\mathrm{MS},\,\a}},
\end{align}
where primes indicate new quantities after the accretion event, and $t_{\mathrm{MS},\,\a}$ is the MS timescale of the accretor (\ssepaper, \eq~5). If the accretor does have a convective core ($\ma \leq 0.35\,\msun$ or $\ma \geq 1.25\,\msun$), then
\begin{align}
t'_{\mathrm{age},\,\a} = t_{\mathrm{age},\,\a} \times \frac{t'_{\mathrm{MS},\,\a}}{t_{\mathrm{MS},\,\a}} \frac{\ma}{\ma + \Delta \ma}.
\end{align}

If the accretor is a giant-like star ($k_\a \in \{2,3,4,5,6\}$), it is assumed to be able to accrete the entire donor material, so $\Delta \ma = \Delta \md$. The accretor's age is updated in the case of a HG star according to
\begin{align}
t'_{\mathrm{age},\,\a} =  t'_{\mathrm{MS},\,\a} + (t_{\mathrm{age},\,\a} - t_{\mathrm{MS},\,\a}) \frac{t'_{\mathrm{BGB},\,\a}}{t_{\mathrm{BGB},\,\a}},
\end{align}
where $t_{\mathrm{BGB},\,\a}$ is the accretor's timescale for the base of the giant branch (\ssepaper, \eq~4). 

For naked He star or WD accretors ($k_\a \in \{7,8,9,10,11,12\}$), we assume that all material is accreted, $\Delta \ma = \Delta \md$, and forms a giant envelope around a degenerate core (\bsepaper). The new stellar type of the accretor, $k'_\a$, is determined according to a stellar type merger table, Table~\ref{table:mergerst} (this table is reproduced here from \bsepaper~in order to provide a self-contained overview). The giant's age and initial mass are determined similarly as described in \S~\ref{sect:bin:col:mass_age:giant} (see also \S~2.7.4 of \bsepaper).

Lastly, for NS or BH accretors ($k_\a \in \{13,14\}$), the accreted mass is limited by the Eddington accretion rate. The latter is generally computed according to
\begin{align}
\label{eq:mdotedd}
\dot{m}_{\mathrm{Edd},\,i} = 4 \pi f_\mathrm{Edd} \, c \frac{\rstar}{\kappa_i},
\end{align}
where $\kappa_i = 0.2 \,(1+X_i)\, \mathrm{cm^2\,g^{-1}}$ is the electron scattering opacity (with $X_i$ the hydrogen fraction), and $f_\mathrm{Edd}$ is the Eddington accretion factor, taken to be $f_\mathrm{Edd}=10$ by default \citep{2004ApJ...613L.129K,2013A&A...552A..24B,2017gacv.workE..56K}. The maximum accreted amount of mass is then 
\begin{align}
\Delta \ma = \mathrm{min} \left ( \dot{m}_{\mathrm{Edd},\,i} \, \tau_\d, \Delta \md \right ).
\end{align}

The donor star is assumed to be destroyed in this process; therefore, it is removed from the code's memory. Given the mass lost, $\Delta \md$, and the accreted matter, $\Delta \ma$, we update the other orbits in the system as described in \S~\ref{sect:bin:mlorbit}. Here, we set the mass-loss timescale to the parameter $\tau_{\dot{m},\,\mathrm{dyn,\,MS}}$; by default $\tau_{\dot{m},\,\mathrm{dyn,\,MS}} = 10^3 \,\yr$.

\subsubsection{Dynamical mass transfer from a WD donor}
\label{sect:bin:mt:dynwd}
Mass transfer from a WD donor ($k_\d \in \{10,11,12\}$) with $q_i > 0.628$ (cf. \S~2.6.5 of \bsepaper) is expected to lead to dynamical mass transfer \citep{1997MNRAS.291..732T}. This case is treated in a similar fashion to \S~\ref{sect:bin:mt:dynlow}, i.e., the evolution is assumed to be fast and not modelled as part of the ODE equations. The donor is expected to transfer all of its mass and the accretor to accept all of it, so $\Delta \ma = \Delta \md = \md$. 

If both the donor and accretor are He WDs, $k_\d = k_\a = 10$, then it is assumed that sufficiently high temperatures are reached to ignite the triple-$\alpha$ reaction and the two stars are destroyed (\bsepaper; we do not consider the possibility of forming a naked He star, see \citealt{1984ApJ...277..355W}). 

If the accretor is an He or CO WD ($k_\a \in \{10,11\}$) with a new mass exceeding the Chandrasekhar mass, $\ma + \Delta \ma > M_\mathrm{Ch} = 1.44\,\msun$, then we assume an SNe Ia event occurs, leaving no remnant. 

If either donor or accretor is an He WD but the companion is a CO or ONe WD ($k_\d = 10$ or $k_\a = 10$, but $k_\d \neq k_\a$), then the He accreted onto the CO or ONe core is assumed to swell up and form a giant envelope, forming an HeGB star ($k=9$). The core mass of the HeGB star is then set to $\Delta \ma$, and the age of the giant is determined similarly as described in \S~\ref{sect:bin:col:mass_age:giant}.

Lastly, dynamical transfer from a CO or ONe WD onto a CO WD is assumed to result into an ONe WD.

Similarly to \S~\ref{sect:bin:mt:dynlow}, we update the other orbits in the system as described in \S~\ref{sect:bin:mlorbit} following mass loss during dynamical mass transfer from a WD donor. Here, we set the mass-loss timescale to the parameter $\tau_{\dot{m},\,\mathrm{dyn,\,WD}}$; by default $\tau_{\dot{m},\,\mathrm{dyn,\,WD}} = 10^3 \,\yr$.

\subsubsection{Stable mass transfer}
\label{sect:bin:mt:stable}
If RLOF between two stars is not identified as dynamical mass transfer from a low-mass WD (\S~\ref{sect:bin:mt:dynlow}) or WD (\S~\ref{sect:bin:mt:dynwd}), nor leading to CE evolution (\S~\ref{sect:bin:ce}), then mass transfer is assumed to be stable and lasting on a longer timescale. In this case, we handle changes associated with mass transfer and pertaining to the masses and orbits as part of the ODE integration (this is different than in \bse, where an Euler scheme is used). Here, we consider the mass time derivatives to be constant during the ODE integration and calculate them beforehand, as described below.

\paragraph{Mass lost from the donor}
\label{sect:bin:mt:stable:loss}
First, we compute the amount of mass lost from the donor ($\Delta \md$; we define $\Delta \md\geq0$) during the ODE timestep ($ \deltatode$), following the prescriptions of \bsepaper. By default, for nuclear timescale mass transfer, we adopt the empirical relation
\begin{align}
\label{eq:mdotnuc}
\frac{\Delta m_{\d, \, \mathrm{nuc}}}{ \deltatode} = 3\times 10^{-6} \, \msun\,\yr^{-1} \left [ \min\left (\frac{\md}{\msun},5 \right ) \right ]^2 \left [ \ln \left (\frac{R_{\star,\,\d}}{ R_{\mathrm{L},\,\d} } \right ) \right ]^3.
\end{align}
For He WD donors, \eq~(\ref{eq:mdotnuc}) is multiplied by the factor $10^3 \, (\md / \msun) \left [ \mathrm{max} \left (R_{\star,\,\d}/\rsun,10^{-4} \right ) \right ]^{-1}$. 

For giant-like stars, $k_\d \in \{2,3,4,5,6,8,9\}$, the amount in \eq~(\ref{eq:mdotnuc}) is limited by the thermal timescale of the donor, i.e.,
\begin{align}
\label{eq:,dotkh}
\Delta m_{\d, \, \mathrm{KH}} = \mathrm{min} \left ( \Delta m_{\d, \, \mathrm{nuc}}, \md \, \frac{\Delta t_{\mathrm{ODE}}}{t_{\mathrm{KH},\,\d}} \right ).
\end{align}

In other cases, mass transfer is limited by the donor's dynamical timescale, i.e.,
\begin{align}
\label{eq:,dotdyn}
\Delta m_{\d, \, \mathrm{dyn}} = \mathrm{min} \left ( \Delta m_{\d, \, \mathrm{nuc}}, \md \, \frac{\Delta t_{\mathrm{ODE}}}{t_{\mathrm{dyn},\,\d}} \right ).
\end{align}

\paragraph{Accreted mass}
\label{sect:bin:mt:stable:accr}
Next, the amount of accreted material during the ODE timestep, $\Delta \ma$, is determined. For MS, HG, and CHeB companions, $k_\a \in \{0,1,2,4\}$, the accreted amount is limited by the companion's thermal timescale according to
\begin{align}
\Delta \ma = \Delta \md \, \mathrm{min} \left (1, 10 \frac{\tau_{\d,\,\dot{m}}}{t_{\mathrm{KH},\,\a}} \right ),
\end{align}
where $\tau_{\d,\,\dot{m}}$ is defined as
\begin{align}
\tau_{\d,\,\dot{m}} =  \deltatode \, \frac{\ma}{\Delta \md}.
\end{align}

If the accretor is a stripped He star ($k_\a \in \{7,8,9\}$) and the donor is not a stripped He star itself, then the accretor is assumed to accrete all the material ($\Delta \ma = \Delta \md$) and to swell up to a CHeB star or AGB star, with the new stellar type $k'_\a = \mathrm{min}(6, 2k_\a-10)$. The age of the accretor is adjusted accordingly (cf. \S~\ref{sect:bin:col:mass_age:giant}).

If the donor has $k_\d \leq 6$ and the accretor is a WD, $k_\a \in \{10,11,12\}$, novae are assumed to occur if the hydrogen mass transfer rate is low, $\Delta \md /  \deltatode < 1.03\times 10^{-7} \, \msun\,\yr^{-1}$ \citep{1997MNRAS.291..732T}. The accretion amount is then limited by the Eddington rate, i.e.,
\begin{align}
\Delta \ma = \epsilon_{\mathrm{nova}} \,\mathrm{min} \left ( \Delta \md, \Delta m_{\mathrm{Edd},\,\a} \right ),
\end{align}
where $\Delta m_{\mathrm{Edd},\,\a} = \dot{m}_{\mathrm{Edd},\,\a} \deltatode$ and with $\dot{m}_{\mathrm{Edd},\,\a}$ given for the accretor by \eq~(\ref{eq:mdotedd}). Here, the nova accretion factor is set to $\epsilon_{\mathrm{nova}} = 10^{-3}$ by default. For higher hydrogen mass transfer rates, $1.03\times 10^{-7} \, \msun\, \yr^{-1} < \Delta \md /  \deltatode < 2.71 \times 10^{-7} \, \msun\, \yr^{-1}$, a supersoft X-ray source is assumed to form, with all the offered material being accreted, $\Delta \ma = \Delta \md$. If $ \Delta \md /  \deltatode \geq 2.71 \times 10^{-7} \, \msun\, \yr^{-1}$, the material is assumed to form a giant envelope around the degenerate core of the accretor, turning the accretor into an RGB, TPAGB, and He GB star (the new $k'_\a = 3,6,9$) if the accretor was originally an He WD, CO WD, and ONe WD, respectively ($k'_\a = 3 k_\a - 27$). The age of the new giant is determined similarly as described in \S~\ref{sect:bin:col:mass_age:giant}.

If the donor is more evolved ($k_\d > 7$) and the accretor is a WD, accretion is limited by the Eddington rate, and we set
\begin{align}
\Delta \ma = \mathrm{min} \left ( \Delta \md, \Delta m_{\mathrm{Edd},\,\a} \right ).
\end{align}

In all other cases, we set $\Delta \md = \Delta \ma$. 

\paragraph{Explosive events}
\label{sect:bin:mt:stable:expl}
If the accretor is a WD ($k_\a \in \{10,11,12\}$), we check if the accreted material might trigger a thermonuclear explosion of the accretor. The accretor's new mass during the ODE timestep would be $\ma' = \ma + \Delta \ma - |\dot{m}_{\a,\,\sse} \deltatode|$, where $\dot{m}_{\a,\,\sse} \deltatode$ is the mass lost in a wind, and where we ignore possible wind accretion from the donor (cf. \S~\ref{sect:bin:wa}). 

Following \bsepaper~and regardless of the donor, if the accretor is an He or CO WD ($k_\a \in \{10,11\}$) and if $\ma' > M_\mathrm{Ch}=1.44\,\msun$, we assume the accretor explodes in an SNe Ia. If $\ma'>M_\mathrm{Ch}$ but the accretor is an ONe WD ($k_\a=12$), we assume that the WD is not destroyed and will become an NS. 

For a donor with $k_\d \leq 10$ and an He WD accretor ($k_\a = 10$), we assume the accretor is destroyed in a possible SNe Ia if $\ma' > 0.7 \, \msun$. 

In the case of an explosion, the accretor star is removed from the code, and the remaining orbits in the system are adjusted according to \S~\ref{sect:bin:mlorbit} with a mass-loss timescale parameter $\tau_{\dot{m},\,\mathrm{expl}}$; by default, $\tau_{\dot{m},\,\mathrm{expl}} = 10^3 \,\yr$.

\paragraph{Ageing and rejuvenation}
\label{sect:bin:mt:stable:age}
Mass loss and mass accretion through RLOF can lead to ageing and rejuvenation of the stars, respectively. 

The donor is aged as follows. Let primed quantities indicate the properties after the amount $\Delta \md$ has been transferred. If the donor is an MS star (including stripped He stars), $k_\d \in \{0,1,7\}$, then its new age is determined by the fractional change in its MS lifetime, i.e.,
\begin{align}
t'_\mathrm{age,\,\d} = t_\mathrm{age,\,\d} \frac{t'_\mathrm{MS,\,\d}}{t_\mathrm{MS,\,\d}}.
\end{align}
Note that the donor's lower new mass implies a longer new MS timescale, hence a higher new age. If the donor is a HG star, $k_\d = 2$, then 
\begin{align}
t'_\mathrm{age,\,\d} = t'_{\mathrm{MS},\,\d} + (t_\mathrm{age,\,\d} -  t_{\mathrm{MS},\,\d}) \frac{t'_\mathrm{BGB,\,\d} - t'_{\mathrm{MS},\,\d}}{t_\mathrm{BGB,\,\d} - t_{\mathrm{MS},\,\d}}.
\end{align}

Regarding rejuvenation of the accretor, if the latter is an MS star, $k_\a \in \{0,1\}$ and has no convective core ($0.35 \,\msun < \ma < 1.25\,\msun$), then
\begin{align}
t'_\mathrm{age,\,\a} = t_\mathrm{age,\,\a} \frac{t'_{\mathrm{MS,\a}}}{t_{\mathrm{MS,\a}}} \frac{\ma}{\ma'}.
\end{align}
For other MS stars (including stripped He stars) with convective cores, 
\begin{align}
t'_\mathrm{age,\,\a} = t_\mathrm{age,\,\a} \frac{t'_\mathrm{MS,\,\a}}{t_\mathrm{MS,\,\a}}.
\end{align}
Lastly, for HG accretors,
\begin{align}
t'_\mathrm{age,\,\a} = t'_{\mathrm{MS},\,\a} + (t_\mathrm{age,\,\a} -  t_{\mathrm{MS},\,\a}) \frac{t'_\mathrm{BGB,\,\a} - t'_{\mathrm{MS},\,\a}}{t_\mathrm{BGB,\,\a} - t_{\mathrm{MS},\,\a}}.
\end{align}

\paragraph{Orbital response}
\label{sect:bin:mt:stable:orb}
The orbital response of stable mass transfer in \mse~is handled during the ODE integration. As mentioned in \S~\ref{sect:introduction}, it is assumed in \bse~that the orbit has circularised by the time of the onset of RLOF due to efficient tides. In \mse, we relax this assumption and adopt the analytic model of \emtpaper~for mass transfer in both circular and eccentric orbits. Here, we make a small modification to the latter model in order to be able to model the case of non-conservative mass transfer (i.e., if not all mass is transferred, but some is lost from the binary orbit). A brief description of the model is given here; for more details, we refer to \emtpaper.

The equations of motion for the relative separation vector $\ve{r}_k$ in a mass-transferring binary system are given by \citep{1969Ap&SS...3...31H,2007ApJ...667.1170S}
\begin{subequations}
\label{eq:mt_eom}
\begin{align}
\label{eq:mt_eom_l1} & \frac{\mathrm{d}^2 \ve{r}_k}{\mathrm{d} t^2} = -\frac{\gconst (m_\d+m_\a)}{r_k^3} \ve{r}_k \\
\label{eq:mt_eom_l2} &\quad + \frac{\ve{f}_\a}{m_\a} - \frac{\ve{f}_\d}{m_\d}  \\
\label{eq:mt_eom_l3} &\quad + \frac{\dot{m}_\a}{m_\a} \left (\ve{w}_\a + \bm{\omega}_{\orb,\,k} \times \ve{r}_\Aa \right ) -  \frac{\dot{m}_\d}{m_\d} \left (\ve{w}_\d + \bm{\omega}_{\orb,\,k} \times \ve{r}_\Ad \right ) \\
\label{eq:mt_eom_l4}&\quad + \frac{\ddot{m}_\a}{m_\a} \ve{r}_\Aa - \frac{\ddot{m}_\d}{m_\d} \ve{r}_\Ad.
\end{align}
\end{subequations}
Here,  $\ve{f}_\a$ and $\ve{f}_\d$ represent perturbations from the ejected mass on the orbit, $\ve{w}_\d$ and $\ve{w}_\a$ are the ejection/accretion velocities relative to the donor and accretor, respectively, $\bm{\omega}_{\orb,\,k}$ is the orbital frequency vector (pointing along the direction of the orbital angular momentum vector), and $\ve{r}_\Ad$ and $\ve{r}_\Aa$ are the ejection and accretion locations, respectively, relative to the donor/accretor.

We make similar assumptions as \emtpaper, with one exception related to the accreted amount of mass. Specifically, we assume that:
\begin{enumerate}[leftmargin=0.5cm]
\item the effects of the mass stream on the orbit are negligible, i.e., we set $\ve{f}_\a = \ve{f}_\d = \ve{0}$;
\item the amount of mass successfully transferred to the companion per unit time is given by
\begin{align}
\label{eq:mt_beta}
\dot{m}_\a = -\betamt \dot{m}_\d,
\end{align}
where $\betamt$ ($0 \leq \betamt \leq 1$) is a dimensionless mass transfer efficiency parameter;
\item the donor ejects mass at a relative velocity of $\ve{w}_\d = \dot{\ve{r}}_k$, and the accretor accretes mass at a relative velocity of $\ve{w}_\a = - \dot{\ve{r}}_k$;
\item $\ve{r}_\Ad$ and $\ve{r}_\Aa$ corotate with the orbit, i.e., they are proportional to $\unit{r}_k$; we take $\ve{r}_\Aa$ to have a constant magnitude, whereas we make two limiting assumptions on the magnitude of $\ve{r}_\Ad$: either a negligible spin frequency of the donor (1), or a large mass ratio (2), $m_\d/m_\a \gg 1$ (see \S~2.2 of \emtpaper~for details);  
\item the donor's mass transfer rate, $\dot{m}_\d$, is sensitively dependent on its `radius excess', $R_{\star,\,\d} - R_{\mathrm{L},\,\d}$; specifically, assuming an $n=3/2$ polytrope \citep{1972AcA....22...73P,1987MNRAS.229..383E}\footnote{An $n=3/2$ polytrope is a reasonable approximation for, e.g., convective stars and low-mass white dwarfs (e.g., \citealt{1939isss.book.....C}) and gas giant planets (e.g., \citealt{2015MNRAS.452.1375W}), but not for all stars. },
\begin{align}
\label{eq:mt_md}
\dot{m}_\d \propto \left ( \frac{R_{\star,\,\d}-R_{\mathrm{L},\,\d}(t-\taumt)}{R_{\star,\,\d}} \right )^3, 
\end{align}
where $R_{\mathrm{L},\,\d}(t)$ is the instantaneous Roche lobe radius of the donor given by \eq~(\ref{eq:rlof}); we include the possibility of a delay between close approach and mass transfer (due to the dynamical response of the donor star) by introducing the delay time parameter $\taumt$.
\end{enumerate}
These assumptions are equivalent to those of \emtpaper~if $\betamt=1$ (no mass lost from the system). They are motivated by the fact that, in the simplest conceivable case of conservative mass transfer ($\betamt=1$, and no orbital angular momentum is lost) in circular orbits and ignoring any finite-size effects, they reduce to the `canonical' relation
\begin{align}
\label{eq:mt_a_dot_can}
\frac{\dot{a}_k}{a_k} = -2\frac{\dot{m}_\d}{m_\d} \left (1-\frac{m_\d}{m_\a} \right ).
\end{align}
(see also \S~2.1.3 of \emtpaper). It is unclear, however, to which extent these assumptions are still valid in more complex situations. An investigation into this will require detailed (hydrodynamical) simulations.

With the above assumptions, the equations of motion for mass transfer reduce to
\begin{align}
\nonumber \frac{\mathrm{d}^2 \ve{r}_k}{\mathrm{d} t^2} &= -\frac{\gconst (m_\d+m_\a)}{r_k^3} \ve{r}_k - \frac{\dot{m}_\d}{m_\d} \left (1-q_k \betamt \right )\dot{\ve{r}}_k \\
&\quad - \frac{\dot{m}_\d}{m_\d} \bm{\omega}_\orb \times \left (\ve{r}_\Ad + q_k \betamt\, \rAa \right ) - \frac{\ddot{m}_\d}{m_\d} \left (\ve{r}_\Ad + q_k \betamt \, \ve{r}_\Aa \right ),
\label{eq:mt_eom_simple}
\end{align}
where $q_k \equiv m_\d/m_\a$\footnote{Note the addition of factors of $\betamt$ in each term involving the mass ratio in \eq~(\ref{eq:mt_eom_simple}) compared to \eq~(6) of \emtpaper.}. Applying standard equations for perturbations to Keplerian orbits and averaging over an orbital period (see \S~2.4.1 of \emtpaper~for details)\footnote{As discussed above, we here allow for non-unity $\betamt$, whereas \emtpaper~assumed $\betamt=1$. Strictly speaking, the fact that the total binary mass, $m_\a+m_\d$, is not constant in this case where $\dot{m}_\d + \dot{m}_\a = \dot{m}_\d(1-\betamt) \neq 0$, should be taken into account when computing the orbital element changes (cf. \eqs~26 and 27 of \emtpaper). However, we expect that $m_\a+m_\d$ does not change significantly during one ODE timestep during stable mass transfer evolution; therefore, we neglect the change of total binary mass and assume that $m_\a+m_\d$ is constant when computing the secular orbital element changes from \eq~(\ref{eq:mt_eom_simple}). In practice, this means that simply terms with $q_k$ need to be multiplied by $\betamt$ to arrive from the `conservative' case of \emtpaper~to the `non-conservative' case here.}, the equations of motion for the secular changes of the orbital elements are given below for the two different assumptions on $\ve{r}_\Ad$. For brevity, we also restrict our presentation here to the case $\taumt=0$ (\mse~allows for non-zero $\taumt$, which can be user specified).

In case (1), negligible spin frequency of the donor,
\begin{subequations}
\label{eq:mt_av_eom_case_1}
\begin{align}
\nonumber \displaystyle\frac{\langle \dot{a}_k \rangle}{a_k} &= \displaystyle -\frac{2 \langle \dot{m}_\d \rangle}{m_\d} \frac{1}{f_{\dot{m}}(e_k,x_k) } \Biggl [ (1-q_k \betamt) f_a(e_k,x_k) + \XLz(q_k) g_a(e_k,x_k)\\
&\qquad \qquad \displaystyle  - q_k \betamt \frac{r_\Aa}{a_k} h_a(e_k,x_k) \Biggl ]; \\
\nonumber \displaystyle \langle \dot{e}_k \rangle &= \displaystyle -\frac{2 \langle \dot{m}_\d \rangle}{m_\d} \frac{1}{f_{\dot{m}}(e_k,x_k) } \Biggl [ (1-q_k \betamt) f_e(e_k,x_k) + \XLz(q_k) g_e(e_k,x_k)\\
&\qquad \qquad \displaystyle  - q_k \betamt \frac{r_\Aa}{a_k} h_e(e_k,x_k) \Biggl ]; \\
\displaystyle \langle \dot{\omega}_k \rangle &= 0.
\end{align}
\end{subequations}
In case (2), a large mass ratio ($q_k\gg 1$), 
\begin{subequations}
\label{eq:mt_av_eom_case_2}
\begin{align}
\nonumber \displaystyle\frac{\langle \dot{a}_k \rangle}{a_k} &= \displaystyle -\frac{2 \langle \dot{m}_\d \rangle}{m_\d} \frac{1}{f_{\dot{m}}(e_k,x_k) } \Biggl [ (1-q_k \betamt) f_a(e_k,x_k) \\
&\qquad \qquad \displaystyle + \left ( \XLz(e_k,\hat{\Omega}_\d) - q_k \betamt \frac{r_\Aa}{a_k} \right ) h_a(e_k,x_k) \Biggl ]; \\
\nonumber \displaystyle \langle \dot{e}_k \rangle &= \displaystyle -\frac{2 \langle \dot{m}_\d \rangle}{m_\d} \frac{1}{f_{\dot{m}}(e_k,x_k) } \Biggl [ (1-q_k \betamt) f_e(e_k,x_k) \\
&\qquad \qquad \displaystyle + \left ( \XLz(e_k,\hat{\Omega}_\d) - q_k \betamt \frac{r_\Aa}{a_k} \right ) h_e(e_k,x_k) \Biggl ]; \\
\displaystyle \langle \dot{\omega}_k \rangle &= 0.
\end{align}
\end{subequations}
Here, $x_k \equiv R_{\mathrm{L},\,\d}(a_k)/R_{\star,\,\d}$ (i.e., the donor's circular Roche lobe radius divided by the stellar radius), the function $\XLz(q_k)$ is given by \eq~(A1) of \emtpaper, $\XLz(e_k,\hat{\Omega}_\d)$ by \eq~(11) of \emtpaper, $f_{\dot{m}}(e_k,x_k)$ by \eq~(B1) of \emtpaper, $f_a(e_k,x_k)$ by \eq~(B3) of \emtpaper, and $f_e(e_k,x_k)$ by \eq~(B4) of \emtpaper. Furthermore, the hat above $\Omega_\d$ indicates the donor's spin frequency normalised to the orbital frequency at periapsis, i.e., $\hat{\Omega}_\d \equiv \Omega_\d/ \omega_{\mathrm{orb,\,peri},\,k}$, where $\omega_{\mathrm{orb,\,peri},\,k} = (2\pi/P_{\orb,\,k}) (1+e_k)^{1/2}(1-e_k)^{-3/2}$. 

Note that there is no secular apsidal motion ($\langle \dot{\omega}_k\rangle=0$) due to mass transfer. This is no longer the case when $\taumt\neq0$ (equations not shown here). 

In \mse, \eqs~(\ref{eq:mt_av_eom_case_1}) and (\ref{eq:mt_av_eom_case_2}) are implemented with $\betamt = -\Delta m_\a/\Delta m_\d$. We implement the set \eqs~(\ref{eq:mt_av_eom_case_1}) if $\hat{\Omega}_\d<0.1$. Otherwise, if $q_k > 10$, we implement the set \eqs~(\ref{eq:mt_av_eom_case_2}). In remaining cases, we do not take into account finite-size terms associated with the donor. The relative accretion distance, $r_\Aa$, is set to $r_\Aa=R_\a$ by default. However, if an accretion disk is expected to form around the accretor, $r_\Aa$ is set to the accretion disk's size, $r_\Aa = r_\mathrm{disk}$, where $r_\mathrm{disk}$ is given by \eq~(\ref{eq:rdisk}). 

The amount of mass not accreted by the companion, $\Delta m_\d - \Delta m_\a \geq 0$, is assumed to be lost from the accretor in an adiabatic wind (cf. \eq~\ref{eq:adotwind}), i.e., the accretor's wind mass loss rate is effectively increased by $-(\Delta m_\d - \Delta m_\a)/\deltatode \leq 0$.

\paragraph{Spin evolution}
\label{sect:bin:mt:stable:spin}
The response of the stellar spins to stable mass transfer is handled in \mse~during the ODE integration. The associated quantities, $\dot{\Omega}_\d$ and $\dot{\Omega}_\a$, contribute to the term $\dot{\Omega}_{i,\,\mathrm{MT}}$ in \eq~(\ref{eq:rmdot}), and are determined as follows. 

First, we discuss how we relate changes in the spin angular momentum to changes in the spin frequency. Generally, the stellar spin angular momentum of star $i$ is given by
\begin{align}
h_{\spin,\,i} = I_i \Omega_i,
\end{align}
where $I_i$ is the moment of inertia of star $i$. Following \ssepaper~and \bsepaper, we compute $I_i$ by considering separately the contributions from the core and envelope, i.e.,
\begin{align}
I_i = k_{2,\,i} \, (m_i - \mc) \rstar^2 + k_{3,\,i} \, \mc \rc^2.
\end{align}
Here, $k_{2,\,i}$ is computed from a routine in \sse, and $k_{3,\,i}$ is fixed as $k_{3,\,i} = 0.21$ (\ssepaper). As was discussed in \S~\ref{sect:stellar:sse}, $m_i$, $\rstar$, and $\Omega_i$ are assumed to vary linearly during the ODE integration, whereas the other stellar-evolution related quantities are assumed to be constant. Consistent with this, we compute $\dot{I}_i$ according to
\begin{align}
\dot{I}_i = k_{2,\,i} \, \dot{m}_i \rstar^2 + 2 k_{2,\,i} \, (m_i-\mc) \rstar \dot{R}_{\star,\,i}.
\end{align}
For a given $\dot{h}_{\mathrm{spin},\,i}$, the spin frequency of star $i$ then changes according to
\begin{align}
\label{eq:omegadotmt}
\dot{\Omega}_i = \frac{ \dot{h}_{\mathrm{spin},\,i} - \dot{I}_\d \Omega_i}{I_i}.
\end{align}

The donor, which loses mass at a rate of $\dot{m}_\d = -\Delta m_\d/\deltatode$ due to mass transfer, is assumed to lose spin angular momentum at a rate given by 
\begin{align}
\dot{h}_{\spin,\,\d} = \dot{m}_\d R_{\star,\,\d}^2 \Omega_\d.
\end{align}
Together with \eq~(\ref{eq:omegadotmt}), this describes the donor's spin response to mass transfer. Here, we include in $\dot{m}_\d$ only the mass change due to mass transfer. 

The spin response of the accretor depends on not whether an accretion disk is expected to form around the accretor. The latter is assumed to be the case if
\begin{align}
R_{\star,\,\a} > r_\mathrm{min},
\end{align}
where $r_\mathrm{min}$ is given by \citep{1976ApJ...206..509U}
\begin{align}
\label{eq:rmin}
r_\mathrm{min} = 0.0425 \, a_k(1-e_k) \left [ \frac{m_\a}{m_\d} \left (1 + \frac{m_\a}{m_\d} \right ) \right ]^{0.25}.
\end{align}

If a disk is present, we assume that the material is accreted near the inner edge of the disk, near the accretor's surface and with local Keplerian rotation. The accreted spin angular momentum is then
\begin{align}
\dot{h}_{\spin,\,\a} = \dot{m}_\a \sqrt{\gconst m_\a R_{\star,\,\a}},
\end{align}
which yields $\dot{\Omega}_\a$ when paired with \eq~(\ref{eq:omegadotmt}). 

If no disk is expected to form, then we follow \citet{1976ApJ...206..509U}, and compute the spin angular momentum of the transferred material  
using the radius of the disk that would have formed if allowed,
\begin{align}
\label{eq:rdisk}
r_\mathrm{disk} = 1.7 \, r_\mathrm{min},
\end{align}
such that
\begin{align}
\dot{h}_{\spin,\,\a} = \dot{m}_\a \sqrt{\gconst m_\a r_\mathrm{disk}}.
\end{align}

\subsection{CE evolution}
\label{sect:bin:ce}
We largely follow \bsepaper~when modelling CE evolution. The main differences arise when taking into account the effects of mass loss during the CE on other (i.e., external) orbits (see \S~\ref{sect:bin:mlorbit}; evidently, the latter effects do not apply to \bse). 

\subsubsection{Energy budget}
\label{sect:bin:ce:energy}
We use the $\alpha$-CE prescription (\citealt{1976IAUS...73...75P,1976IAUS...73...35V,1984ApJ...277..355W,1988ApJ...329..764L,1993PASP..105.1373I}; see, e.g., \citealt{2013A&ARv..21...59I} for a review) to parameterise the efficiency at which the energy of the orbit in which the CE event occurs is used to expel the donor's envelope. Let $E_{\orb,\,\init}$ ($E_{\orb,\,\fin}$) and $E_{\bind,\,\init}$ ($E_{\bind,\,\fin}$) denote the initial (final) orbital and donor's envelope binding energies, respectively. The initial binding energy is given by
\begin{align}
\label{eq:ce_bind_init}
E_{\bind,\,\init} = - \frac{\gconst m_\d (m_\d - m_{\core,\,\d})}{\lambda_{\CE,\,\d} R_{\star,\,\d} },
\end{align}
which, if the accretor is also giant-like ($k_\a \in \{2,3,4,5,6,8,9\}$), is increased by
\begin{align}
- \frac{\gconst m_\a (m_\a - m_{\core,\,\a})}{\lambda_{\CE,\,\a} R_{\star,\,\a} }.
\end{align}
Here, $\lambda_{\CE,\,i}$ denotes the standard dimensionless binding energy parameter, which is computed a routine provided with \sse. The initial orbital energy is given by
\begin{align}
E_{\orb,\,\init} = -\frac{\gconst m_\d m_\a}{2 a_{\init}},
\end{align}
with $a_{\init}$ the initial semimajor axis of the orbit in which the CE event occurs. 

The final orbital energy is computed from
\begin{align}
\label{eq:ce_energy1}
E_{\orb,\,\fin} = E_{\orb,\,\init} + \frac{E_{\bind,\,\init}}{\alpha_{\CE}},
\end{align}
where $\alpha_{\CE}$ is the CE $\alpha$ parameter, which can be user specified for any orbit (by default, $\alpha_{\CE}=1$). If the stars would not merge, this corresponds to a final orbital separation of
\begin{align}
\label{eq:ce_afin}
a_{\fin} = - \frac{\gconst m_{\core,\,\d} m'_\a}{2 E_{\orb,\,\fin}},
\end{align}
where $m'_\a$ is either $m'_\a = m_\a$ in a main sequence accretor without a core; it is assumed to survive without mass loss, or $m'_\a = m_{\core,\,\a}$ in a degenerate accretor or a giant-like accretor, which is assumed to lose its envelope. The donor is always assumed to lose its entire envelope mass during the CE event. 

Given $a_{\fin}$, we determine whether or not the binary coalesces. Coalescence is assumed if, in the new orbit with semimajor axis $a_{\fin}$, either of the stars would fill their Roche lobe assuming a circular orbit (cf. \eq~\ref{eq:mt_crit}). Here, the core radius of the donor is used to assess if it would be Roche lobe overflowing, whereas for the accretor, the stellar radius is used to check for RLOF if the accretor was on the MS including stripped He stars ($k_\a \in \{0,1,7\}$), or the accretor's core radius if the accretor was a degenerate star or giant-like ($2 \leq k_\a \leq 6$ or $8 \leq k_\a \leq 14$). 

In the case of coalescence, the merger remnant properties need to be established. If the binary survives, then the new two stellar properties need to be specified, as well as the new orbital properties. 

\subsubsection{Merger}
\label{sect:bin:ce:merger}
The merger remnant of a `failed' CE event (i.e., in which the two stars merge) can have interesting properties such as enhanced rotation, peculiar abundances, and enhanced luminosity; also, mass loss during the merger event can be substantial (e.g., \citealt{1987ApJ...323..614B,1995ApJ...445L.117L,1996ApJ...468..797L,2008A&A...488.1007G,2008A&A...488.1017G,2008MNRAS.383L...5G}). In particular, stellar mergers have been extensively considered as potential pathways to forming blue straggler stars in dense stellar systems (e.g., \citealt{1993PASP..105.1081S,1995ARA&A..33..133B,1997ApJ...487..290S,1999ApJ...513..428S,2001ApJ...548..323S,2002MNRAS.332...49S,2013ApJ...777..106C}), and the formation of very massive stars and intermediate-mass BHs (e.g., \citealt{1999A&A...348..117P,2010MNRAS.402..105G,2013MNRAS.430.1018F,2016MNRAS.459.3432M,2021MNRAS.501.5257R}).

The (presently) most accurate method to determine the properties of the merger remnant would be to carry out detailed hydrodynamical simulations, but these are evidently computationally expensive and therefore not suitable for a population synthesis code such as \mse. We therefore implement a number of prescriptions which are adopted mostly from \bsepaper, and are outlined below. 

\paragraph{Stellar type}
\label{sect:bin:ce:merger:st}
The stellar type of the merger remnant, $k'$ (in this section, we denote properties of the merged object with a prime), is determined according to a merger stellar type table, Table~\ref{table:mergerst}.

\paragraph{Core mass}
\label{sect:bin:ce:merger:core}
If the accretor was on the MS including stripped He stars ($k_\a \in \{0,1,7\}$), the merger remnant core mass is given by $m'_\core = m_{\core,\,\d}$. The latter is increased by $m_\a$ if the new star is a CHeB star ($k'=4$) and the accretor was a stripped He-burning star ($k_\a =7$). Otherwise, when the accretor was a degenerate star or giant-like ($2 \leq k_\a \leq 6$ or $8 \leq k_\a \leq 14$), the new core mass is given by the combined core mass, $m'_\core = m_{\core,\,\d} + m_{\core,\,\a}$, unless the accretor was an NS or BH ($k_\a \in \{13,14\}$), in which case it is assumed that the outcome is an unstable Thorne-$\mathrm{\dot{Z}}$ytkow object \citep{1977ApJ...212..832T}, leaving only the accretor's core so $m'_\core = m_{\core,\,\a}$, $m' = m_{\core,\,\d}$, and $k' = k_\a$. 

\paragraph{Mass}
\label{sect:bin:ce:merger:mass}
The merger product mass is determined by considering its remaining envelope binding energy, $E_{\bind,\,\fin}$. The latter is calculated by assuming that the binary merged at a separation corresponding to the moment of RLOF of either core or star, i.e.,
\begin{align}
\label{eq:e_orb_fin_rlof}
E_{\orb,\,\fin} = - \frac{\gconst m_{\core,\,\d} m'_\a}{2 R_{\star,\,i}},
\end{align}
where $R_{\star,\,i} = R_{\core,\,\d}$ if the donor filled its Roche lobe first; if the accretor filled its Roche lobe first, then $R_{\star,\,i} = R_{\star,\,\a}$ for an MS accretor, and $R_{\star,\,i} = R_{\core,\,\a}$ for a degenerate or giant-like accretor. The remaining envelope binding energy is then computed similarly to \eq~(\ref{eq:ce_energy1}), i.e.,
\begin{align}
\label{eq:ce_energy2}
E_{\bind,\,\fin} - E_{\bind,\,\init} = \alpha_{\CE} \, (E_{\orb,\,\init} - E_{\orb,\,\fin} ).
\end{align}
The final binding energy is related to the merged star's equilibrium radius, $R'_{\star}$, according to
\begin{align}
\label{eq:ce_ebind_fin_merge}
E_{\bind,\,\fin} = - \frac{\gconst m'(m'-m'_{\core})}{\lambda'_{\CE} R'_{\star}}.
\end{align}
Its radius immediately after merger (before reaching equilibrium), $R_{\star}$, can be estimated according to
\begin{align}
\label{eq:ce_ebind_init_merge}
E_{\bind,\,\init} = - \frac{\gconst (m_\d+m_\a)(m_\d+m_\a - m'_{\core})}{\lambda'_{\CE} R_{\star}}
\end{align}
(here, we neglect changes in $\lambda_{\CE}$, i.e., we set $\lambda'_{\CE} = \lambda_{\CE}$). Assuming that the merger remnant readjusts on a dynamical timescale, its radius is expected to scale with mass as $R_\star \propto m^{-x}$, with $x$ given by \eq~(47) of \ssepaper. Therefore, 
\begin{align}
\label{eq:ce_rad_ratio}
\frac{R'_\star}{R_\star} = \left ( \frac{m_\d + m_\a}{m'} \right )^x.
\end{align}
Combining \eqs~(\ref{eq:ce_ebind_fin_merge}), (\ref{eq:ce_ebind_init_merge}), and (\ref{eq:ce_rad_ratio}), gives
\begin{align}
\label{eq:ce_mf}
\frac{E_{\bind,\,\fin}}{E_{\bind,\,\init}} = \left ( \frac{m'}{m_\d + m_\a} \right )^{1+x} \frac{m' - m'_{\core}}{m_\d + m_\a - m'_{\core}}.
\end{align}
With $E_{\bind,\,\init}$ given by \eq~(\ref{eq:ce_bind_init}), $E_{\bind,\,\fin}$ given by \eq~(\ref{eq:ce_energy2}), and the new core mass $m'_{\core}$ prescribed as above, \eq~(\ref{eq:ce_mf}) is solved (by Newton-Raphson iteration) for the final merger remnant mass $m'$. 

\paragraph{Initial mass and age} 
\label{sect:bin:ce:merger:init_age}
If the new star is an HG star ($k'=2$), then the initial mass is set to $m'_{\init} = m'$, and the age is calculated according to
\begin{align}
t'_\age = t_{\MS,\,\a} + (t_{\age,\,\d} - t_{\MS,\,\d}) \frac{t_{\BGB,\,\a} - t_{\MS,\,\a}}{t_{\BGB,\,\d} - t_{\MS,\,\d}}.
\end{align}

If the new star is a stripped He star ($k'=7$), then $m'_{\init} = m'$, and the new age depends on the amount of He that has been burnt in the progenitor stars,
\begin{align}
\label{eq:tage_merge_hg}
t'_\age = t_{\MS,\,\d} \frac{y_\d \, m_{\core,\,\d} + y_\a \, m_{\core,\,\a}}{m_{\core,\,\d} + m_{\core,\,\a}}.
\end{align}
Here, $m_{\core,\,\a}$ is replaced with $m_\a$ if the accretor was a stripped He star ($k_\a=7$). The `age factors' $y_\d$ and $y_\a$ are determined according to
\begin{align}
y_\d = \left \{ \begin{array}{cc}
\displaystyle 0, & k_\d \in \{ 0,1,2,3\}; \\
\displaystyle 1, & k_\d \in \{ 6,7,8,9,10,11,12,13,14\}; \\
\displaystyle \frac{t_{\age,\,\d} - t_{\BGB,\,\d}}{t_{\mathrm{DU},\,\d} - t_{\BGB,\,\d}}, & k_\d \in \{4, 5\},
\end{array} \right.
\end{align}
for the donor, where $t_{\mathrm{DU}}$ is the time of second dredge-up at the start of the TPAGB phase (cf. \eq~70 of \ssepaper), and
\begin{align}
y_\a = \left \{ \begin{array}{cc}
\displaystyle 0, & k_\a \in \{ 0,1,2,3, 10\}; \\
\displaystyle 1, & k_\a \in \{ 6,8,9,11,12,13,14\}; \\
\displaystyle \frac{t_{\age,\,\a}}{t_{\MS,\,\a}}, & k_\a \in \{7\}; \\
\displaystyle \frac{t_{\age,\,\a} - t_{\BGB,\,\a}}{t_{\mathrm{DU},\,\a} - t_{\BGB,\,\a}}, & k_\a \in \{4, 5\},
\end{array} \right.
\end{align}
for the accretor. 

For giant-like merger remnant stars ($k' \in \{3,4,5,6,9\}$), the new initial mass and age are determined similarly as described in \S~\ref{sect:bin:col:mass_age:giant}.

\paragraph{Spin}
\label{sect:bin:ce:merger:spin}
By default, the merger remnant's spin is assumed to be aligned with the previous orbital angular momentum, $\hat{\ve{\Omega}}' = \hat{\ve{h}}$, and its magnitude is set to the orbital angular frequency corresponding to the orbit just prior to coalescence (cf. \eq~\ref{eq:e_orb_fin_rlof}), unless the latter would imply above-critical rotation ($\Omega' > \Omega'_\mathrm{crit}$, where $\Omega_\mathrm{crit}$ is generally given by
\begin{align}
\label{eq:omega_crit}
\Omega_\mathrm{crit} = \sqrt{\frac{ \gconst m}{R^3_{\star}}}.
\end{align} 
Optionally, the spin of the surviving star can be let unaffected, i.e., the remnant spin frequency is equal to the spin frequency of the donor star just before the CE event. 

\subsubsection{Surviving binary}
\label{sect:bin:ce:survive}
If both stars do not fill their Roche lobe in the new orbit with semimajor axis $a_{\fin}$ (cf. \eq~\ref{eq:ce_afin}), then the binary is assumed to survive with its orbital angular-momentum vector unaffected in its direction. 

\paragraph{Stellar properties}
\label{sect:bin:ce:survive:star}
The donor's envelope is assumed to be stripped, so $m'_\d = m_{\core,\,\d}$. The secondary star is assumed to be unaffected in its mass if it was an MS star ($k_\a \in  \{0,1,7\}$) so $m'_\a = m_\a$, whereas, otherwise, its envelope is assumed to stripped as well so $m'_\a = m_{\core,\,\a}$. Other stellar properties such as the new age are determined by calling the relevant routines from \sse~to the new stars with the adjusted masses. The spins of the two stars are either assumed to be unaffected by the CE event, or they are assumed to become aligned and cororating with the new orbit. 

\paragraph{New orbital properties}
\label{sect:bin:ce:survive:orb}
The orbital angular-momentum and eccentricity vectors are assumed to be not affected in their direction by the CE event. We assign a new eccentricity to the binary orbit as follows.

Generally, the orbital angular momentum $h$ is related to the orbital energy $E_{\orb}$ according to 
\begin{align}
h^2 \propto \frac{1-e^2}{E_{\orb}},
\end{align}
where we ignored the mass dependence. Assuming that orbital energy is dissipated prior to orbital angular momentum, $h$ can be taken to be constant, such that the initial and final orbital eccentricity are related according to
\begin{align}
\label{eq:ce_surv_ecc}
\frac{1-e_{\init}^2}{E_{\orb,\,\init}} = \frac{1-e_{\fin}^2}{E_{\orb,\,\fin}}.
\end{align}
If $E_{\orb,\,\fin}< E_{\orb,\,\init}$, then we compute the final eccentricity $e_\fin$ from \eq~(\ref{eq:ce_surv_ecc}). Otherwise, we assume that the final orbit is circular. 

Similarly to \Ss~\ref{sect:bin:mt:dynlow} and \ref{sect:bin:mt:dynwd}, we update any orbits exterior to the orbit undergoing CE evolution as described in \S~\ref{sect:bin:mlorbit} following mass loss during the CE, with a default mass loss timescale $\tau_{\dot{m},\,\CE} = 10^3\,\yr$ (e.g., \citealt{2019MNRAS.484.4711M}).

\subsection{Wind accretion}
\label{sect:bin:wa}
Similarly to \bsepaper, we adopt the Bondi-Hoyle-Lyttleton formalism \citep{1939PCPS...35..405H,1944MNRAS.104..273B} to model accretion of material ejected as a wind from the companion star. In \mse, the wind accretion rate of an object $i$ in an orbit $k$ with a wind-losing companion star $j$ is given by (e.g., \citealt{1988A&A...205..155B})
\begin{align}
\label{eq:wa}
\dot{m}_{i,\,\WA} = - \dot{m}_{j,\,\sse} \times \mathrm{min} \left [1, \frac{1}{\sqrt{1-e_k^2}} \left ( \frac{\gconst m_i}{v_{\mathrm{W},\,j}^2} \right )^2 \frac{\alpha_\WA}{2 a_k^2} \left ( 1 + \frac{v_{\orb,\,k}^2}{v_{\mathrm{W},\,j}^2} \right )^{-3/2} \right ],
\end{align}
where $\alpha_\WA=3/2$ is a (user-adjustable) wind accretion parameter,
\begin{align}
v^2_{\orb,\,k} = \frac{\gconst (m_i+m_j)}{a_k}
\end{align}
is the squared (circular) orbital speed, and the wind speed from the wind-losing star $j$ is set to be proportional to the escape speed from its surface,
\begin{align}
v_{\mathrm{W},\,j}^2 = 2 \beta_\mathrm{W} \frac{\gconst m_j}{R_{\star,\,j}},
\end{align}
with $\beta_\mathrm{W}=0.125$ by default. The minimum operator in \eq~(\ref{eq:wa}) ensures that the rate of wind accretion by the companion does not exceed the rate at which the mass is lost by the wind-losing star. The wind accretion rate is computed in \mse~according to \eq~(\ref{eq:wa}) before the ODE integration, and it is assumed to be constant during the integration (cf. \S~\ref{sect:stellar:sse}). To compute the orbital response to wind accretion, we simply add $\dot{m}_{i,\,\WA}$ for each object due to wind accretion to the wind mass loss rate. The total mass rate change related to stellar winds, 
\begin{align}
\dot{m}_{i,\,\wind} = \dot{m}_{i,\,\sse} + \dot{m}_{i,\,\WA},
\end{align}
is used to determine $\dot{M}_{k,\,\mathrm{wind}}$ (cf. \eq~\ref{eq:adotwind}). 

Currently, we implement wind accretion of any physical star losing mass in a wind onto its companion. The companion can be another star, or a binary orbit. Here, we implicitly assume that wind accretion onto a single star occurs similarly as accretion onto a binary, which is likely not accurate. Regardless, if the companion is a binary, then the orbit of the wind mass losing star is relatively wide by necessity, in order to guarantee dynamical stability. The wind accretion rate is therefore small (mostly due to the factor $1/a_k^2$ in \eq~\ref{eq:wa}).

\subsection{Collisions}
\label{sect:bin:col}
In addition to `failed' CE events (cf. \S~\ref{sect:bin:ce:merger}), mergers can also occur as a result of direct physical collisions, e.g., following a dynamical instability in the system. In the latter case, the collision is likely `hard' (with high relative impact speed), whereas mergers during CE events are typically expected to be more `soft' (lower relative impact speed). We adopt similar prescriptions for physical collisions as \bsepaper, and remark that similar caveats apply when using such simplified prescriptions as for mergers following CE evolution (see \S~\ref{sect:bin:ce:merger}). 

We check for physical collisions between two stars during the gravitational dynamical evolution\footnote{This excludes dynamical disruptions such as dynamical mass transfer from a low-mass MS donor (see \S~\ref{sect:bin:mt:dynlow}), and dynamical mass transfer from a WD donor (see \S~\ref{sect:bin:mt:dynwd}).}. Specifically, when in secular integration mode, we check for collision between two stars at periapsis, 
\begin{align}
\label{eq:colsec}
a_k(1-e_k) \leq R_{\star,\,\eff,\,i} + R_{\star,\,\eff,\,j},
\end{align}
where $i$ and $j$ refer to the two stars in orbit $k$. For computational reasons and for the purposes of collision handling only, instead of the physical radii, we use the `effective' radii $R_{\star,\,\eff,\,i}$, which are defined as
\begin{align}
R_{\star,\,\eff,\,i} \equiv f_\mathrm{col} \,R_{\star,\,i}.
\end{align}
The parameter $f_\mathrm{col}$ depends on the integration mode and stellar type.

In the secular integration mode, $f_\mathrm{col}=1$, except for compact objects, when we set $f_\mathrm{col}=10^3$ by default. BHs and NSs in \sse~and \mse~are assigned their Schwarzschild radii which are small compared to the orbital separation, except for merging compact objects in the lasts moments before coalescence. However, integrating the secular equations of motion just before merger is computationally very expensive since the 1PN apsidal motion rate diverges as $a_k(1-e_k) \rightarrow 0$ (cf. \eq~\ref{eq:1PN}). In practice, a tight binary with compact objects becomes decoupled from external secular dynamical excitation well before it merges. For example, for a circular orbit $k$ with $a_k = f_\mathrm{col} \, G M_k/c^2$, the remaining merger time is \citep{1964PhRv..136.1224P}
\begin{align}
\nonumber t_{\mathrm{GW},\,k}  &= \frac{5}{256} f_\mathrm{col}^4 \frac{M_k^2}{m_i m_{\mathrm{comp},\,i}} \frac{GM_k}{c^3} \\
&\simeq 167 \,\yr \, \left ( \frac{f_\mathrm{col}}{1000} \right )^4 \left ( \frac{M_k}{40\,\msun} \right )^3 \left ( \frac{m_i}{20 \, \msun} \right )^{-1} \left ( \frac{m_{\mathrm{comp},\,i}}{20 \, \msun} \right )^{-1}
\end{align}
(this estimate is conservative, since the merger time would be shorter for an eccentric orbit). Therefore, when the condition \eq~(\ref{eq:colsec}) is met, the orbit is effectively decoupled from secular evolution. The remaining evolution of the binary $k$ until coalescence\footnote{Strictly speaking, the equations of \citet{1964PhRv..136.1224P} apply in the PN limit, which breaks down at separations on the order of the gravitational radii.} follows from the orbit-averaged equations for $a_k$ and $e_k$ from \citet{1964PhRv..136.1224P}.

When integrating directly, we check for the condition
\begin{align}
\left | \left | \ve{R}_i - \ve{R}_j \right | \right | \leq R_{\star,\,\eff,\,i} + R_{\star,\,\eff,\,j} 
\end{align}
for each pair of stars $(i,j)$. Since tidal evolution is not (yet) included in the direct $N$-body mode, we set $f_\mathrm{col}=3$ for non-compact objects ($k_i < 10$). For compact objects, we set  $f_\mathrm{col}=10^3$, similarly to the case of secular integration. 

When two stars collide, we stop the dynamical integration, and handle the collision depending on the stellar properties. If the collision involves a giant-like star ($k\in \{2,3,4,5,6,8,9\}$) with any other star, we invoke CE evolution (\S~\ref{sect:bin:ce}) with the giant-like star as the donor, and the other star as the accretor. Otherwise, a separate routine is used to handle the collision (mostly following \bsepaper, Section 2.7.3). The latter is described below (remainder of \S~\ref{sect:bin:col}).

\begin{table*}
\begin{center}
\begin{tabular}{lccccccccccccccccc}
\toprule 
 & \multicolumn{15}{c}{$k_i$} \\
& & 0 & 1 & 2 & 3 & 4 & 5 & 6 & 7 & 8 & 9 &10 & 11 & 12 & 13 & 14 \\
\midrule
	& 0 & 1 & 1 & 2 & 3 & 4 & 5 & 6 & 4 & 6 & 6 & 3 & 6 & 6 & 13 & 14 \\
	& 1 & 1 & 1 & 2 & 3 & 4 & 5 & 6 & 4 & 6 & 6 & 3 & 6 & 6 & 13 & 14 \\
	& 2 & 2 & 2 & 3 & 3 & 4 & 4 & 5 & 4 & 4 & 4 & 3 & 5 & 5 & 13 & 14 \\
	& 3 & 3 & 3 & 3 & 3 & 4 & 4 & 5 & 4 & 4 & 4 & 3 & 5 & 5 & 13 & 14 \\
	& 4 & 4 & 4 & 4 & 4 & 4 & 4 & 4 & 4 & 4 & 4 & 4 & 4 & 4 & 13 & 14 \\
	& 5 & 5 & 5 & 4 & 4 & 4 & 4 & 4 & 4 & 4 & 4 & 4 & 4 & 4 & 13 & 14 \\
	& 6 & 6 & 6 & 5 & 5 & 4 & 4 & 6 & 4 & 6 & 6 & 5 & 6 & 6 & 13 & 14 \\
$k_j$& 7 & 4 & 4 & 4 & 4 & 4 & 4 & 4 & 7 & 7 & 7 & 15 & 9 & 9 & 13 & 14 \\
	& 8 & 6 & 6 & 4 & 4 & 4 & 4 & 6 & 8 & 8 & 9 & 7 & 9 & 9 & 13 & 14 \\
	& 9 & 6 & 6 & 4 & 4 & 4 & 4 & 6 & 9 & 9 & 9 & 7 & 9 & 9 & 13 & 14 \\
	& 10 & 3 & 3 & 3 & 3 & 4 & 4 & 5 & 7 & 7 & 7 & 15 & 9 & 9 & 13 & 14 \\
	& 11 & 6 & 6 & 5 & 5 & 4 & 4 & 6 & 9 & 9 & 9 & 9 & 11 & 12 & 13 & 14 \\
	& 12 & 6 & 6 & 5 & 5 & 4 & 4 & 6 & 9 & 9 & 9 & 9 & 12 & 12 & 13 & 14 \\
	& 13 & 13 & 13 & 13 & 13 & 13 & 13 & 13 & 13 & 13 & 13 & 13 & 13 & 13 & 13 &14 \\
	& 14 & 14 & 14 & 14 & 14 & 14 & 14 & 14 & 14 & 14 & 14 & 14 & 14 & 14 & 14 & 14 \\
\bottomrule
\end{tabular}
\end{center}
\caption{\small Merger table giving the stellar type of the merger remnant for a merger between two stars with stellar types $k_i$ and $k_j$. Reproduced from \bsepaper. }
\label{table:mergerst}
\end{table*}

\subsubsection{Stellar type}
\label{sect:bin:col:st}
The stellar type of the remnant object, $k'$, is determined according to Table~\ref{table:mergerst}. Note that $k'$ in some cases deviates from this table (see below).

\subsubsection{New mass and age}
\label{sect:bin:col:mass_age}
\paragraph{MS-MS}
\label{sect:bin:col:mass_age:MS}
When two MS stars including stripped He stars (both $k \in \{0,1,7\}$) collide, no mass is assumed to be lost so $m' = m_i+m_j$, and the new initial mass is $m'_{\init} = m'$. The age of the remnant MS star is determined according to \citep{1997MNRAS.291..732T}
\begin{align}
t'_{\age} = 0.1 \, \frac{t'_{\MS}}{m'} \left ( \frac{m_i t_{\age,\,i}}{t_{\MS,\,i}} + \frac{m_j t_{\age,\,j}}{t_{\MS,\,j}} \right ).
\end{align}

\paragraph{New giant star}
\label{sect:bin:col:mass_age:giant}
If an MS star ($k \in \{0,1\}$) collides with a WD ($k \in \{10,11,12\}$), a giant star is assumed to form with a core mass $m'_\core$ given by the compact object's mass (with the additional assumption of no mass loss in the envelope). The new age and initial mass of the giant are determined as follows (see \S~2.7.4 of \bsepaper~for more details). If $k'=3$, then the new initial mass, $m'_\mathrm{init}$, and new age, $t'_\mathrm{age}$, are determined by an iterative process such that the new giant star with core mass $m'_\core$ is placed at the base of the RGB. If $k'=4$, then a bisection method is used as described in \S~2.7.4 of \bsepaper.  If $k'\in\{5,6\}$, the initial age and mass are determined such that the star is placed at the base of the AGB. If $k'\in \{8,9\}$, the initial mass and age are such that the new giant starts at the end of the stripped He star MS phase. 

\paragraph{MS-He star}
An MS star (including He star, $k \in \{0,1,7\}$ colliding with an NS or BH is assumed to form an unstable Thorne-$\mathrm{\dot{Z}}$ytkow object \citep{1977ApJ...212..832T} or quasi-star (e.g., \citealt{2008MNRAS.387.1649B}), respectively, removing all mass of the MS star and leaving only the compact object. 

\paragraph{He star-He WD}
If a stripped He star ($k=7$) collides with an He WD ($k=10$), then the He star is rejuvenated by absorbing the He WD without mass loss, $m' = m_i+m_j$. The new age of the He star is given by
\begin{align}
t'_{\age} = t'_{\MS} \frac{m_{\mathrm{He\,star}}}{m'} \frac{t_{\age,\,\mathrm{He\,star}}}{t_{\MS,\,\mathrm{He\,star}}}.
\end{align}

\paragraph{He star-CO/ONe WD}
If a stripped He star ($k=7$) collides with a CO or ONe WD ($k\in\{11,12\}$), then an evolved He star is formed ($k'=9$; possibly resembling an R Coronae Borealis star, e.g., \citealt{2011ApJ...737L..34L}) with a core mass given by the CO or ONe WD mass. The new age and initial mass of the evolved He star are determined similarly as in \S~\ref{sect:bin:col:mass_age:giant}. 

\paragraph{He WDs}
Following \bsepaper, a collision of two He WDs is assumed to lead to a nuclear runaway explosion, destroying the two stars. 

\paragraph{He WD - CO/ONe WD}
An He WD colliding with a CO or ONe WD is assumed to lead to an evolved He star with the core mass determined by the CO or ONe WD. The new age and initial mass of the evolved He star are determined similarly as in \S~\ref{sect:bin:col:mass_age:giant}. 

\paragraph{CO WDs}
Two colliding CO WDs are assumed to result in single CO WD with no mass loss, except if the new mass $m'=m_i+m_j > M_\mathrm{Ch} = 1.44\,\msun$ (e.g., \citealt{2016MNRAS.463.3461S,2019Natur.569..684G}).

\paragraph{CO WD - ONe WD}
An ONe WD colliding with a CO or ONe WD is assumed to result in an ONe WD without mass loss if $m'=m_i+m_j \leq M_\mathrm{Ch}$; if $m'=m_i+m_j > M_\mathrm{Ch}$, then accretion-induced collapse is assumed to result in an NS without mass loss (e.g., \citealt{1991ApJ...367L..19N,1997A&A...317L...9V}).

\paragraph{NS/BH}
When two NS and/or BHs collide, we use the empirical fits to numerical relativity simulations of \citet{2010CQGra..27k4006L} to compute the remnant compact object mass, spin, and recoil velocity, as a function of the initial masses and spins.

\subsubsection{New position and velocity}
\label{sect:bin:col:pos_vel}
The position vector of the collision product is set to the centre of mass position of the two colliding objects just prior to collision. The velocity is computed by assuming linear momentum conservation of the initial two colliding objects. Any potential kicks (e.g., due to BH merger recoil) are added as well to the collision product's velocity. 

We assume that any mass lost during the collision event occurs instantaneously (this is in contrast to mass loss on a finite lifetime, see \S~\ref{sect:bin:mlorbit}). Directly after the collision, the integration mode is set to direct $N$-body integration, since dynamical stability is not generally guaranteed after the mass loss and/or kicks in the system.

\subsubsection{New spin}
\label{sect:bin:col:pos_vel}
The spin frequency of the collision product is assumed to be aligned with the orbital angular momentum vector just prior to collision. Its magnitude is set to the orbital mean motion just prior to collision, unless the latter would imply critical rotation (cf. \eq~\ref{eq:omega_crit}). In the case of BH/NS mergers, the new spin is determined by the fits of \citet{2010CQGra..27k4006L}.

\subsection{Effect of non-instantaneous mass changes on external orbits}
\label{sect:bin:mlorbit}
Some of the processes described in the above sections (e.g., CE evolution) involve mass loss from an inner binary system consisting of two single stars, occurring on a finite timescale (i.e., not instantaneous as in \S~\ref{sect:bin:col}). Such mass loss will affect any orbits exterior to this inner binary system. Here, we describe how these effects are taken into account in \mse. 

Generally, one can identify between different regimes of mass loss affecting the exterior orbits depending on the relation between the timescale of the mass loss, $\tau_{\dot{m}}$, to the exterior orbital timescale, $P_{\orb,\,k}$. If $\tau_{\dot{m}} \ll P_{\orb,\,k}$ for an orbit $k$, then the mass loss is effectively instantaneous and the new orbital elements can be determined by generating a realisation of the positions and velocities of the old system, updating the masses of the now merged object in the inner orbit (possibly combined with a recoil velocity), and determining the new orbits. The latter is identical to our approach to model the orbital effects of SNe in \mse~(\S~\ref{sect:stellar:sne:gen}). On the other hand, if $\tau_{\dot{m}} \gg P_{\orb\,k}$, then mass loss is adiabatic, such that $M_k a_k$ is conserved (cf. \eq~\ref{eq:adotwind}). 

In the intermediate regime, $\tau_{\dot{m}} \sim P_{\orb,\,k}$, the effects of mass loss on external orbits are difficult to describe analytically (see, e.g., \citealt{2011MNRAS.417.2104V}). For generality, we therefore follow the following scheme to handle mass loss which is assumed to occur on a timescale $\tau_{\dot{m}}$. This scheme, while being more computationally demanding, captures the two analytic regimes discussed above but also describes the intermediate regime. 

The positions and velocities of all bodies in the system are integrated directly for a duration of $\tau_{\dot{m}}$. For the purposes of this integration and where applicable, we replace the two mass-losing objects (e.g., the two stars undergoing CE evolution) with a point mass. Mass loss of the mass-losing object is taken into account by splitting the integration over $\tau_{\dot{m}}$ into $N_{\mathrm{mass\,loss,\,split}}$ equal time segments (default value $N_{\mathrm{mass\,loss,\,split}}=100$), and adjusting the mass of the body representing the two mass-losing objects in equal steps. 

After the integration, if a CE occurred and the binary survives, we update the positions and velocities of the two stars in the new orbit computed according to the CE prescription (cf. \S~\ref{sect:bin:ce:survive:orb}), with the center of mass given by the body that represented the two stars during the $N$-body integration, and assuming that $\unit{h}$ and $\unit{e}$ of the orbit undergoing the CE event did not change. 

We motivate the replacement of the two mass-losing objects with a point mass during the $N$-body integration by noting that surviving post-CE orbits are typically very compact (with separations significantly smaller than, say, $1\,\au$), and their binarity is therefore expected to be unimportant for the dynamics of exterior bodies. Moreover, integration of such a tight orbit is computationally expensive. Also, in our approach we assume that mass loss occurs linearly with time\footnote{Other dependencies of mass loss rate on time could be implemented in the future.}, and any mass lost immediately escapes the system (and does not interact with other bodies). 

In the case of a direct collision between two stars which does not lead to CE evolution (\S~\ref{sect:bin:col}), we assume that the mass loss is always instantaneous, i.e., $\tau_{\dot{m}} \ll P_{\orb\,k}$ for all exterior orbits, and the direct $N$-body integration into $N_{\mathrm{mass\,loss,\,split}}$ steps with changing masses does not apply.

\section{Triple evolution}
\label{sect:triple}
Here, we discuss several evolutionary processes associated with a single star orbiting around and interacting with a companion inner binary. This `triple subsystem' could be part of a larger, more complicated hierarchical system. Currently, we only model a subset of possible interactions in triple subsystems. 

\subsection{Triple CE evolution}
\label{sect:triple:ce}
If the criterion for CE evolution is satisfied (\S~\ref{sect:bin:mt}; item ii) but the companion object is a binary instead of a single star (the binary is required to consist of two physical stars, i.e., no further nesting is allowed), then we invoke `circumstellar triple CE evolution', i.e., an outer tertiary star fills its Roche lobe around the inner binary and undergoes unstable mass transfer. Generally, many uncertainties remain in CE evolution in {\it binary} stars, with the details of how the envelope can be efficiently ejected being unclear (see, e.g., \citealt{2013A&ARv..21...59I} for a review). CE evolution in {\it triple} stars is even more uncertain. Some pioneering studies of the circumstellar case (e.g., \citealt{2021MNRAS.500.1921G,2020MNRAS.498.2957C}) indicate that a variety of outcomes is possible, including the merger of the binary inside the tertiary's envelope, or the dynamical disruption of the binary leading to a chaotic interaction between the binary components and the core of the tertiary star. Furthermore, when compared with an equivalent binary CE case, the inspiral of the binary is typically slower, more mass is ejected, and the remnant is more aspherical.

Modelling such aspects in detail with hydrodynamical simulations is beyond the scope and capabilities of \mse. Instead, we here devise a simple ad hoc scheme to be able to at least qualitatively capture some of the effects that are expected to occur. We emphasise that this scheme is extremely simplified. Guided by detailed simulations, a more sophisticated scheme could be implemented in the future. 

Let properties associated with the tertiary star, the donor, be indicated with `$\d$'. We assume that the inner binary (component masses $m_1$ and $m_2$, with $m\equiv m_1+m_2$)  is relatively compact. First, assuming that the tertiary star's envelope is shed entirely, we estimate the separation of the outer orbit, $a_{\mathrm{out}}$, using the $\alpha$ CE formalism, similarly to the binary CE case. Here, we do not consider the orbital energy of the inner binary and assume that the latter is unaffected, since, without more detailed modelling, it is unclear how the inner orbital energy changes during the inspiral. The initial tertiary star's binding energy is
\begin{align}
\label{eq:tce_bind_init}
E_{\bind,\,\init} = - \frac{\gconst m_\d (m_\d - m_{\core,\,\d})}{\lambda_{\CE,\,\d} R_{\star,\,\d} }.
\end{align}
The initial outer orbital energy is given by
\begin{align}
E_{\orb,\,\mathrm{out},\,\init} = -\frac{\gconst m_\d (m_1+m_2)}{2 a_{\mathrm{out},\,\init}},
\end{align}
with $a_{\mathrm{out},\,\init}$ the initial semimajor axis of the outer orbit. The final orbital energy is given by
\begin{align}
\label{eq:tce_energy1}
E_{\orb,\,\mathrm{out},\,\fin} = E_{\orb,\,\mathrm{out},\,\init} + \frac{E_{\bind,\,\init}}{\alpha_{\TCE}},
\end{align}
where $\alpha_{\TCE}$ is the triple CE $\alpha$ parameter (in \mse, $\alpha_{\TCE}$ is distinct from the binary CE parameter $\alpha_\CE$). The corresponding putative final outer orbital separation is
\begin{align}
\label{eq:tce_afin}
a_{\mathrm{out},\,\fin} = - \frac{\gconst m_{\core,\,\d} (m_1+m_2)}{2 E_{\orb,\,\mathrm{out},\,\fin}}.
\end{align}
Following similar arguments as in \S~(\ref{sect:bin:ce:survive:orb}), we assign a final eccentricity to the outer orbit determined by
\begin{align}
\label{eq:tce_surv_ecc}
\frac{1-e_{\mathrm{out},\,\init}^2}{E_{\orb,\mathrm{out},\,\,\init}} = \frac{1-e_{\mathrm{out},\,\fin}^2}{E_{\orb,\,\mathrm{out},\,\fin}},
\end{align}
if $E_{\orb,\,\mathrm{out},\,\fin}< E_{\orb,\,\mathrm{out},\,\init}$. Otherwise, we assume that the final outer orbit is circular. 

Next, we check if the new outer orbit would be wide enough for the triple (sub)system to be dynamically stable, assuming that the inner binary is not affected in its separation during its inspiral, and ignoring all hydrodynamical effects. We use \eq~(\ref{eq:dynstab}) to evaluate stability; here, we assume that the donor star always fully loses its envelope, so $m'_\d = m_{\core,\,\d}$ and $q_\mathrm{out} = m'_\d/(m_1+m_2)$. The donor's new properties are determined from the relevant \sse~routines with the new mass. 

If the putative outer orbit is such that the triple subsystem would be dynamically stable, then we adjust the outer orbital properties according to \eqs~(\ref{eq:tce_afin}) and (\ref{eq:tce_surv_ecc}). The donor's spin is assumed to be either unaffected, or aligned with and corotating with the new outer orbit (bounded by critical rotation). The effect of mass loss in the triple system on possible external orbits is taken into account similarly to the case of binary CE, with the same mass-loss timescale (cf. \S~\ref{sect:bin:mlorbit}). 

Otherwise, i.e., if the outer orbit would be unstable, then the the outer orbital semimajor axis is set to the critical value for dynamical instability (cf. \eq~\ref{eq:dynstab}), with $e_{\mathrm{out},\,\fin} = 0$. The integration mode is switched to direct integration, with the positions and velocities of the three stars updated (including the new donor star mass). Effects of mass loss on the other orbits are, in this case, taken into account assuming instantaneous mass loss. The dynamical interaction between the inner binary and the core of the donor star are taken into account in \mse~in subsequent evolution by means of direct $N$-body integration. Evidently, we here neglect any hydrodynamical effects. Nevertheless, this simplified approach is able to capture some of the effects seen in detailed simulations (e.g., \citealt{2021MNRAS.500.1921G}), such as collisions or ejections of components during the triple CE event. We note that our approach is similar to the prescription proposed by \citet{2020MNRAS.498.2957C}.

\subsection{Triple mass transfer}
\label{sect:triple:mt}
If the criterion for stable mass transfer evolution is satisfied (\S~\ref{sect:bin:mt}; item iv) but the companion is a binary instead of a physical star, we invoke `circumstellar triple mass transfer', i.e., mass transfer of a tertiary star onto an inner binary. Such evolution can occur in a subset of systems and produce a variety of phenomena such as SNe Ia and GW sources \citep{2020MNRAS.496.1819L}. Currently, only transfer onto a companion binary consisting itself of two physical stars is modelled in \mse. 

Similarly to triple CE, many uncertainties remain in the evolution of mass transfer in triple systems. The circumstellar case was studied with detailed hydrodynamical simulations by \citet{2014MNRAS.438.1909D}. Here, we use the results of \citet{2014MNRAS.438.1909D} to motivate a highly simplified prescription for circumstellar triple mass transfer, as described below.

\subsubsection{Mass transfer amounts}
\label{sect:triple:mt:delta}
First, we estimate the amount of mass lost from the donor, $\Delta m_\d$, using the same scheme which is used for binary mass transfer (see \S~\ref{sect:bin:mt:stable:loss}).  The accreted amount onto the inner binary is assumed to depend on whether or not an accretion disk can form around the inner binary. Letting the effective size of the inner binary be its apoapsis distance, we assume that an accretion disk forms if
\begin{align}
a_\mathrm{in}(1+e_\mathrm{in}) > r_\mathrm{min,\,triple},
\end{align}
where $r_\mathrm{min,\,triple}$ is given by \citep{1976ApJ...206..509U}
\begin{align}
\label{eq:trmin}
r_\mathrm{min,\,triple} = 0.0425 \, a_{\mathrm{out}}(1-e_\mathrm{out}) \left [ \frac{m_1+m_2}{m_\d} \left (1 + \frac{m_1+m_2}{m_\d} \right ) \right ]^{0.25}.
\end{align}
In the simulations \citet{2014MNRAS.438.1909D}, no accretion disk formed, and only little mass was accreted onto the inner binary, with most mass ejected from the system. Therefore, if no disk forms, we assume that the accretion efficiency is low, and
\begin{align}
\left \{
\begin{array}{llc}
\Delta m_1 &= \alpha_{\mathrm{TMT,\,no\,disk,\,prim}} \, \Delta m_\d; & \\
\Delta m_2 &= \alpha_{\mathrm{TMT,\,no\,disk,\,sec}} \, \Delta m_\d, & \\
\end{array}
\right. 
\end{align}
where $ \alpha_{\mathrm{TMT,\,no\,disk,\,prim}}$ and $\alpha_{\mathrm{TMT,\,no\,disk,\,sec}}$ are user-adjustable parameters which are set to 0.1 by default. If, however, a disk forms, we assume that the accretion efficiency is much higher, and 
\begin{align}
\left \{
\begin{array}{llc}
\Delta m_1 &= \alpha_{\mathrm{TMT,\,disk,\,prim}} \, \Delta m_\d; & \\
\Delta m_2 &= \alpha_{\mathrm{TMT,\,disk,\,sec}} \, \Delta m_\d, & \\
\end{array}
\right. 
\end{align}
where the user-adjustable $\alpha_{\mathrm{TMT,\,disk,\,prim}}$ and $\alpha_{\mathrm{TMT,\,disk,\,sec}}$ are set to 0.9 by default. 

\subsubsection{Inner binary evolution}
\label{sect:triple:mt:in}
In their simulations, \citet{2014MNRAS.438.1909D} found that the inner binary evolution can be described as CE-like events, in which the inner binary orbital energy is used to expel the `binding' energy of the material entering the inner binary. They estimate the latter as
\begin{align}
E_{\bind,\,\init} = - \frac{\gconst (m_1+m_2) \Delta m_\d}{\lambda_{\mathrm{TMT}} a_{\mathrm{in},\,\init}},
\end{align}
where $\lambda_{\mathrm{TMT}}$ is an effective `structure' parameter of the material entering the inner binary. With the usual definition of $\alpha_\mathrm{TMT}$ for CE evolution,
\begin{align}
\nonumber E_{\bind,\,\init} &= \alpha_{\mathrm{TMT}} \left ( E_{\orb,\,\mathrm{in},\,\fin} - E_{\orb,\,\mathrm{in},\,\init} \right ) \\
\quad &= \alpha_{\mathrm{TMT}} \left ( \frac{\gconst m_1 m_2}{2 a_{\mathrm{in},\,\init}} - \frac{\gconst (m_1+\Delta m_1)(m_2 + \Delta m_2)}{2 a_{\mathrm{in},\,\fin}} \right ).
\end{align}
This implies
\begin{align}
\label{eq:tmt_ainfin}
\displaystyle a_{\mathrm{in},\,\fin} = a_{\mathrm{in},\,\init} \frac{ (m_1+\Delta m_1)(m_2 + \Delta m_2)}{m_1 m_2 + \displaystyle \frac{2 (m_1+m_2) \Delta m_\d}{\alpha_{\mathrm{TMT}} \lambda_{\mathrm{TMT}}}}.
\end{align}
From their simulations, \citet{2014MNRAS.438.1909D} inferred the corresponding CE parameters in the form of the product $(\alpha \lambda)_{\mathrm{TMT}} \equiv \alpha_{\mathrm{TMT}} \lambda_{\mathrm{TMT}}$ for several cases. Typically, $(\alpha \lambda)_{\mathrm{TMT}} \sim 5$. In \mse, we compute the rate of change of $a_\mathrm{in}$ according to
\begin{align}
\dot{a}_\mathrm{in} = \frac{a_{\mathrm{in},\,\fin} - a_{\mathrm{in},\,\init}}{\deltatode},
\end{align}
with $a_{\mathrm{in},\,\fin}$ given by \eq~(\ref{eq:tmt_ainfin}). This rate of change is used to evolve the inner orbit during the ODE integration. As a simplification, we assume that the inner orbit eccentricity is not affected by the triple mass transfer.

Any mass not accreted by the inner binary is assumed to leave the inner binary in an adiabatic wind. Therefore, to the inner binary, the wind mass loss rate is increased by
\begin{align}
- \frac{\Delta m_\d - (\Delta m_1 + \Delta m_2)}{\deltatode} \leq 0.
\end{align}

\subsubsection{Outer binary evolution}
\label{sect:triple:mt:out}
To describe the outer orbital evolution, \citet{2014MNRAS.438.1909D} used standard expressions for the orbital response to non-conservative mass transfer in circular orbits. We follow this approach, but adopt a slightly different formulation. Specifically (e.g., \citealt{1997A&A...327..620S}),
\begin{align}
\label{eq:tmt_aout}
\nonumber \frac{\dot{a}_\mathrm{out}}{a_\mathrm{out}} &= -2 \frac{\dot{m}_\d}{m_\d} \left [ 1 - \beta_\mathrm{TMT} \frac{m_\d}{m_1+m_2} \right. \\
&\qquad \left. - \left(1-\beta_\mathrm{TMT} \right ) \left (\gamma_\mathrm{TMT} + \frac{1}{2} \right ) \frac{m_\d}{m_\d + m_1 + m_2} \right ].
\end{align}
Here, $\beta_\mathrm{TMT}$ and $\gamma_\mathrm{TMT}$ parameterise the mass transfer efficiency and angular-momentum loss, respectively. In \mse, we include \eq~(\ref{eq:tmt_aout}) in the ODE integration, with $\dot{m}_\d = - \Delta m_\d/\deltatode$, $\beta_\mathrm{TMT} = -(\Delta m_1 + \Delta m_2)/\Delta m_\d$, and $\gamma_\mathrm{TMT} = m_\d/(m_1+m_2)$. The choice of $\gamma_\mathrm{TMT}$ corresponds to isotropic re-emission of the material which is not accreted by the inner binary \citep{1997A&A...327..620S}. When applying \eq~(\ref{eq:tmt_aout}), we implicitly assume that the outer orbit is circular. We ignore any possible complications when the outer orbit is not circular (e.g., due to secular eccentricity excitation from exterior companions).

\section{Fly-bys}
\label{sect:fly}
\subsection{Regimes}
\label{sect:fly:regime}
In \mse, we take into account the effects of stars passing by the multiple system. We restrict such fly-bys to the class of impulsive encounters, for which the relative motion of the passing star is much faster than the orbital motion. In addition, we currently do not account for the possibility of the perturber being a binary instead of a single star. 

The impulsive approximation is well justified for wide orbits. More quantitatively, the Keplerian orbital speed of an orbit $k$, assuming a circular orbit, is given by
\begin{align}
v_{\orb,\,k} = \sqrt{\frac{\gconst M_k}{a_k}} \simeq 0.94\,\kms\, \left (\frac{M_k}{1\,\msun} \right )^{1/2} \left ( \frac{a_k}{10^3\,\au} \right )^{-1/2},
\end{align}
which, for the numbers adopted, is significantly lower than the typical velocity dispersion in the Galactic disk, $\sim 40\,\kms$ \citep{2008gady.book.....B}. Orbits much more compact than $10^3\,\au$ can have orbital speeds comparable to the relative encounter speed. However, given the comparatively low density in the Galactic neighborhood, close encounters with such compact orbits are exceedingly rare, and the much more common distant encounters have a negligible effect on compact orbits. Quantitatively, in the regime where the perturber's orbit periapsis distance $\qper$ to the binary's center of mass is much larger than the binary orbit (semimajor axis $a_k$), the secular approximation applies \citep{1975MNRAS.173..729H,1996MNRAS.282.1064H,2018MNRAS.476.4139H,2019MNRAS.487.5630H}, and the change in the binary's eccentricity can be estimated as
\begin{align}
\label{eq:delta_ek_sec}
\nonumber \Delta e_k &\sim \varepsilon_{\mathrm{SA},\,k}\, \eper = \left [ \frac{\mper^2}{M_k(M_k+\mper)} \left ( \frac{a_k}{\qper} \right )^3 \left(1+\eper \right )^{-3} \right ]^{1/2}\, \eper \\
\nonumber &\simeq 7 \times 10^{-11} \, \left ( \frac{M_k}{1\,\msun} \right )^{-1/2} \left ( \frac{\mper}{1\,\msun} \right ) \left ( \frac{a_k}{1\,\au} \right )^{3/2} \left ( \frac{\qper}{10^5\,\au} \right )^{-2} \\
&\quad \times \left ( \frac{V_\infty}{40\,\kms} \right )^{-1},
\end{align}
where $V_\infty$ is the relative speed at infinity, and where we used in the second and third lines that the perturber's eccentricity $\eper$ for the numerical values adopted is $\eper = 1 + \qper V_\infty^2/[\gconst (M_k + \mper)] \sim 10^5 \gg 1$. \Eq~(\ref{eq:delta_ek_sec}) shows that the secular effects of distant encounters in the Galactic neighborhood on compact orbits are negligible.

\subsection{Impulsive encounters}
\label{sect:fly:imp}
In the impulsive approximation, the objects in the `internal' system are assumed to be stationary as the perturber passes by in a straight trajectory. An orbit $k$ can be affected in all its orbital elements. In \mse, we compute the new orbital elements by applying a velocity kick $\Delta \ve{V}_i$ to all bodies in the system, where $\Delta \ve{V}_i$ is computed as follows. The trajectory of the perturber relative to the multiple system's center of mass is given by
\begin{align}
\label{eq:straight_line}
\ve{R}_\mathrm{per}(t) = \ve{b} + \ve{V}_\mathrm{per} \, t,
\end{align}
where $\ve{b}$ is the impact parameter vector, and $\ve{V}_\mathrm{per}$ is the perturber's velocity. The perturber then imparts a velocity kick $\Delta \ve{V}_i$ on each body $i$ given by integrating the acceleration on body $i$, i.e.,
\begin{align}
\label{eq:imp}
\nonumber \Delta \ve{V}_i &= \int_{-\infty}^\infty \mathrm{d} t \, \gconst \mper \frac{\ve{b} + \ve{V}_\mathrm{per} \, t - \ve{R}_i}{ \left [ \left (\ve{b} - \ve{R}_i \right )^2 + 2\left(\ve{b}-\ve{R}_i \right ) \cdot \ve{V}_\mathrm{per} \, t + \ve{V}_\mathrm{per}^2 \, t^2 \right ]^{3/2}} \\
&= 2 \frac{\gconst \mper}{V_\mathrm{per}} \frac{\unit{b}_i}{b_i},
\end{align}
where we defined the impact parameter vector with respect to body $i$,
\begin{align}
\label{eq:bi_def}
\ve{b}_i \equiv \ve{b}-\ve{R}_i - \unit{V}_\mathrm{per} \left [ \left (\ve{b}-\ve{R}_i \right ) \cdot \unit{V}_\mathrm{per} \right ].
\end{align}

\subsection{Encounter sampling}
\label{sect:fly:sampl}
In \mse, we sample each encounter separately and compute its effect on all orbits as described above in \S~\ref{sect:fly:imp}. We adopt a similar methodology as used by \citet{2017AJ....154..272H} to generate encounters; for completeness, we repeat some of the material from \citet{2017AJ....154..272H} here.

\subsubsection{Distribution function and the encounter rate}
\label{sect:fly:sampl:distr}
We assume a locally homogeneous stellar background with stellar number density $n_\star$ and a one-dimensional velocity dispersion $\sigma$ independent of stellar mass. We assume a Maxwellian stellar velocity distribution at large distances from the multiple system and take into account gravitational focusing induced by the multiple system's gravity, such that the distribution function (DF) is given by
\begin{align}
\label{eq:DF2}
\mathrm{DF} &\propto n_\star f(\mper) \exp \left ( -\frac{v^2}{2\srel^2} \right ) \exp \left ( \frac{ \gconst (\mint+\mper)}{r \srel^2} \right )
\end{align}
for $v > \sqrt{2\gconst (\mint+\mper)/r}$ with $r$ the distance of the perturber to the multiple system's center of mass, and the $\mathrm{DF}$ is zero otherwise. Here, $f(\mper)\,\mathrm{d}\mper$ is the fraction of stars with masses in the interval $\mathrm{d} \mper$, $\srel= \sqrt{2} \, \sigma$ is the relative velocity dispersion \citep{2008gady.book.....B}, and $\mint$ is the total mass of the multiple-star system.

Consider an imaginary `encounter sphere' centered at the multiple system's center of mass, with a radius $\renc \gg a_\mathrm{int}$, and where $a_\mathrm{int} \equiv \mathrm{max}_k (a_k)$ is the largest semimajor axis in the multiple system. We assume that $\renc$ is large enough to satisfy $\renc \gg a_\mathrm{int}$, but it should not be too large for computational reasons as the rate of encounters approximately grows as $\renc^2$ (cf. \eq~\ref{eq:Gamma}). Encountering stars impinging on the encounter sphere are considered as perturbers. From equation~(\ref{eq:DF2}), the number density of perturbers at the encounter sphere within a mass range $\mathrm{d} \mper$ and with velocities between $(v_x,v_y,v_z)$ and $(v_x+\mathrm{d}v_x,v_y+\mathrm{d}v_y,v_z+\mathrm{d}v_z)$ can be derived to be
\begin{align}
\label{eq:dn}
\nonumber &\mathrm{d} {n_{\star,\,\mathrm{enc}}} = \frac{ n_\star}{(2\pi \srel^2 )^{3/2}} f(\mper) \, \mathrm{d} \mper \exp \left ( \frac{\gconst (\mint+\mper)}{\renc \srel^2} \right ) \\
&\quad \times H\left(v^2-\frac{2\gconst (\mint+\mper)}{\renc} \right ) \exp \left ( -\frac{v^2}{2 \srel^2} \right ) \, \mathrm{d} v_x \, \mathrm{d} v_y \, \mathrm{d} v_z,
\end{align}
where $H(x)$ is the Heaviside step function. Integration of equation~(\ref{eq:dn}) over all perturber masses and velocities gives
\begin{align}
\label{eq:n}
n_{\star,\mathrm{enc}} = n_\star \int \, \mathrm{d} \mper f(\mper) \, W \left ( \frac{\gconst (\mint + \mper)}{\renc \srel^2} \right ),
\end{align}
where
\begin{align}
\label{eq:W}
W(x) \equiv 2\sqrt{x/\pi} + \exp(x) \, \mathrm{erfc}\left (\sqrt{x} \right ),
\end{align}
and $\mathrm{erfc}(x) = 1 - \mathrm{erf}(x)$ is the complementary error function. The fraction of perturbers at the encounter sphere with mass $\mper$ is proportional to $W(x)f(\mper)$. Equation~(\ref{eq:n}) shows that the stellar number density at the encounter sphere, $n_{\star,\,\mathrm{enc}}$, is larger than $n_\star$ due to gravitational focusing. 

Consider a point on the encounter sphere with position vector $\ve{R}_\enc$ relative to the multiple system's center of mass. Next, define a local coordinate system centered on this point in which the $z$ axis is directed toward the host star, i.e., $\hat{\boldsymbol{z}} = - \unit{R}_\enc$, and the $x$ and $y$ axes lie on the tangent plane of $\ve{R}_\enc$ on the encounter sphere. The differential flux of stars into the encounter sphere is given by $\mathrm{d} F = v_z H(v_z) \,\mathrm{d} n_{\star,\,\mathrm{enc}}$ \citep{1972A&A....19..488H}, i.e.,
\begin{align}
\label{eq:vel_distr}
\nonumber &\mathrm{d} F = \frac{ n_\star}{(2\pi \srel^2 )^{3/2}} f(\mper) \, \mathrm{d} \mper \exp \left ( \frac{\gconst (\mint + \mper)}{\renc \srel^2} \right ) H(v_z) v_z \\
&\quad \times H\left(v^2-\frac{2\gconst (\mint + \mper)}{\renc} \right ) \exp \left ( -\frac{v^2}{2 \srel^2} \right ) \, \mathrm{d} v_x \, \mathrm{d} v_y \, \mathrm{d} v_z.
\end{align}
Integrating the differential flux over all perturber masses, velocities, and the entire encounter sphere, we obtain a total encounter rate of
\begin{align}
\label{eq:Gamma}
\nonumber \Gamma &= 2 \sqrt{2\pi} \renc^2 n_{\star} \srel \\
&\quad \times \int \mathrm{d} \mper f(\mper ) \left [1 + \frac{\gconst (\mint + \mper)}{\renc \srel^2} \right ].
\end{align}
In the limit of large $\renc$ ($\renc \gg \gconst [\mint + \mper]/\srel^2$, i.e., negligible gravitational focusing), equation~(\ref{eq:Gamma}) reduces to
\begin{align}
\label{eq:Gamma_limit}
\nonumber \Gamma &\approx 2 \sqrt{2\pi} \renc^2 n_{\star} \srel \int \mathrm{d} \mper f(\mper ) \\
&= 2 \sqrt{2\pi} \renc^2 n_{\star} \srel,
\end{align}
independent of the perturber mass function. 

\subsubsection{Generation procedure}
The following scheme is adopted in \mse~to incorporate the effects of passing stars.

\begin{enumerate}[leftmargin=0.5cm]
\item We sample an initial perturber mass, $\mper$, from either a Salpeter distribution, $\mathrm{d}N/\mathrm{d}\mper \propto \mper^{-2.35}$ \citep{1955ApJ...121..161S}, or a Kroupa distribution \citep{1993MNRAS.262..545K},
\begin{align}
\frac{\mathrm{d}N}{\mathrm{d}\mper} \propto \left \{ 
\begin{array}{ll}
\mper^{-\alpha_1}, & m_{\mathrm{Kr},1} < \mper < m_{\mathrm{Kr},2}; \\
\mper^{-\alpha_2}, & m_{\mathrm{Kr},2} < \mper < m_{\mathrm{Kr},3}; \\
\mper^{-\alpha_3}, & m_{\mathrm{Kr},3} < \mper < m_{\mathrm{Kr},4}, \\
\end{array} \right.
\label{eq:kroupaimf}
\end{align}
where $\alpha_j = \{1.3,2.2,2.7\}$ and $m_{\mathrm{Kr},j} /\msun = \{0.1,0.5,1,100\}$. The assumed mass range for both distributions is $0.1\,\msun < \mper < 100\,\msun$. 

We modify the sampled mass as described above to account for gravitational focusing by the multiple system implied by equation~(\ref{eq:n}). Specifically, given the initially sampled value of $\mper$, we compute the associated value of 
\begin{align}
x \equiv \frac{\gconst (\mint+\mper)}{\renc \srel^2},
\end{align}
as well as $W(x)$ (eq.~\ref{eq:W}). We reject the sampled mass if $W(x)/W_\mathrm{max} < y$, where $y$ is a random number between 0 and 1, and where $W_\mathrm{max}$ is the maximum value of $W$ over the allowed range of $\mper$.

\item We sample a random position of the perturber impinging on the encounter sphere relative to the multiple system's centre of mass, $\ve{R}_\enc$ (no preferred direction). The perturber velocity $\ve{V}_\enc$ relative to the centre of mass is then sampled from the distribution implied in equation~(\ref{eq:vel_distr}). Considering that $\ve{V}_\enc$ and the impact parameter $\ve{b}$ for the perturber relative to the centre of mass are perpendicular by definition, we compute $\ve{b}$ according to
\begin{align}
\ve{b} = \ve{R}_\enc - \unit{V}_\enc \left ( \ve{R}_\enc \cdot \unit{V}_\enc \right ).
\end{align}

We reject a perturber if it is not impulsive relative to the outermost orbit of the multiple system. Specifically, we associate a (highly) hyperbolic orbit of the perturber relative to the centre of mass of the multiple system with a periapsis distance $Q=b$ and an eccentricity $E_\per= 1 + b V_\per^2/[\gconst (\mint+\mper)]$, such that the perturber's angular speed at periapsis is 
\begin{align}
\dot{\theta}_\per = \sqrt{(1+E_\per) \gconst (\mint+\mper)/b^3},
\end{align}
while the mean motion of the outermost orbit of the multiple system is 
\begin{align}
n_\mathrm{int} = \sqrt{\gconst \mint/a_\mathrm{int}^3}.
\end{align} 
We reject the perturber if $\dot{\theta}_\per < n_\mathrm{int}$ (see, e.g., \citealt{2018MNRAS.476.4139H}). 

\item The imparted velocity, $\Delta \ve{V}_i$, is computed for all bodies $i$ (cf. \eq~\ref{eq:imp}). Here, $V_\mathrm{per} = \left |\left | \ve{V}_\enc \right | \right |$, and the $\ve{R}_i$ used to determine $\ve{b}_i$ (cf. \eq~\ref{eq:bi_def}) is the position vector relative to the centre of mass. New orbits are computed (using the routines described in \citealt{2018MNRAS.476.4139H}). If one or more of the new orbits are unbound, the integration mode is changed to direct $N$-body integration in the future.

\item The time of the next encounter is generated assuming that the probability for the time delay between encounters to exceed $\Delta t$ is $\mathrm{exp}(- \Gamma \Delta t)$, where $\Gamma$ is given by equation~(\ref{eq:Gamma}).  
\end{enumerate}

\section{Main evolution algorithm}
\label{sect:alg}
The initial state of the system is described, at minimum, by the initial masses and all orbital parameters (see \S~\ref{sect:usage} for a description of how to use the code in practice). It is possible to start the simulation with evolved stars (assuming these stars would have evolved in complete isolation), which is achieved by specifying an initial stellar type $k_i>1$. 

In the main evolution algorithm, a timestep $\Delta t$ is known at the beginning of each iteration loop. The timestep at the first iteration loop is equal to the minimum allowed timestep, $\Delta t_\mathrm{min} = 1\,\yr$ (default value). The timestep for the next iteration loop, $\Delta t'$, is determined during the current loop and can be affected by several processes. 

\subsection{Single stellar evolution}
\label{sect:alg:single}
Each loop starts with single stellar evolution with \sse~for a duration of $\Delta t$ (cf. \S~\ref{sect:stellar}). Stellar evolution is carried out in both the secular and direct integration modes. The stellar evolution algorithm returns a new timestep, $\Delta t_{\sse}$ (see \S~\ref{sect:stellar:sse}), which is used in part to determine $\Delta t'$ (see \S~\ref{sect:alg:fly_time} below). During this stage, all stellar quantities are updated immediately, except for those that are assumed to vary during the ODE integration (masses, radii and spins, cf. \S~\ref{sect:stellar:sse}). 

\subsection{Binary evolution}
\label{sect:alg:bin}
Next, if the current integration mode is secular, then binary evolution is handled (cf. \S~\ref{sect:bin}), which includes dynamical mass transfer events, CE evolution, collisions, and determining properties for stable mass transfer evolution and wind accretion. In the case of mass transfer, the binary timestep, $\Delta t_\mathrm{bin}$, is set according to
\begin{align}
\Delta t_\mathrm{bin} = \alpha_\mathrm{bin} \Delta t_{\mathrm{ODE}} \times \mathrm{min} \left ( \frac{m_\d}{\Delta m_\d},  \frac{m_\a}{\Delta m_\a} \right ),
\end{align}
where the user-controllable parameter $\alpha_\mathrm{bin}$ (default value 0.05) ensures that the masses would not change due to mass transfer more than a given fraction. In the other RLOF-related cases (dynamical mass transfer events, CE evolution, and collisions), which are followed by direct $N$-body integration, $\Delta t_\mathrm{bin}=\Delta t_\mathrm{min}$.

\subsection{Orbital evolution}
\label{sect:alg:orb}
\subsubsection{Secular}
\label{sect:alg:orb:sec}
The orbital evolution is modelled next. When the integration mode is secular (cf. \S~\ref{sect:dyn:sec}), a set of ODE equations is evolved which describes the secular Newtonian and PN dynamics, tidal evolution, adiabatic orbital response to wind mass loss or wind accretion or mass lost during non-conservative mass transfer, and orbital response to mass transfer (either between two single stars, cf. \S~\ref{sect:bin:mt:stable:orb}, or between a star and a binary, cf. \S~\ref{sect:triple:mt}). As described in \S~\ref{sect:stellar:sse}, in addition to the orbital vectors ($\ve{h}_k$ and $\ve{e}_k$), the stellar masses, radii, and spins are assumed to vary during $\Delta t$, and are updated in the code after the ODE integration. 

Several stopping conditions are checked for during the integration. The included stopping conditions are as follows.
\begin{enumerate}[leftmargin=0.5cm]
\item RLOF of a star onto another star or binary (cf. \eq~\ref{eq:mt_crit}). RLOF will be taken into account starting in the binary evolution part of the code during the next loop.
\item Dynamical instability according to an analytic stability criterion (cf. \S~\ref{sect:dyn:Nbody:to}). If unstable according to this criterion, the integration mode is switched to direct, so the dynamical gravitational evolution will be carried out using direct $N$-body integration at the next loop.
\item Entering the semisecular regime (cf. \S~\ref{sect:dyn:Nbody:to}). The integration mode is switched to direct $N$-body for the next loop.
\item A direct collision between two objects, $a_k(1-e_k) \leq R_{\star,\,\eff,\,i} + R_{\star,\,\eff,\,j}$, where $i$ and $j$ are two stars in an orbit $k$. Collisions are handled immediately and are followed by direct $N$-body integration, as described in \S~\ref{sect:bin:col}.
\end{enumerate}

\subsubsection{$N$-body}
\label{sect:alg:orb:Nbody}
If the system was not deemed to be amenable to the secular approximation in a previous loop, the integration mode was switched to direct. The gravitational dynamics are integrated directly in this loop (cf. \S~\ref{sect:dyn:Nbody}). After the system has evolved for $\Delta t$, a new `$N$-body timestep', $\Delta t_N$, is determined from the orbital periods in the system (cf. \eq~\ref{eq:deltatn}). Although the masses, radii, and spins are assumed to be constant during the $N$-body integration (in contrast to secular integration), they are updated at the end of the $N$-body evolution since stellar evolution changes would otherwise not be taken into account when integrating directly. 

Furthermore, after evolving for $\Delta t$, the system is analysed for stability by determining the orbital elements and requiring that the orbital energies are constant on short timescales (cf. \S~\ref{sect:dyn:Nbody:from}). If deemed stable, the integration mode is switched to secular, i.e., the gravitational dynamical evolution in the next loop will be carried out using the secular approximation.

\subsection{Fly-bys and the new timestep}
\label{sect:alg:fly_time}
After the above, a preliminary next timestep is determined according to
\begin{align}
\Delta t' = \mathrm{min} \left ( \Delta t_\mathrm{min}, \, \Delta t_{\sse}, \, \Delta t_\mathrm{bin},\,\Delta t_N \right ).
\end{align}
If the new time, $t+\Delta t'$, would be larger than the time at which the next fly-by encounter would occur, $t_\mathrm{fly\,next}$, then $\Delta t'$ is adjusted to reach precisely the time of the next encounter. Also, the effects of the encounter are taken into account as described in \S~\ref{sect:fly}.

The next timestep $\Delta t'$ is also adjusted when the next time would exceed the desired integration time, $t_\mathrm{end}$. In the latter case, $\Delta t'$ is adjusted to reach precisely $t_\mathrm{end}$. 

The next loop is entered, unless the end integration time has been reached.

\section{Using the code}
\label{sect:usage}
In this section, we briefly discuss some aspects on how to use \mse~in practice. For detailed information, we refer to the user guide which is included with the online release of the code.

\subsection{Prerequisites}
As mentioned in \S~\ref{sect:introduction:code}, the main programming language of \mse~is \textsc{C++}, with some linking to \textsc{Fortran} routines from \sse~(\ssepaper)~and \bse~(\bsepaper) and to the \textsc{C}-code \textsc{MSTAR} \citep{2020MNRAS.492.4131R}. For convenience of the user, a comprehensive \textsc{Python} interface is included which allows the code to be used within \textsc{Python} (\S~\ref{sect:usage:python}). A \textsc{C++} and \textsc{Fortran} compiler are required. In order to use the \textsc{Python} interface, a \textsc{Python} installation with \textsc{Numpy} is also needed (\textsc{Matplotlib} is needed to produce plots in some of the scripts provided with the code).

\subsection{\textsc{Python} interface}
\label{sect:usage:python}
Here, we briefly describe the most important aspects of running \mse~within \textsc{Python}.

\subsubsection{Command line usage, or a wrapper function}
\label{sect:usage:python:clw}
The easiest method to run a system in \mse~is via the included \texttt{run\_system.py} \textsc{Python} script. This script runs a single system, with its properties specified using the command line. It produces basic plots with the time evolution of some quantities, mobile diagrams \citep{1968QJRAS...9..388E}, and positions. It supports a number of configurations of hierarchical systems, including `fully nested' systems, in which the number of levels in the system is maximised. For example, a `fully nested' triple is simply a hierarchical triple, whereas a `fully nested' quadruple is a 3+1 quadruple. 

\begin{lstlisting}[caption={Example usage of \texttt{run\_system.py} for a triple.},label={code:simpletriple},language=bash]
python3 run_system.py --configuration "fully_nested"
--masses 40 10 2 --metallicities 0.02 0.02 0.02 
--smas 15 120 --es 0.1 0.2 --is 0.001 1.4 
--LANs 0.01 0.01 --APs 0.01 0.01
--tend 2e7 --Nsteps 2000 --plot_filename "figs/mytriple"
\end{lstlisting}
An example of running a hierarchical triple within the command line is shown in \code~\ref{code:simpletriple}. Note that arrays are to be input using spaces to separate values. All masses should be given in units of $\msun$, all semimajor axes in units of $\au$, times should be given in yr, and all orbital angles in rad. When \texttt{--configuration} is given as \texttt{"fully\_nested"}, the number of bodies in the system is inferred from the number of values given to the \texttt{--masses} argument. 

\begin{figure}
  \center
  \includegraphics[width=0.48\textwidth,trim=0mm 0mm 0mm 0mm]{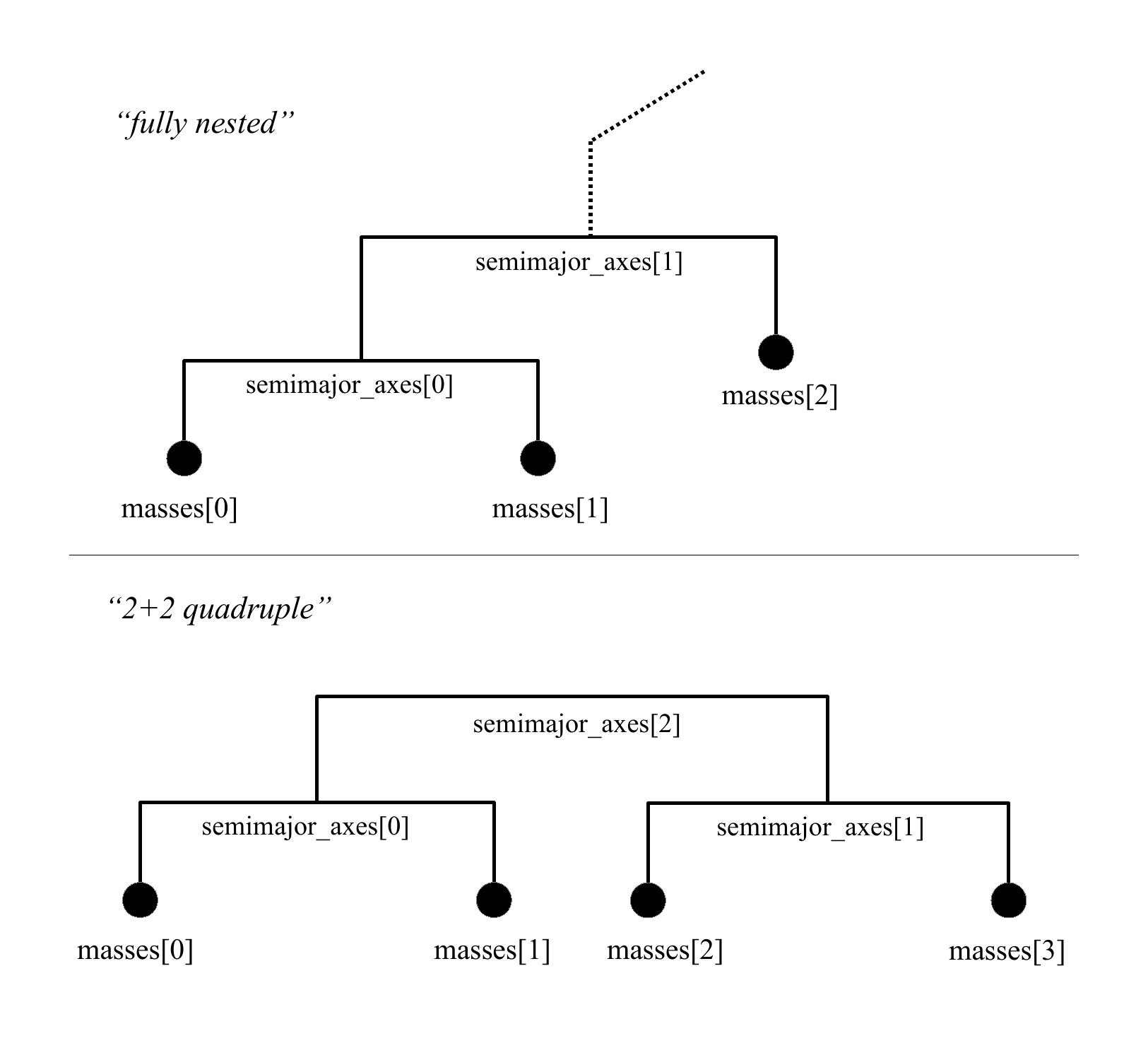}
  \caption{Illustration of configurations that can be used within the \texttt{run\_system.py} script (and the \texttt{evolve\_system()} wrapper function). Indicated are how the orders within \textsc{Python} lists such as \texttt{masses} and \texttt{semimajor\_axes} correspond to the stars and orbits, respectively. Top: a `fully nested' configuration (explicitly showing the simplest case of a triple). Bottom: a 2+2 quadruple system. }
\label{fig:wrapper}
\end{figure}

In the above example \code~\ref{code:simpletriple}, \texttt{--configuration} is ``\texttt{fully\_nested}'' and three masses are given; therefore, the system is assumed to be a hierarchical triple. The order of the arrays should be consistent with the order as indicated in the top panel of \F~\ref{fig:wrapper}. In the example given, the initial masses of the two stars in the inner binary are 40 and 10 $\msun$, respectively, and the tertiary mass is 2 $\msun$. The metallicities are 0.02 for all stars. The inner and outer orbital semimajor axes (\texttt{--smas}) are 15 and 120 $\au$, respectively, and the eccentricities (\texttt{--es}) are 0.1 and 0.2 for the inner and outer orbits, respectively. The inclinations of the inner and outer orbits (\texttt{--is}) are 0.001 and 1.4 rad, respectively. The longitudes of the ascending node (\texttt{--LANs}) and the arguments of periapsis (\texttt{--APs}) are all 0.01 for the inner and outer orbits, respectively. 

By default, the bodies in the system are assumed to be stars. Planets (and other objects) can be specified with the optional argument \texttt{--object\_types}, which should be a list of integers corresponding to \texttt{--masses}; an integer value of 1 (default value) indicates a star, and an integer value of 2 indicates a planet.

Other hierarchical configurations than `fully nested' can be specified with the \texttt{--configuration} command line option (a text string). In particular, a 2+2 quadruple can be specified by setting \texttt{--configuration} to be \texttt{2+2\_quadruple}. More generally, any hierarchical system can be specified using a bracket notation, i.e., with `1' indicating bodies, higher integers representing fully-nested subsystems, and square brackets denoting the structure. For example, a `(2+2)+1' quintuple can be specified with  \texttt{--configuration "[[2,2],1]"}. More information on this generalised input can be found in the online documentation. 

The duration of the integration is specified with the \texttt{--tend} argument. The number of output steps (this is not the same as the number of steps taken internally within \mse) is given by \texttt{--Nsteps}. After completing the simulation, various plots showing various aspects of the evolution are saved, using filenames based on the \texttt{plot\_filename} argument.

Alternatively to calling \texttt{run\_system.py} directly from the command line and inputting the system's parameters as command line arguments, the code can also be used in a similar fashion as above but within a user-created \textsc{Python} script, i.e., with the supplied function \texttt{evolve\_system()} wrapper function. We refer to the online documentation for more details.

\subsubsection{Custom usage}
\label{sect:usage:python:custom}
Although the command line-based \texttt{run\_system.py} script or the \texttt{evolve\_system()} wrapper function can be convenient, they do not give full control over the simulation. If more customisation is required, it is possible to call \mse~directly. We briefly describe this use case here.

\begin{lstlisting}[caption={Custom usage example within \textsc{Python} for a triple.},label={code:custom},language=Python,numbers=left]
from mse import MSE,Tools,Particle

N_bodies=3
masses = [40,10,2]
metallicities = [0.02,0.02,0.02]
semimajor_axes = [15 ,120] 
eccentricities = [0.1,0.2]
inclinations = [0.001,1.4]
arguments_of_pericentre = [0.01,0.01]
longitudes_of_ascending_node = [0.01,0.01]

particles = Tools.create_fully_nested_multiple(N_bodies,masses,\
    semimajor_axes,eccentricities,inclinations,\
    arguments_of_pericentre,longitudes_of_ascending_node,\
    metallicities=metallicities)

code = MSE()
code.add_particles(particles)

Nsteps = 2000
tend = 2.0e7
t = 0.0
dt = tend/float(Nsteps)

while t<tend:
    t+=dt
    code.evolve_model(t)

    particles = code.particles
    orbits = [x for x in particles if x.is_binary==True]
    bodies = [x for x in particles if x.is_binary==False]
    
    print( 't/Myr',t*1e-6,'es',[o.e for o in orbits],'smas',\
         [o.a for o in orbits])

print("log",code.log)
code.reset()
\end{lstlisting}
We give a simple example in \code~\ref{code:custom}, initialising and running the same system that was given in \S~\ref{sect:usage:python:clw}. The function \texttt{create\_fully\_nested\_multiple()} from the \texttt{Tools} class is used (lines 12-15) to generate the \texttt{particles}. Here, \texttt{particles} is a \textsc{Python} \texttt{list} containing \texttt{Particle} objects; the \texttt{particles} represent the entire system. An instance of the code is made (line 17), and the \texttt{particles} are added to the code (line 18).   

Lines 25-34 represent a time loop in which the code is called repeatedly until reaching the end time. In line 29, the \texttt{particles} set local to the user is updated from the \texttt{particles} set in \mse. The subsequent lines, 30-31, separate the particles out into \texttt{orbits}, i.e., particles representing orbits, and \texttt{bodies}, i.e., particles representing bodies/stars. In lines 33-34, the eccentricities and semimajor axes of all orbits are printed. Any other time-dependent data from the \texttt{particles} can also be accessed at this point. 

In line 36, the \texttt{code.log} is accessed. This contains detailed information on important events during the evolution. Its purpose is to be able to track important events, without the risk of missing these by running the high-level time loop (lines 25-34 in \code~\ref{code:custom}), which is particularly useful for population synthesis studies. The contents of the code logs are described in detail in the online documentation.

Lastly, when finished with the code, care should be taken to call \texttt{code.reset()}. Failing to do so can result in unexpected behaviour when running the code repeatedly.

\subsection{Direct access through \textsc{C++}}
Running the code within \textsc{C++} gives most customisability and minimises any potential overhead losses generated by the \textsc{Python} interface. However, we only recommend it for more advanced use cases, and refer to the online documentation for more details.

\section{Example systems}
\label{sect:ex}
In this section, we discuss the evolution of a few systems evolved with \mse. The assumed initial conditions for the different systems\footnote{The examples presented in this section were generated with the first release version of \mse~(version 0.79). The same initial conditions could lead to different outcomes with future versions of the code. See also \S~\ref{sect:discussion:sens}.} are summarised in Table~\ref{table:exICs}. In all examples, we assume $Z_i=0.02$ for all stars. Note that we evolve the examples for less than a Hubble time and focus on early aspects of the evolution.

\begin{table*}
\begin{center}
\begin{tabular}{cccccccccc}
\toprule 
S & Configuration & $m_i/\msun$ & $a_k/\au$ & $e_k$ & $i_k$ & $\Omega_k$ & $\omega_k$ & $t_\mathrm{end}/\yr$ & $N_\mathrm{steps}$ \\
\midrule
\ref{sect:ex:ex1} & Triple & 3, 2, 1 & 15, 500 & 0.1, 0.8 & 0.01, 1.5 & 3.45, 4.49 & 3.79, 3.42 & $5\times 10^8$ & $10^3$ \\
\ref{sect:ex:ex2} & 3+1 quadruple & 20, 15, 10, 20 & 20, 800, 8000 & 0.1, 0.2, 0.5 & 0.01, 1.5, 0.01 & 3.45, 4.49, 3.79 & 3.42, 2.66, 4.06 & $10^7$ & $10^2$  \\
\ref{sect:ex:ex3} & 2+2 quadruple & 4, $10^{-4}$, 5, 2 & 18, 10, 1000 & 0.01, 0.01, 0.41 & 0.01, 0.01, 1.0 & 3.45, 4.49, 3.79 & 3.42, 2.66, 4.06 & $5\times 10^8$ & $10^3$ \\
\bottomrule
\end{tabular}
\end{center}
\caption{\small Initial conditions for the examples systems discussed in \S~\ref{sect:ex}. Here, $t_\mathrm{end}$ is the integration time, and $N_\mathrm{steps}$ is the number of output steps for plotting purposes. }
\label{table:exICs}
\end{table*}

\begin{figure}
  \center
  \includegraphics[width=0.5\textwidth,trim=8mm 0mm 0mm 0mm]{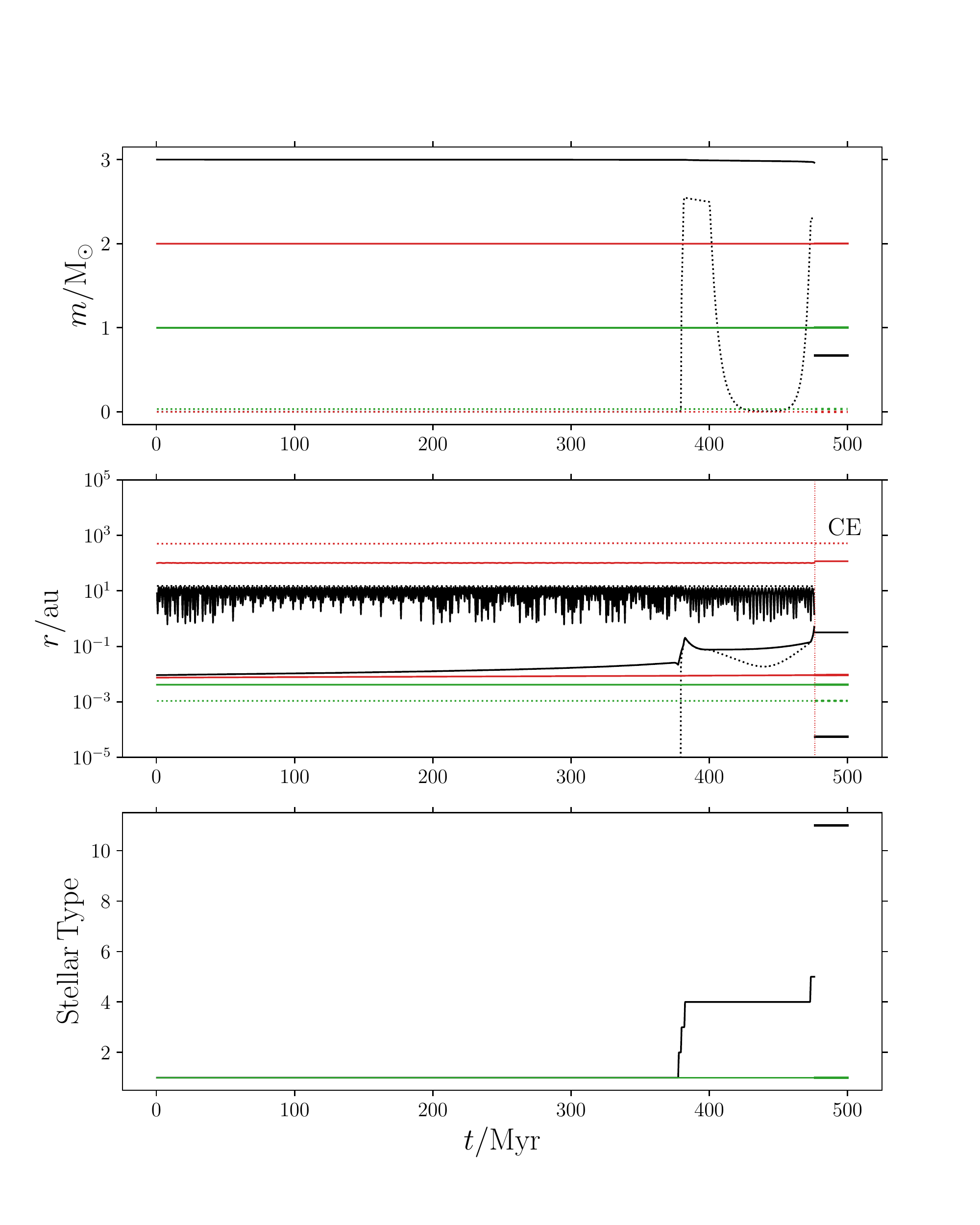}
  \caption{Evolution of the masses (top row), orbital separations and stellar radii (middle row), and stellar types (bottom row) for the example triple system discussed in \S~\ref{sect:ex:ex1}. See Table~\ref{table:exICs} for the initial conditions. In the top panel, the three masses are shown with solid black, red, and green lines, respectively. The convective core radii of the corresponding stars are shown with dotted lines. In the middle panel, the bottom solid lines show the stellar radii, with the same colours used as in the top panel. The black and red lines in the top and middle part of the panel show the orbital separations (solid: periapsis distances; dotted: semimajor axes) of the inner and outer orbits, respectively. The onset of a CE event in the inner binary is indicated. The bottom panel shows the evolution of the stellar types (cf. Table~\ref{table:st}), with the same colours used as in the top panel.  }
\label{fig:ex1}
\end{figure}

\begin{figure*}
  \center
  \includegraphics[width=0.98\textwidth,trim=8mm 0mm 0mm 0mm]{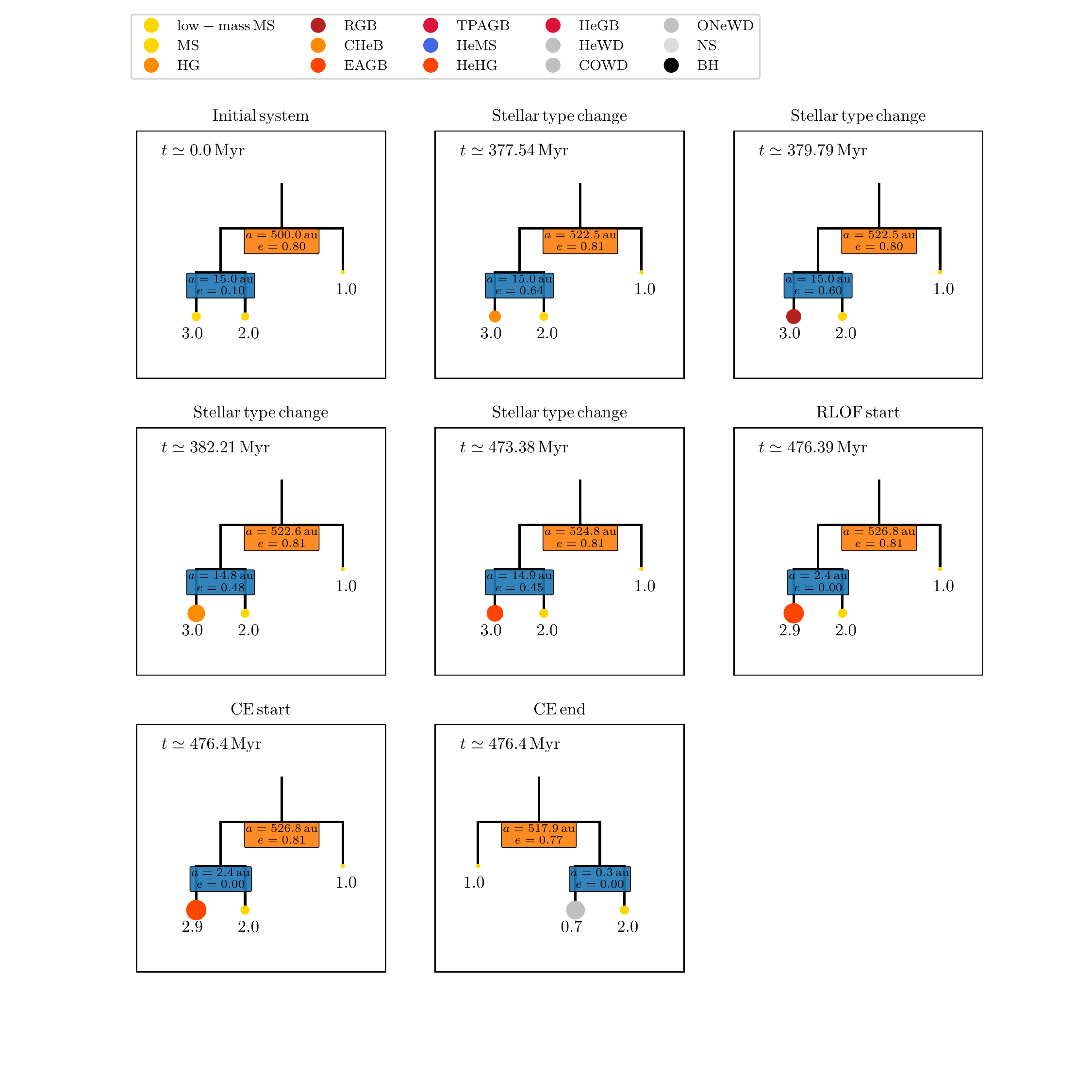}
  \caption{ Mobile diagram for the example triple system discussed in \S~\ref{sect:ex:ex1}. The title of each panel gives a description of the event that occurred. The semimajor axes and eccentricities are indicated at each orbit. Numbers next to stars show the masses of the objects (in $\msun$). The colours of the stars depend on the stellar type; see the legend at the top of the figure. }
\label{fig:ex1_mobile}
\end{figure*}

\subsection{RLOF and CE evolution in a stellar triple}
\label{sect:ex:ex1}
We begin with an example of RLOF and CE evolution occurring in a stellar triple. In \F~\ref{fig:ex1}, we show the evolution of the masses (top row), orbital separations and stellar radii (middle row), and stellar types (bottom row). We also show important events during the evolution of the system in the form of a mobile diagram \citep{1968QJRAS...9..388E} in \F~\ref{fig:ex1_mobile}. 

During the MS, high-amplitude ZLK oscillations are induced in the inner binary, but they are not sufficiently strong to induce interaction. As the $3\,\msun$ primary star evolves to an AGB star, it fills its Roche lobe around its $2\,\msun$ companion. The donor is then stripped of its envelope, and the core turns into a CO WD which is orbiting the companion (still an MS star) in a more compact orbit. In this example, we focus on the early evolution. However, at later times, the inner orbit could undergo further interaction (e.g., produce a cataclysmic variable).

\begin{figure}
  \center
  \includegraphics[width=0.5\textwidth,trim=8mm 0mm 0mm 0mm]{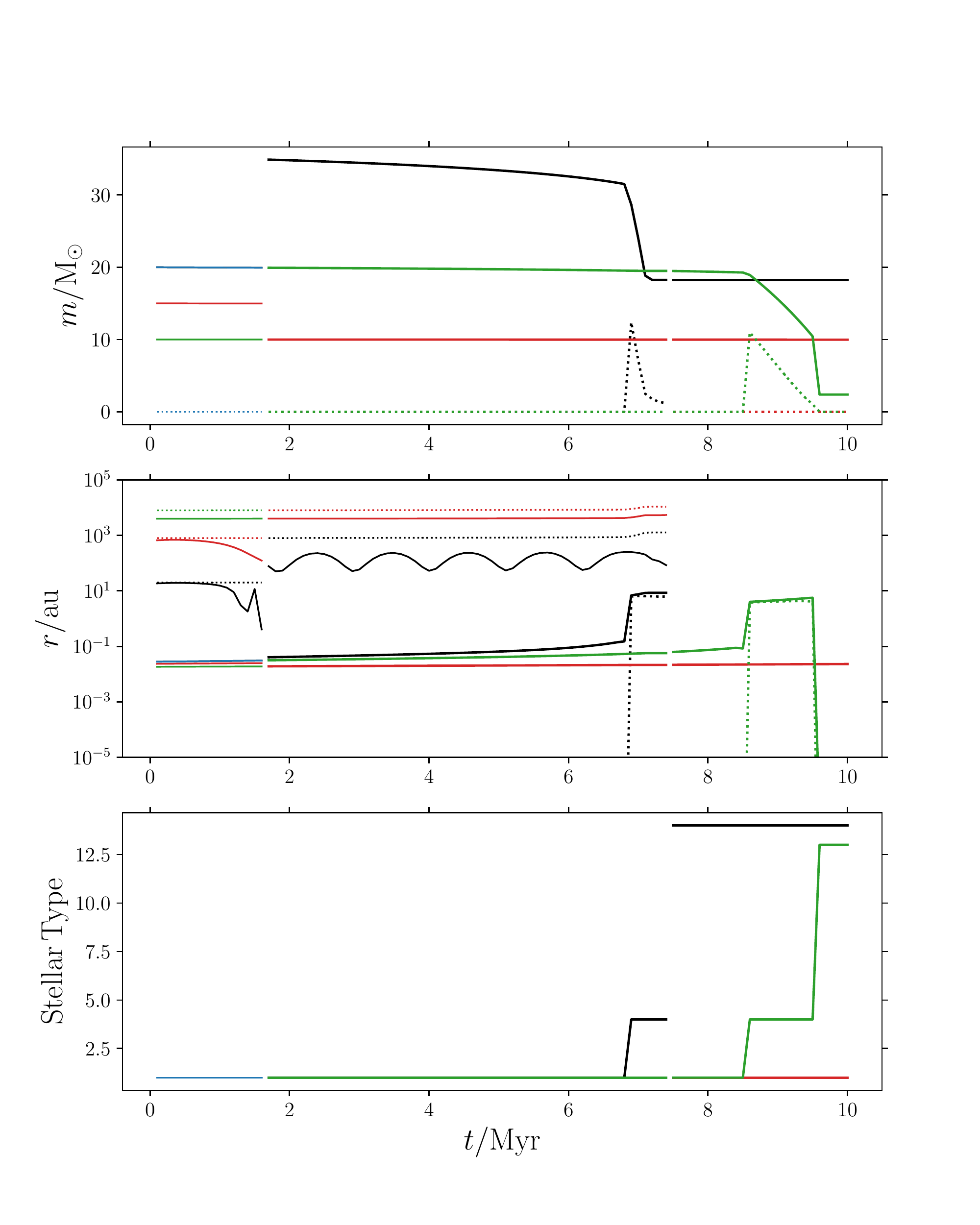}
  \caption{ Evolution of the masses (top row), orbital separations and stellar radii (middle row), and stellar types (bottom row) for the example 3+1 quadruple system discussed in \S~\ref{sect:ex:ex2}. Refer to the caption in \F~\ref{fig:ex1} for the meaning of the different lines. In the middle panel, orbital parameters are shown with black, red, and green lines for the innermost, intermediate, and outermost orbit, respectively (where applicable). }
\label{fig:ex2}
\end{figure}

\begin{figure*}
  \center
  \includegraphics[width=0.98\textwidth,trim=8mm 0mm 0mm 0mm]{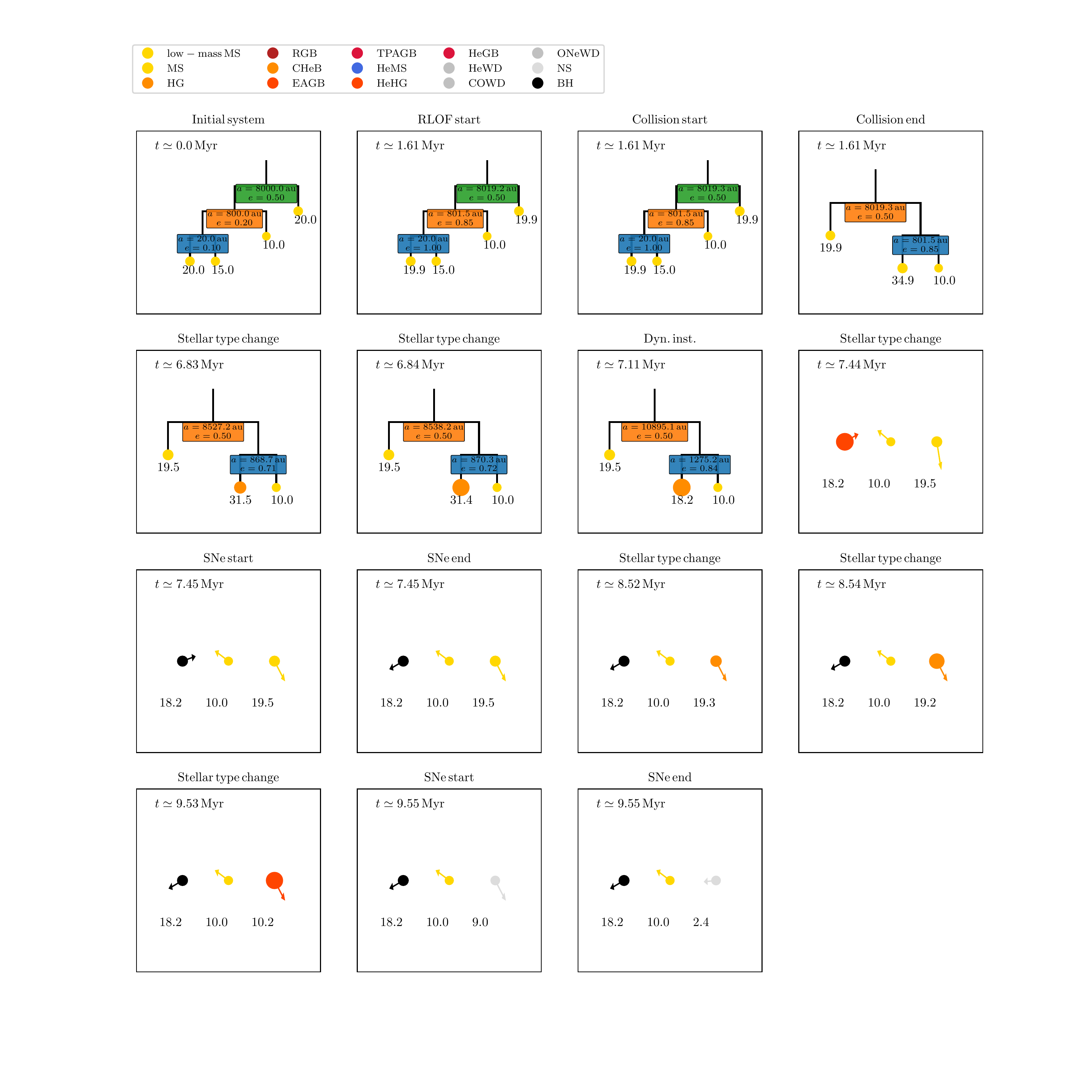}
  \caption{ Mobile diagram for the example triple system discussed in \S~\ref{sect:ex:ex2}. }
\label{fig:ex2_mobile}
\end{figure*}

\subsection{Destruction of a massive 3+1 quadruple}
\label{sect:ex:ex2}
Next, we consider a massive 3+1 quadruple system. The time evolution is shown in \F~\ref{fig:ex2}, and a mobile diagram is shown in \F~\ref{fig:ex2_mobile}. 

Owing to secular evolution, the innermost orbital eccentricity is highly excited during the MS, and the innermost two MS stars merge into a single MS star. The resulting triple is temporarily dynamically stable, and ZLK oscillations continue to be excited in the (now) inner orbit. As the merger remnant star evolves, the resulting mass loss triggers a dynamical instability (e.g., \citealt{2012ApJ...760...99P}), and the code switches to direct integration. During the instability phase, the merger remnant continues to evolve, and explodes in an SNe event at $t \simeq 7.45\,\myr$. The kick causes a complete disruption of the triple system, and the remaining stars continue to evolve in isolation.

\begin{figure}
  \center
  \includegraphics[width=0.5\textwidth,trim=8mm 0mm 0mm 0mm]{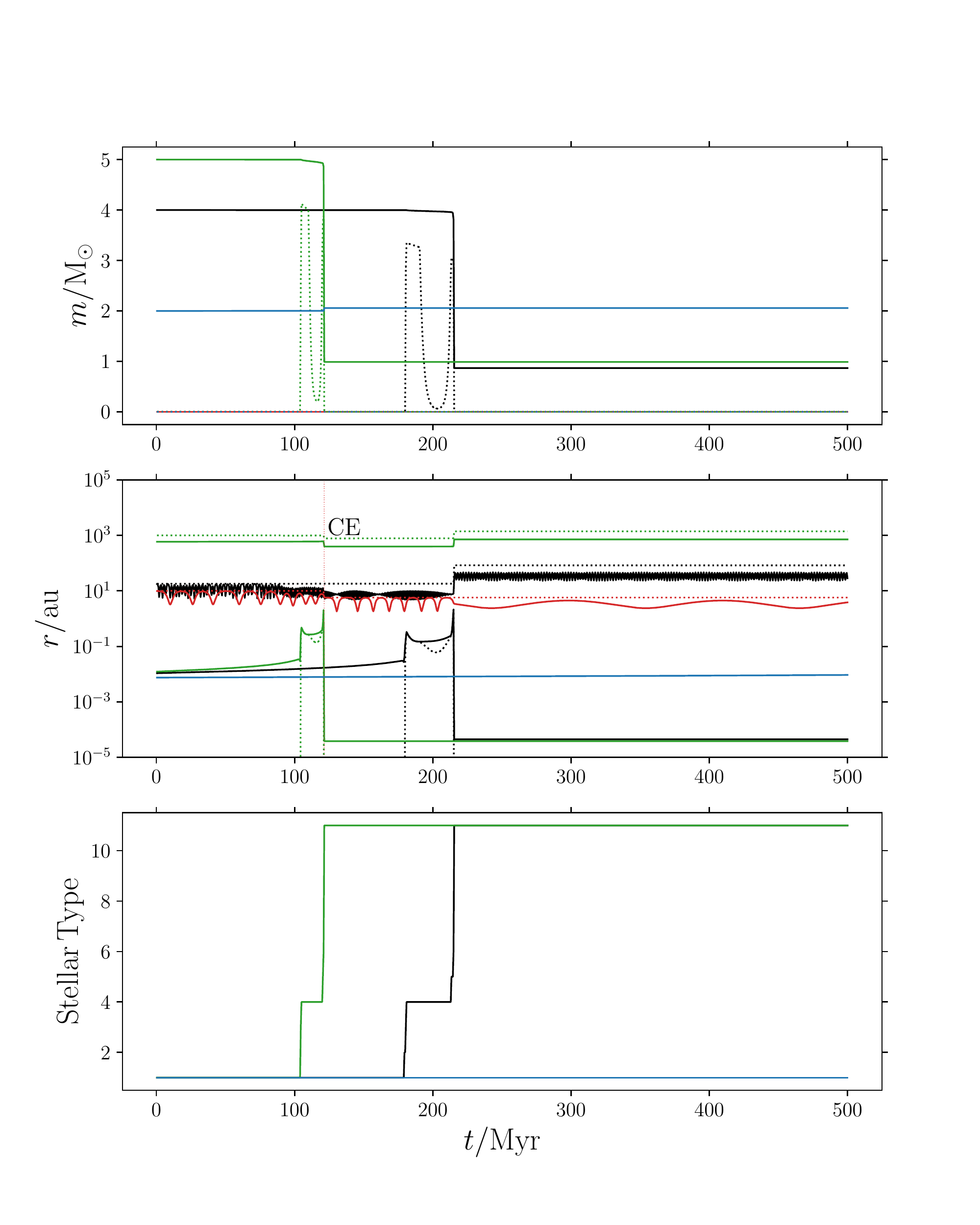}
  \caption{Evolution of the masses (top row), orbital separations and stellar radii (middle row), and stellar types (bottom row) for the example planet-in-triple system discussed in \S~\ref{sect:ex:ex3}. Refer to the caption in \F~\ref{fig:ex1} for the meaning of the different lines. In the middle panel, orbital parameters are shown with black, red, and green lines for the star-planet orbit, stellar binary companion orbit, and outer orbit, respectively. }
\label{fig:ex3}
\end{figure}

\begin{figure*}
  \center
  \includegraphics[width=0.98\textwidth,trim=8mm 0mm 0mm 0mm]{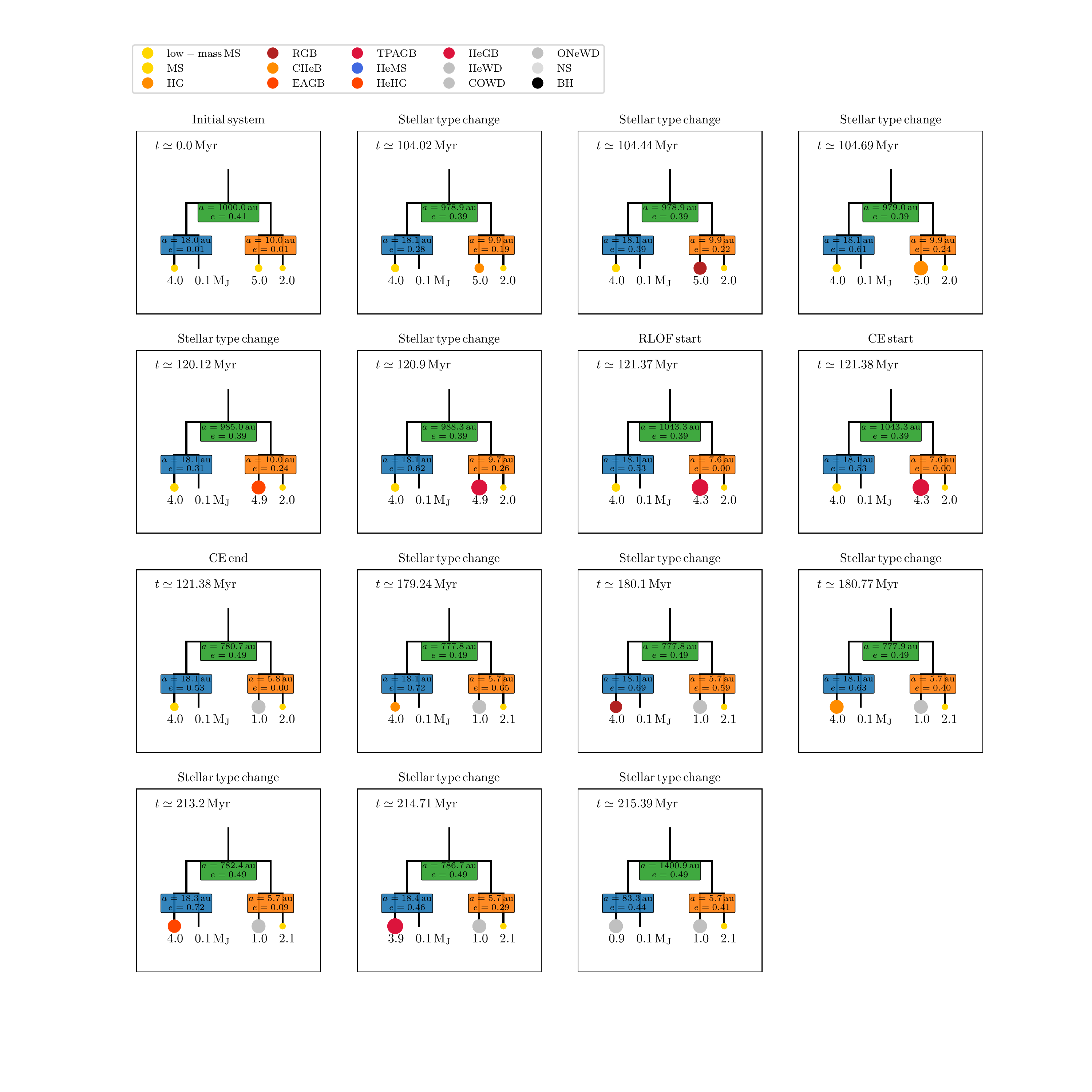}
  \caption{ Mobile diagram for the example of a planet in an evolving triple-star system discussed in \S~\ref{sect:ex:ex3}. }
\label{fig:ex3_mobile}
\end{figure*}

\subsection{A planet in an evolving triple}
\label{sect:ex:ex3}
Lastly, we show in \F~\ref{fig:ex3} the time evolution of a planet in an evolving triple-star system (the mobile diagram is shown in \F~\ref{fig:ex3_mobile}). In particular, we consider a planet in an S-type orbit \citep{1982OAWMN.191..423D} around a $4\,\msun$ star which forms the tertiary star in a triple (the companion binary has component masses of 5 and 2 $\msun$).

In this example, the planet survives during the first $500 \, \myr$, although its orbital evolution (cf. the middle black lines in the middle panel of \F~\ref{fig:ex3}) is affected by the evolution of the stellar binary. During the MS, the planet's orbit is excited in eccentricity due to secular excitation from the companion binary. Around $120\,\myr$, the primary star in the companion binary fills its Roche lobe onto its companion, and CE evolution is triggered. The companion binary survives in a more compact orbit, but the mass loss during the CE event decreases the semimajor axis and periapsis distance of the outer orbit, affecting the planet-star eccentricity oscillations. At $\sim 215\,\myr$, the planet's host star evolves, but the planet survives the stellar expansion as its orbit widens due to mass loss. This also affects the outer orbit, and thereby the eccentricity oscillations of the companion stellar binary. Eccentricity oscillations in the planet-star system continue, but the amplitude is not sufficiently high to induce strong interactions of the planets with its host star (now a WD).

\section{Discussion}
\label{sect:discussion}

\subsection{Sensitivity on initial conditions and other factors}
\label{sect:discussion:sens}
Dynamical systems often display chaotic behaviour (e.g., \citealt{1993cds..book.....O}). In \mse, this is compounded by the fact that the code not only includes gravitational dynamical evolution, but also combines it with other physical processes, some of which can be sensitive to initial conditions. It may therefore come to no surprise that \mse~is generally very sensitive to the initial conditions. Nevertheless, a given system will produce the same outcome when run multiple times on the same machine, provided that the initial random seed (which can be specified) is the same. 

However, we also remark that, when running on different machines or (virtual) environments (or even using different compiler options), results of the code can differ because of the way that floating points are represented and rounded in computers. Although these effects are tiny for single operations, they can add up and become significant when the number of floating point operations is very large, which can occur if a system is integrated in \mse~for a relatively long time. For example, if a triple system undergoes a very large number of ZLK oscillations, then the exact phase of the ZLK cycle in the inner binary at an arbitrary time can differ from machine to machine. This can make the difference whether or not a star in the inner binary will fill its Roche lobe as it evolves, and such a difference will propagate further in the evolution of the entire system. 

In practice, both the chaotic nature of multiple systems and numerical difficulties imply that the exact outcome of one specific system should not be given too much emphasis or weight. \mse~is meant to be used as a population synthesis code; uncertainties due to chaos or numerical difficulties as described above are expected to average out when a large enough sample of systems is considered\footnote{What constitutes `large enough' should be investigated on a case-by-case basis, by considering convergence of the statistical properties of interest with respect to the number of included systems. Furthermore, if all parameters of a system are to be kept fully fixed, then uncertainties due to chaos or numerics could be investigated by running systems with different initial random seeds.}. We furthermore remark that, apart from the numerical difficulties described above, the exact outcome of a specific system also sensitively depends on physical processes included in the code, many of which are uncertain (for example, the precise way that an eccentric orbit responds to mass transfer, or the kick velocity during an SNe event).

\subsection{Future directions}
\label{sect:discussion:fut}
Although offering a more complete numerical description of the evolution of multiple-star systems compared to previous codes, limitations remain within \mse~and we plan to address them in the future. We briefly touch on future directions here.

\subsubsection{Tides}
\label{sect:discussion:fut:tide}
The implementation of tidal evolution (in the secular context) was described in \S~\ref{sect:dyn:sec:tide}. Currently, the equilibrium tide model is assumed in \mse, and the prescription of \bsepaper~is used to compute the efficiency of tidal dissipation as a function of evolutionary stage. In future work, we plan to improve on this treatment of tides by (1) implementing more consistently dynamical tides, which are important for highly eccentric orbits (possibly following a similar approach as \citealt{2018ApJ...854...44M}), and (2) given the large uncertainties in the prescription of \bsepaper, developing an improved prescription for the efficiency of tidal dissipation for stars at any evolutionary stage. 

In addition, tides are currently not included in the $N$-body integrations (cf. \S~\ref{sect:dyn:Nbody}), and this will be amended in future versions of the code.

\subsubsection{Updated stellar evolution}
The stellar evolution tracks in \sse~which \mse~relies upon are based on somewhat dated detailed models. More recent detailed models can give different results, especially for high-mass stars and their remnants given the large uncertainties in wind mass loss rates. The \textsc{METISSE} code \citep{2020MNRAS.497.4549A}, which is based on interpolation instead of analytic fits, has a similar interface as the original \sse~code, making an implementation of \textsc{METISSE} into \mse~straightforward. We will also consider implementing other population synthesis-oriented codes into \mse~in the future, allowing for direct comparison between different stellar evolution models and investigating their impact on multiple-star evolution. 

\subsubsection{Mass transfer and common envelope evolution}
The current assumptions underlying the orbital response to mass transfer in binary systems (\S~\ref{sect:bin:mt:stable:orb}) are restrictive. Detailed hydrodynamical simulations could be used in the future to investigate their validity, and to improve upon the analytic model of HD19 which is used in \mse. Even more uncertainties exist for triple mass transfer (\S~\ref{sect:triple:mt}), for which we assumed an ad hoc model. This model should be improved based on detailed simulations. Also, other modes of mass transfer in multiple-star systems could be included (e.g., transfer of mass from the outer Lagrangian points in an inner binary to exterior stars). 

Similar limitations apply to binary CE evolution, of which the details remain elusive (see, e.g., \citealt{2013A&ARv..21...59I}). Triple CE is even more uncertain; here, we formulated a simplified model for the case of a tertiary star undergoing CE with an inner binary (\S~\ref{sect:triple:ce}). More detailed hydrodynamical simulations will be required to inform a more accurate prescription.

\subsubsection{Dense stellar systems}
The current version of \mse~applies to multiple-star systems in the field (i.e., low-density environments), where perturbations from passing stars are relatively unimportant (except for wide orbits). However, triple and higher-order systems can form through dynamical interactions in dense stellar systems such as open and globular clusters (e.g., \citealt{2007MNRAS.379..111V}). Although they may not be long-lived because the same strong perturbations that induced their formation can lead to their destruction, they are interesting since secular interactions during their lifetime can lead to stellar and/or compact object mergers (e.g., \citealt{2016ApJ...816...65A,2020ApJ...903...67M,2020ApJ...900...16F}). In the future, we will consider implementing into \mse~the effects from both weak (i.e., `secular' but non-impulsive) perturbations (e.g., \citealt{1996MNRAS.282.1064H,2009ApJ...697..458S,2018MNRAS.476.4139H,2019ApJ...872..165G,2019MNRAS.487.5630H,2019MNRAS.488.5192H}), as well as strong perturbations (with direct $N$-body integration). In addition, \mse~could be implemented in and extend codes that model dense stellar systems which currently rely on binary population synthesis codes, such as the \textsc{Nbody} series (e.g., \citealt{1999JCoAM.109..407S,2003gnbs.book.....A,2008LNP...760..377S,2015MNRAS.450.4070W}), and Monte Carlo codes such as \textsc{Mocca} (e.g., \citealt{2013MNRAS.429.1221H,2013MNRAS.431.2184G}), and \textsc{CMC} (e.g., \citealt{2000ApJ...540..969J,2001ApJ...550..691J,2003ApJ...593..772F,2007ApJ...658.1047F,2010ApJ...719..915C,2012ApJ...750...31U}).

\subsubsection{Planetary evolution}
Planets in \mse~are currently considered as bodies with fixed radius that do not undergo stellar evolution. In the future, we plan to implement various aspects of planetary evolution such as contraction of the planet on a thermal timescale.

\section{Conclusions}
\label{sect:conclusions}
We have presented a new population synthesis algorithm, \mse, which can be used to quickly model the stellar, binary, and gravitational dynamical evolution of hierarchical multiple systems with any number of stars and any structure, provided the system is composed of nested binary orbits. The gravitational dynamics are taken into account using either a secular approach \citep{2016MNRAS.459.2827H,2018MNRAS.476.4139H,2020MNRAS.494.5492H}, or a direct $N$-body integrator \citep{2020MNRAS.492.4131R}. Stellar evolution is taken into account by adopting the \sse~fitting functions (\ssepaper), whereas binary evolution is modeled using semi-analytic models and prescriptions. 

New features of \mse~in comparison to previous population synthesis codes include (1) an arbitrary number of stars, as long as the initial system is hierarchical, (2) dynamic switching between secular and direct $N$-body integration for efficient computation of the gravitational dynamics, (3) treatment of mass transfer in eccentric orbits, which occurs commonly in multiple-star systems, (4) a simple treatment of tidal, common-envelope, and mass transfer evolution in which the accretor is a binary instead of a single star, (5) taking into account planets within the stellar system, and (6) including gravitational perturbations from passing field stars.

\mse, written primarily in \textsc{C++} (with some linking to \textsc{Fortran} code), will be made publicly available and has few prerequisites; a convenient \textsc{Python} interface is provided. We have given a number of examples illustrating how the code can be used in a variety of hierarchical systems.

\section*{Acknowledgements}
We thank Achim Wei\ss, Thomas Janka, and Scott Tremaine for discussions and comments on the manuscript, and the referee for a helpful report. A.S.H. thanks the Max Planck Society for support through a Max Planck Research Group.

\section*{Data availability}
At the time of writing, the code presented in this paper is part of a private repository on \texttt{GitHub} (see the link given in \S~\ref{sect:introduction}). Access to this repository can be requested by contacting the authors. In the future, the repository will be made publicly available. 

\bibliographystyle{mnras}
\bibliography{literature}

\begin{thebibliography}{}
\makeatletter
\relax
\def\mn@urlcharsother{\let\do\@makeother \do\$\do\&\do\#\do\^\do\_\do\%\do\~}
\def\mn@doi{\begingroup\mn@urlcharsother \@ifnextchar [ {\mn@doi@}
  {\mn@doi@[]}}
\def\mn@doi@[#1]#2{\def\@tempa{#1}\ifx\@tempa\@empty \href
  {http://dx.doi.org/#2} {doi:#2}\else \href {http://dx.doi.org/#2} {#1}\fi
  \endgroup}
\def\mn@eprint#1#2{\mn@eprint@#1:#2::\@nil}
\def\mn@eprint@arXiv#1{\href {http://arxiv.org/abs/#1} {{\tt arXiv:#1}}}
\def\mn@eprint@dblp#1{\href {http://dblp.uni-trier.de/rec/bibtex/#1.xml}
  {dblp:#1}}
\def\mn@eprint@#1:#2:#3:#4\@nil{\def\@tempa {#1}\def\@tempb {#2}\def\@tempc
  {#3}\ifx \@tempc \@empty \let \@tempc \@tempb \let \@tempb \@tempa \fi \ifx
  \@tempb \@empty \def\@tempb {arXiv}\fi \@ifundefined
  {mn@eprint@\@tempb}{\@tempb:\@tempc}{\expandafter \expandafter \csname
  mn@eprint@\@tempb\endcsname \expandafter{\@tempc}}}

\bibitem[\protect\citeauthoryear{{Aarseth}}{{Aarseth}}{2003}]{2003gnbs.book.....A}
{Aarseth} S.~J.,  2003, {Gravitational N-Body Simulations}

\bibitem[\protect\citeauthoryear{{Abate}, {Pols}, {Stancliffe}, {Izzard},
  {Karakas}, {Beers}  \& {Lee}}{{Abate} et~al.}{2015}]{2015A&A...581A..62A}
{Abate} C.,  {Pols} O.~R.,  {Stancliffe} R.~J.,  {Izzard} R.~G.,  {Karakas}
  A.~I.,  {Beers} T.~C.,   {Lee} Y.~S.,  2015, \mn@doi [\aap]
  {10.1051/0004-6361/201526200}, \href
  {https://ui.adsabs.harvard.edu/abs/2015A&A...581A..62A} {581, A62}

\bibitem[\protect\citeauthoryear{{Agrawal}, {Hurley}, {Stevenson}, {Sz{\'e}csi}
   \& {Flynn}}{{Agrawal} et~al.}{2020}]{2020MNRAS.497.4549A}
{Agrawal} P.,  {Hurley} J.,  {Stevenson} S.,  {Sz{\'e}csi} D.,   {Flynn} C.,
  2020, \mn@doi [\mnras] {10.1093/mnras/staa2264}, \href
  {https://ui.adsabs.harvard.edu/abs/2020MNRAS.497.4549A} {497, 4549}

\bibitem[\protect\citeauthoryear{{Alexander}}{{Alexander}}{1973}]{1973Ap&SS..23..459A}
{Alexander} M.~E.,  1973, \mn@doi [\apss] {10.1007/BF00645172}, \href
  {https://ui.adsabs.harvard.edu/abs/1973Ap&SS..23..459A} {23, 459}

\bibitem[\protect\citeauthoryear{{Anderson}, {Storch}  \& {Lai}}{{Anderson}
  et~al.}{2016}]{2016MNRAS.456.3671A}
{Anderson} K.~R.,  {Storch} N.~I.,   {Lai} D.,  2016, \mn@doi [\mnras]
  {10.1093/mnras/stv2906}, \href
  {http://cdsads.u-strasbg.fr/abs/2016MNRAS.456.3671A} {456, 3671}

\bibitem[\protect\citeauthoryear{{Andrews}, {Zezas}  \& {Fragos}}{{Andrews}
  et~al.}{2018}]{2018ApJS..237....1A}
{Andrews} J.~J.,  {Zezas} A.,   {Fragos} T.,  2018, \mn@doi [\apjs]
  {10.3847/1538-4365/aaca30}, \href
  {https://ui.adsabs.harvard.edu/abs/2018ApJS..237....1A} {237, 1}

\bibitem[\protect\citeauthoryear{{Antognini}}{{Antognini}}{2015}]{2015MNRAS.452.3610A}
{Antognini} J.~M.~O.,  2015, \mn@doi [\mnras] {10.1093/mnras/stv1552}, \href
  {http://adsabs.harvard.edu/abs/2015MNRAS.452.3610A} {452, 3610}

\bibitem[\protect\citeauthoryear{{Antonini} \& {Perets}}{{Antonini} \&
  {Perets}}{2012}]{2012ApJ...757...27A}
{Antonini} F.,  {Perets} H.~B.,  2012, \mn@doi [\apj]
  {10.1088/0004-637X/757/1/27}, \href
  {http://adsabs.harvard.edu/abs/2012ApJ...757...27A} {757, 27}

\bibitem[\protect\citeauthoryear{{Antonini}, {Murray}  \& {Mikkola}}{{Antonini}
  et~al.}{2014}]{2014ApJ...781...45A}
{Antonini} F.,  {Murray} N.,   {Mikkola} S.,  2014, \mn@doi [\apj]
  {10.1088/0004-637X/781/1/45}, \href
  {http://adsabs.harvard.edu/abs/2014ApJ...781...45A} {781, 45}

\bibitem[\protect\citeauthoryear{{Antonini}, {Chatterjee}, {Rodriguez},
  {Morscher}, {Pattabiraman}, {Kalogera}  \& {Rasio}}{{Antonini}
  et~al.}{2016}]{2016ApJ...816...65A}
{Antonini} F.,  {Chatterjee} S.,  {Rodriguez} C.~L.,  {Morscher} M.,
  {Pattabiraman} B.,  {Kalogera} V.,   {Rasio} F.~A.,  2016, \mn@doi [\apj]
  {10.3847/0004-637X/816/2/65}, \href
  {https://ui.adsabs.harvard.edu/abs/2016ApJ...816...65A} {816, 65}

\bibitem[\protect\citeauthoryear{{Antonini}, {Toonen}  \& {Hamers}}{{Antonini}
  et~al.}{2017}]{2017ApJ...841...77A}
{Antonini} F.,  {Toonen} S.,   {Hamers} A.~S.,  2017, \mn@doi [\apj]
  {10.3847/1538-4357/aa6f5e}, \href
  {http://adsabs.harvard.edu/abs/2017ApJ...841...77A} {841, 77}

\bibitem[\protect\citeauthoryear{{Antonini}, {Rodriguez}, {Petrovich}  \&
  {Fischer}}{{Antonini} et~al.}{2018}]{2018MNRAS.480L..58A}
{Antonini} F.,  {Rodriguez} C.~L.,  {Petrovich} C.,   {Fischer} C.~L.,  2018,
  \mn@doi [\mnras] {10.1093/mnrasl/sly126}, \href
  {https://ui.adsabs.harvard.edu/abs/2018MNRAS.480L..58A} {480, L58}

\bibitem[\protect\citeauthoryear{{Bailyn}}{{Bailyn}}{1995}]{1995ARA&A..33..133B}
{Bailyn} C.~D.,  1995, \mn@doi [\araa] {10.1146/annurev.aa.33.090195.001025},
  \href {https://ui.adsabs.harvard.edu/abs/1995ARA&A..33..133B} {33, 133}

\bibitem[\protect\citeauthoryear{{Barker} \& {O'Connell}}{{Barker} \&
  {O'Connell}}{1975}]{1975PhRvD..12..329B}
{Barker} B.~M.,  {O'Connell} R.~F.,  1975, \mn@doi [\prd]
  {10.1103/PhysRevD.12.329}, \href
  {https://ui.adsabs.harvard.edu/abs/1975PhRvD..12..329B} {12, 329}

\bibitem[\protect\citeauthoryear{{Barker} \& {Ogilvie}}{{Barker} \&
  {Ogilvie}}{2009}]{2009MNRAS.395.2268B}
{Barker} A.~J.,  {Ogilvie} G.~I.,  2009, \mn@doi [\mnras]
  {10.1111/j.1365-2966.2009.14694.x}, \href
  {http://adsabs.harvard.edu/abs/2009MNRAS.395.2268B} {395, 2268}

\bibitem[\protect\citeauthoryear{{Barrett}, {Gaebel}, {Neijssel},
  {Vigna-G{\'o}mez}, {Stevenson}, {Berry}, {Farr}  \& {Mandel}}{{Barrett}
  et~al.}{2018}]{2018MNRAS.477.4685B}
{Barrett} J.~W.,  {Gaebel} S.~M.,  {Neijssel} C.~J.,  {Vigna-G{\'o}mez} A.,
  {Stevenson} S.,  {Berry} C. P.~L.,  {Farr} W.~M.,   {Mandel} I.,  2018,
  \mn@doi [\mnras] {10.1093/mnras/sty908}, \href
  {https://ui.adsabs.harvard.edu/abs/2018MNRAS.477.4685B} {477, 4685}

\bibitem[\protect\citeauthoryear{{Begelman}, {Rossi}  \& {Armitage}}{{Begelman}
  et~al.}{2008}]{2008MNRAS.387.1649B}
{Begelman} M.~C.,  {Rossi} E.~M.,   {Armitage} P.~J.,  2008, \mn@doi [\mnras]
  {10.1111/j.1365-2966.2008.13344.x}, \href
  {https://ui.adsabs.harvard.edu/abs/2008MNRAS.387.1649B} {387, 1649}

\bibitem[\protect\citeauthoryear{{Belczynski}, {Bulik}  \&
  {Kalogera}}{{Belczynski} et~al.}{2002a}]{2002ApJ...571L.147B}
{Belczynski} K.,  {Bulik} T.,   {Kalogera} V.,  2002a, \mn@doi [\apjl]
  {10.1086/341365}, \href
  {https://ui.adsabs.harvard.edu/abs/2002ApJ...571L.147B} {571, L147}

\bibitem[\protect\citeauthoryear{{Belczynski}, {Kalogera}  \&
  {Bulik}}{{Belczynski} et~al.}{2002b}]{2002ApJ...572..407B}
{Belczynski} K.,  {Kalogera} V.,   {Bulik} T.,  2002b, \mn@doi [\apj]
  {10.1086/340304}, \href
  {https://ui.adsabs.harvard.edu/abs/2002ApJ...572..407B} {572, 407}

\bibitem[\protect\citeauthoryear{{Belczynski}, {Kalogera}, {Rasio}, {Taam},
  {Zezas}, {Bulik}, {Maccarone}  \& {Ivanova}}{{Belczynski}
  et~al.}{2008}]{2008ApJS..174..223B}
{Belczynski} K.,  {Kalogera} V.,  {Rasio} F.~A.,  {Taam} R.~E.,  {Zezas} A.,
  {Bulik} T.,  {Maccarone} T.~J.,   {Ivanova} N.,  2008, \mn@doi [\apjs]
  {10.1086/521026}, \href
  {https://ui.adsabs.harvard.edu/abs/2008ApJS..174..223B} {174, 223}

\bibitem[\protect\citeauthoryear{{Benz} \& {Hills}}{{Benz} \&
  {Hills}}{1987}]{1987ApJ...323..614B}
{Benz} W.,  {Hills} J.~G.,  1987, \mn@doi [\apj] {10.1086/165857}, \href
  {https://ui.adsabs.harvard.edu/abs/1987ApJ...323..614B} {323, 614}

\bibitem[\protect\citeauthoryear{{Binney} \& {Tremaine}}{{Binney} \&
  {Tremaine}}{2008}]{2008gady.book.....B}
{Binney} J.,  {Tremaine} S.,  2008, {Galactic Dynamics: Second Edition}.
Princeton University Press

\bibitem[\protect\citeauthoryear{{Blaes}, {Lee}  \& {Socrates}}{{Blaes}
  et~al.}{2002}]{2002ApJ...578..775B}
{Blaes} O.,  {Lee} M.~H.,   {Socrates} A.,  2002, \mn@doi [\apj]
  {10.1086/342655}, \href {http://adsabs.harvard.edu/abs/2002ApJ...578..775B}
  {578, 775}

\bibitem[\protect\citeauthoryear{{Bobrick}, {Davies}  \& {Church}}{{Bobrick}
  et~al.}{2017}]{2017MNRAS.467.3556B}
{Bobrick} A.,  {Davies} M.~B.,   {Church} R.~P.,  2017, \mn@doi [\mnras]
  {10.1093/mnras/stx312}, \href
  {http://adsabs.harvard.edu/abs/2017MNRAS.467.3556B} {467, 3556}

\bibitem[\protect\citeauthoryear{{Boffin} \& {Jorissen}}{{Boffin} \&
  {Jorissen}}{1988}]{1988A&A...205..155B}
{Boffin} H.~M.~J.,  {Jorissen} A.,  1988, \aap, \href
  {https://ui.adsabs.harvard.edu/abs/1988A&A...205..155B} {205, 155}

\bibitem[\protect\citeauthoryear{{Bonavita} \& {Desidera}}{{Bonavita} \&
  {Desidera}}{2020}]{2020Galax...8...16B}
{Bonavita} M.,  {Desidera} S.,  2020, \mn@doi [Galaxies]
  {10.3390/galaxies8010016}, \href
  {https://ui.adsabs.harvard.edu/abs/2020Galax...8...16B} {8, 16}

\bibitem[\protect\citeauthoryear{{Bondi} \& {Hoyle}}{{Bondi} \&
  {Hoyle}}{1944}]{1944MNRAS.104..273B}
{Bondi} H.,  {Hoyle} F.,  1944, \mn@doi [\mnras] {10.1093/mnras/104.5.273},
  \href {https://ui.adsabs.harvard.edu/abs/1944MNRAS.104..273B} {104, 273}

\bibitem[\protect\citeauthoryear{{Bours}, {Toonen}  \& {Nelemans}}{{Bours}
  et~al.}{2013}]{2013A&A...552A..24B}
{Bours} M.~C.~P.,  {Toonen} S.,   {Nelemans} G.,  2013, \mn@doi [\aap]
  {10.1051/0004-6361/201220692}, \href
  {https://ui.adsabs.harvard.edu/abs/2013A&A...552A..24B} {552, A24}

\bibitem[\protect\citeauthoryear{{Breivik} et~al.,}{{Breivik}
  et~al.}{2020}]{2020ApJ...898...71B}
{Breivik} K.,  et~al., 2020, \mn@doi [\apj] {10.3847/1538-4357/ab9d85}, \href
  {https://ui.adsabs.harvard.edu/abs/2020ApJ...898...71B} {898, 71}

\bibitem[\protect\citeauthoryear{{Broekgaarden} et~al.,}{{Broekgaarden}
  et~al.}{2019}]{2019MNRAS.490.5228B}
{Broekgaarden} F.~S.,  et~al., 2019, \mn@doi [\mnras] {10.1093/mnras/stz2558},
  \href {https://ui.adsabs.harvard.edu/abs/2019MNRAS.490.5228B} {490, 5228}

\bibitem[\protect\citeauthoryear{{Brooker} \& {Olle}}{{Brooker} \&
  {Olle}}{1955}]{1955MNRAS.115..101B}
{Brooker} R.~A.,  {Olle} T.~W.,  1955, \mn@doi [\mnras]
  {10.1093/mnras/115.1.101}, \href
  {https://ui.adsabs.harvard.edu/abs/1955MNRAS.115..101B} {115, 101}

\bibitem[\protect\citeauthoryear{{Bulirsch} \& {Stoer}}{{Bulirsch} \&
  {Stoer}}{1966}]{bulirsch_stoer_1966}
{Bulirsch} R.,  {Stoer} J.,  1966, \mn@doi [Numerische Mathematik]
  {10.1007/BF02165234}, 8

\bibitem[\protect\citeauthoryear{{Campbell}}{{Campbell}}{1984}]{1984MNRAS.207..433C}
{Campbell} C.~G.,  1984, \mn@doi [\mnras] {10.1093/mnras/207.3.433}, \href
  {https://ui.adsabs.harvard.edu/abs/1984MNRAS.207..433C} {207, 433}

\bibitem[\protect\citeauthoryear{{Chandrasekhar}}{{Chandrasekhar}}{1939}]{1939isss.book.....C}
{Chandrasekhar} S.,  1939, {An introduction to the study of stellar structure}

\bibitem[\protect\citeauthoryear{{Chatterjee}, {Fregeau}, {Umbreit}  \&
  {Rasio}}{{Chatterjee} et~al.}{2010}]{2010ApJ...719..915C}
{Chatterjee} S.,  {Fregeau} J.~M.,  {Umbreit} S.,   {Rasio} F.~A.,  2010,
  \mn@doi [\apj] {10.1088/0004-637X/719/1/915}, \href
  {https://ui.adsabs.harvard.edu/abs/2010ApJ...719..915C} {719, 915}

\bibitem[\protect\citeauthoryear{{Chatterjee}, {Rasio}, {Sills}  \&
  {Glebbeek}}{{Chatterjee} et~al.}{2013}]{2013ApJ...777..106C}
{Chatterjee} S.,  {Rasio} F.~A.,  {Sills} A.,   {Glebbeek} E.,  2013, \mn@doi
  [\apj] {10.1088/0004-637X/777/2/106}, \href
  {https://ui.adsabs.harvard.edu/abs/2013ApJ...777..106C} {777, 106}

\bibitem[\protect\citeauthoryear{{Church}, {Dischler}, {Davies}, {Tout},
  {Adams}  \& {Beer}}{{Church} et~al.}{2009}]{2009MNRAS.395.1127C}
{Church} R.~P.,  {Dischler} J.,  {Davies} M.~B.,  {Tout} C.~A.,  {Adams} T.,
  {Beer} M.~E.,  2009, \mn@doi [\mnras] {10.1111/j.1365-2966.2009.14619.x},
  \href {http://adsabs.harvard.edu/abs/2009MNRAS.395.1127C} {395, 1127}

\bibitem[\protect\citeauthoryear{{Claeys}, {Pols}, {Izzard}, {Vink}  \&
  {Verbunt}}{{Claeys} et~al.}{2014}]{2014A&A...563A..83C}
{Claeys} J.~S.~W.,  {Pols} O.~R.,  {Izzard} R.~G.,  {Vink} J.,   {Verbunt}
  F.~W.~M.,  2014, \mn@doi [\aap] {10.1051/0004-6361/201322714}, \href
  {https://ui.adsabs.harvard.edu/abs/2014A&A...563A..83C} {563, A83}

\bibitem[\protect\citeauthoryear{{Claret}}{{Claret}}{2004}]{2004A&A...424..919C}
{Claret} A.,  2004, \mn@doi [\aap] {10.1051/0004-6361:20040470}, \href
  {https://ui.adsabs.harvard.edu/abs/2004A&A...424..919C} {424, 919}

\bibitem[\protect\citeauthoryear{{Cohen}, {Hindmarsh}  \& {Dubois}}{{Cohen}
  et~al.}{1996}]{1996ComPh..10..138C}
{Cohen} S.~D.,  {Hindmarsh} A.~C.,   {Dubois} P.~F.,  1996, Computers in
  Physics, \href {https://ui.adsabs.harvard.edu/abs/1996ComPh..10..138C} {10,
  138}

\bibitem[\protect\citeauthoryear{{Comerford} \& {Izzard}}{{Comerford} \&
  {Izzard}}{2020}]{2020MNRAS.498.2957C}
{Comerford} T.~A.~F.,  {Izzard} R.~G.,  2020, \mn@doi [\mnras]
  {10.1093/mnras/staa2539}, \href
  {https://ui.adsabs.harvard.edu/abs/2020MNRAS.498.2957C} {498, 2957}

\bibitem[\protect\citeauthoryear{{Counselman}}{{Counselman}}{1973}]{1973ApJ...180..307C}
{Counselman} Charles~C. I.,  1973, \mn@doi [\apj] {10.1086/151964}, \href
  {https://ui.adsabs.harvard.edu/abs/1973ApJ...180..307C} {180, 307}

\bibitem[\protect\citeauthoryear{{Damineli}, {Conti}  \& {Lopes}}{{Damineli}
  et~al.}{1997}]{1997NewA....2..107D}
{Damineli} A.,  {Conti} P.~S.,   {Lopes} D.~F.,  1997, \mn@doi [\na]
  {10.1016/S1384-1076(97)00008-0}, \href
  {https://ui.adsabs.harvard.edu/abs/1997NewA....2..107D} {2, 107}

\bibitem[\protect\citeauthoryear{{Darwin}}{{Darwin}}{1879}]{1879RSPT..170....1D}
{Darwin} G.~H.,  1879, Philosophical Transactions of the Royal Society of
  London Series I, \href
  {https://ui.adsabs.harvard.edu/abs/1879RSPT..170....1D} {170, 1}

\bibitem[\protect\citeauthoryear{{Davidson} \& {Humphreys}}{{Davidson} \&
  {Humphreys}}{1997}]{1997ARA&A..35....1D}
{Davidson} K.,  {Humphreys} R.~M.,  1997, \mn@doi [\araa]
  {10.1146/annurev.astro.35.1.1}, \href
  {https://ui.adsabs.harvard.edu/abs/1997ARA&A..35....1D} {35, 1}

\bibitem[\protect\citeauthoryear{{De Donder} \& {Vanbeveren}}{{De Donder} \&
  {Vanbeveren}}{2004}]{2004NewAR..48..861D}
{De Donder} E.,  {Vanbeveren} D.,  2004, \mn@doi [\nar]
  {10.1016/j.newar.2004.07.001}, \href
  {https://ui.adsabs.harvard.edu/abs/2004NewAR..48..861D} {48, 861}

\bibitem[\protect\citeauthoryear{{Dewey} \& {Cordes}}{{Dewey} \&
  {Cordes}}{1987}]{1987ApJ...321..780D}
{Dewey} R.~J.,  {Cordes} J.~M.,  1987, \mn@doi [\apj] {10.1086/165671}, \href
  {https://ui.adsabs.harvard.edu/abs/1987ApJ...321..780D} {321, 780}

\bibitem[\protect\citeauthoryear{{Di Stefano}}{{Di
  Stefano}}{2020}]{2020MNRAS.491..495D}
{Di Stefano} R.,  2020, \mn@doi [\mnras] {10.1093/mnras/stz2572}, \href
  {https://ui.adsabs.harvard.edu/abs/2020MNRAS.491..495D} {491, 495}

\bibitem[\protect\citeauthoryear{{Dosopoulou} \& {Kalogera}}{{Dosopoulou} \&
  {Kalogera}}{2016a}]{2016ApJ...825...70D}
{Dosopoulou} F.,  {Kalogera} V.,  2016a, \mn@doi [\apj]
  {10.3847/0004-637X/825/1/70}, \href
  {http://adsabs.harvard.edu/abs/2016ApJ...825...70D} {825, 70}

\bibitem[\protect\citeauthoryear{{Dosopoulou} \& {Kalogera}}{{Dosopoulou} \&
  {Kalogera}}{2016b}]{2016ApJ...825...71D}
{Dosopoulou} F.,  {Kalogera} V.,  2016b, \mn@doi [\apj]
  {10.3847/0004-637X/825/1/71}, \href
  {http://adsabs.harvard.edu/abs/2016ApJ...825...71D} {825, 71}

\bibitem[\protect\citeauthoryear{{Doyle} et~al.,}{{Doyle}
  et~al.}{2011}]{2011Sci...333.1602D}
{Doyle} L.~R.,  et~al., 2011, \mn@doi [Science] {10.1126/science.1210923},
  \href {https://ui.adsabs.harvard.edu/abs/2011Sci...333.1602D} {333, 1602}

\bibitem[\protect\citeauthoryear{{Duch{\^e}ne} \& {Kraus}}{{Duch{\^e}ne} \&
  {Kraus}}{2013}]{2013ARA&A..51..269D}
{Duch{\^e}ne} G.,  {Kraus} A.,  2013, \mn@doi [\araa]
  {10.1146/annurev-astro-081710-102602}, \href
  {https://ui.adsabs.harvard.edu/abs/2013ARA&A..51..269D} {51, 269}

\bibitem[\protect\citeauthoryear{{Dvorak}}{{Dvorak}}{1982}]{1982OAWMN.191..423D}
{Dvorak} R.,  1982, Oesterreichische Akademie Wissenschaften Mathematisch
  naturwissenschaftliche Klasse Sitzungsberichte Abteilung, \href
  {http://adsabs.harvard.edu/abs/1982OAWMN.191..423D} {191, 423}

\bibitem[\protect\citeauthoryear{{Edwards} \& {Pringle}}{{Edwards} \&
  {Pringle}}{1987}]{1987MNRAS.229..383E}
{Edwards} D.~A.,  {Pringle} J.~E.,  1987, \mn@doi [\mnras]
  {10.1093/mnras/229.3.383}, \href
  {http://adsabs.harvard.edu/abs/1987MNRAS.229..383E} {229, 383}

\bibitem[\protect\citeauthoryear{{Eggleton}}{{Eggleton}}{1983}]{1983ApJ...268..368E}
{Eggleton} P.~P.,  1983, \mn@doi [\apj] {10.1086/160960}, \href
  {http://adsabs.harvard.edu/abs/1983ApJ...268..368E} {268, 368}

\bibitem[\protect\citeauthoryear{{Eggleton} \& {Kiseleva-Eggleton}}{{Eggleton}
  \& {Kiseleva-Eggleton}}{2001}]{2001ApJ...562.1012E}
{Eggleton} P.~P.,  {Kiseleva-Eggleton} L.,  2001, \mn@doi [\apj]
  {10.1086/323843}, \href
  {http://ui.adsabs.harvard.edu/abs/2001ApJ...562.1012E} {562, 1012}

\bibitem[\protect\citeauthoryear{{Eggleton} \& {Kisseleva-Eggleton}}{{Eggleton}
  \& {Kisseleva-Eggleton}}{2006}]{2006Ap&SS.304...75E}
{Eggleton} P.~P.,  {Kisseleva-Eggleton} L.,  2006, \mn@doi [\apss]
  {10.1007/s10509-006-9078-z}, \href
  {http://ui.adsabs.harvard.edu/abs/2006Ap%26SS.304...75E} {304, 75}

\bibitem[\protect\citeauthoryear{{Eggleton}, {Kiseleva}  \& {Hut}}{{Eggleton}
  et~al.}{1998}]{1998ApJ...499..853E}
{Eggleton} P.~P.,  {Kiseleva} L.~G.,   {Hut} P.,  1998, \apj, \href
  {http://adsabs.harvard.edu/abs/1998ApJ...499..853E} {499, 853}

\bibitem[\protect\citeauthoryear{{El-Badry} \& {Rix}}{{El-Badry} \&
  {Rix}}{2018}]{2018MNRAS.480.4884E}
{El-Badry} K.,  {Rix} H.-W.,  2018, \mn@doi [\mnras] {10.1093/mnras/sty2186},
  \href {https://ui.adsabs.harvard.edu/abs/2018MNRAS.480.4884E} {480, 4884}

\bibitem[\protect\citeauthoryear{{Eldridge}}{{Eldridge}}{2012}]{2012MNRAS.422..794E}
{Eldridge} J.~J.,  2012, \mn@doi [\mnras] {10.1111/j.1365-2966.2012.20662.x},
  \href {https://ui.adsabs.harvard.edu/abs/2012MNRAS.422..794E} {422, 794}

\bibitem[\protect\citeauthoryear{{Eldridge} \& {Stanway}}{{Eldridge} \&
  {Stanway}}{2009}]{2009MNRAS.400.1019E}
{Eldridge} J.~J.,  {Stanway} E.~R.,  2009, \mn@doi [\mnras]
  {10.1111/j.1365-2966.2009.15514.x}, \href
  {https://ui.adsabs.harvard.edu/abs/2009MNRAS.400.1019E} {400, 1019}

\bibitem[\protect\citeauthoryear{{Eldridge} \& {Stanway}}{{Eldridge} \&
  {Stanway}}{2016}]{2016MNRAS.462.3302E}
{Eldridge} J.~J.,  {Stanway} E.~R.,  2016, \mn@doi [\mnras]
  {10.1093/mnras/stw1772}, \href
  {https://ui.adsabs.harvard.edu/abs/2016MNRAS.462.3302E} {462, 3302}

\bibitem[\protect\citeauthoryear{{Eldridge}, {Stanway}, {Xiao}, {McClelland },
  {Taylor}, {Ng}, {Greis}  \& {Bray}}{{Eldridge}
  et~al.}{2017}]{2017PASA...34...58E}
{Eldridge} J.~J.,  {Stanway} E.~R.,  {Xiao} L.,  {McClelland } L.~A.~S.,
  {Taylor} G.,  {Ng} M.,  {Greis} S.~M.~L.,   {Bray} J.~C.,  2017, \mn@doi
  [\pasa] {10.1017/pasa.2017.51}, \href
  {https://ui.adsabs.harvard.edu/abs/2017PASA...34...58E} {34, e058}

\bibitem[\protect\citeauthoryear{{Evans}}{{Evans}}{1968}]{1968QJRAS...9..388E}
{Evans} D.~S.,  1968, \qjras, \href
  {http://adsabs.harvard.edu/abs/1968QJRAS...9..388E} {9, 388}

\bibitem[\protect\citeauthoryear{{Fabrycky} \& {Tremaine}}{{Fabrycky} \&
  {Tremaine}}{2007}]{2007ApJ...669.1298F}
{Fabrycky} D.,  {Tremaine} S.,  2007, \mn@doi [\apj] {10.1086/521702}, \href
  {http://cdsads.u-strasbg.fr/abs/2007ApJ...669.1298F} {669, 1298}

\bibitem[\protect\citeauthoryear{{Fang}, {Thompson}  \& {Hirata}}{{Fang}
  et~al.}{2018}]{2018MNRAS.476.4234F}
{Fang} X.,  {Thompson} T.~A.,   {Hirata} C.~M.,  2018, \mn@doi [\mnras]
  {10.1093/mnras/sty472}, \href
  {https://ui.adsabs.harvard.edu/abs/2018MNRAS.476.4234F} {476, 4234}

\bibitem[\protect\citeauthoryear{{Fellhauer}, {Lin}, {Bolte}, {Aarseth}  \&
  {Williams}}{{Fellhauer} et~al.}{2003}]{2003ApJ...595L..53F}
{Fellhauer} M.,  {Lin} D.~N.~C.,  {Bolte} M.,  {Aarseth} S.~J.,   {Williams}
  K.~A.,  2003, \mn@doi [\apjl] {10.1086/379005}, \href
  {https://ui.adsabs.harvard.edu/abs/2003ApJ...595L..53F} {595, L53}

\bibitem[\protect\citeauthoryear{{Ford}, {Kozinsky}  \& {Rasio}}{{Ford}
  et~al.}{2000}]{2000ApJ...535..385F}
{Ford} E.~B.,  {Kozinsky} B.,   {Rasio} F.~A.,  2000, \mn@doi [\apj]
  {10.1086/308815}, \href {http://adsabs.harvard.edu/abs/2000ApJ...535..385F}
  {535, 385}

\bibitem[\protect\citeauthoryear{{Fragione} \& {Antonini}}{{Fragione} \&
  {Antonini}}{2019}]{2019MNRAS.488..728F}
{Fragione} G.,  {Antonini} F.,  2019, \mn@doi [\mnras] {10.1093/mnras/stz1723},
  \href {https://ui.adsabs.harvard.edu/abs/2019MNRAS.488..728F} {488, 728}

\bibitem[\protect\citeauthoryear{{Fragione} \& {Kocsis}}{{Fragione} \&
  {Kocsis}}{2019}]{2019MNRAS.486.4781F}
{Fragione} G.,  {Kocsis} B.,  2019, \mn@doi [\mnras] {10.1093/mnras/stz1175},
  \href {https://ui.adsabs.harvard.edu/abs/2019MNRAS.486.4781F} {486, 4781}

\bibitem[\protect\citeauthoryear{{Fragione} \& {Loeb}}{{Fragione} \&
  {Loeb}}{2019}]{2019MNRAS.486.4443F}
{Fragione} G.,  {Loeb} A.,  2019, \mn@doi [\mnras] {10.1093/mnras/stz1131},
  \href {https://ui.adsabs.harvard.edu/abs/2019MNRAS.486.4443F} {486, 4443}

\bibitem[\protect\citeauthoryear{{Fragione}, {Loeb}  \& {Rasio}}{{Fragione}
  et~al.}{2020a}]{2020ApJ...895L..15F}
{Fragione} G.,  {Loeb} A.,   {Rasio} F.~A.,  2020a, \mn@doi [\apjl]
  {10.3847/2041-8213/ab9093}, \href
  {https://ui.adsabs.harvard.edu/abs/2020ApJ...895L..15F} {895, L15}

\bibitem[\protect\citeauthoryear{{Fragione} et~al.,}{{Fragione}
  et~al.}{2020b}]{2020ApJ...900...16F}
{Fragione} G.,  et~al., 2020b, \mn@doi [\apj] {10.3847/1538-4357/aba89b}, \href
  {https://ui.adsabs.harvard.edu/abs/2020ApJ...900...16F} {900, 16}

\bibitem[\protect\citeauthoryear{{Franchini}, {Martin}  \& {Lubow}}{{Franchini}
  et~al.}{2019}]{2019MNRAS.485..315F}
{Franchini} A.,  {Martin} R.~G.,   {Lubow} S.~H.,  2019, \mn@doi [\mnras]
  {10.1093/mnras/stz424}, \href
  {https://ui.adsabs.harvard.edu/abs/2019MNRAS.485..315F} {485, 315}

\bibitem[\protect\citeauthoryear{{Fregeau} \& {Rasio}}{{Fregeau} \&
  {Rasio}}{2007}]{2007ApJ...658.1047F}
{Fregeau} J.~M.,  {Rasio} F.~A.,  2007, \mn@doi [\apj] {10.1086/511809}, \href
  {https://ui.adsabs.harvard.edu/abs/2007ApJ...658.1047F} {658, 1047}

\bibitem[\protect\citeauthoryear{{Fregeau}, {G{\"u}rkan}, {Joshi}  \&
  {Rasio}}{{Fregeau} et~al.}{2003}]{2003ApJ...593..772F}
{Fregeau} J.~M.,  {G{\"u}rkan} M.~A.,  {Joshi} K.~J.,   {Rasio} F.~A.,  2003,
  \mn@doi [\apj] {10.1086/376593}, \href
  {https://ui.adsabs.harvard.edu/abs/2003ApJ...593..772F} {593, 772}

\bibitem[\protect\citeauthoryear{{Fryer}, {Belczynski}, {Wiktorowicz},
  {Dominik}, {Kalogera}  \& {Holz}}{{Fryer} et~al.}{2012}]{2012ApJ...749...91F}
{Fryer} C.~L.,  {Belczynski} K.,  {Wiktorowicz} G.,  {Dominik} M.,  {Kalogera}
  V.,   {Holz} D.~E.,  2012, \mn@doi [\apj] {10.1088/0004-637X/749/1/91}, \href
  {https://ui.adsabs.harvard.edu/abs/2012ApJ...749...91F} {749, 91}

\bibitem[\protect\citeauthoryear{{Fu}, {Lubow}  \& {Martin}}{{Fu}
  et~al.}{2015}]{2015ApJ...813..105F}
{Fu} W.,  {Lubow} S.~H.,   {Martin} R.~G.,  2015, \mn@doi [\apj]
  {10.1088/0004-637X/813/2/105}, \href
  {https://ui.adsabs.harvard.edu/abs/2015ApJ...813..105F} {813, 105}

\bibitem[\protect\citeauthoryear{{Fujii} \& {Portegies Zwart}}{{Fujii} \&
  {Portegies Zwart}}{2013}]{2013MNRAS.430.1018F}
{Fujii} M.~S.,  {Portegies Zwart} S.,  2013, \mn@doi [\mnras]
  {10.1093/mnras/sts673}, \href
  {https://ui.adsabs.harvard.edu/abs/2013MNRAS.430.1018F} {430, 1018}

\bibitem[\protect\citeauthoryear{{Gaburov}, {Lombardi}  \& {Portegies
  Zwart}}{{Gaburov} et~al.}{2008}]{2008MNRAS.383L...5G}
{Gaburov} E.,  {Lombardi} J.~C.,   {Portegies Zwart} S.,  2008, \mn@doi
  [\mnras] {10.1111/j.1745-3933.2007.00399.x}, \href
  {https://ui.adsabs.harvard.edu/abs/2008MNRAS.383L...5G} {383, L5}

\bibitem[\protect\citeauthoryear{{Gaburov}, {Lombardi}  \& {Portegies
  Zwart}}{{Gaburov} et~al.}{2010}]{2010MNRAS.402..105G}
{Gaburov} E.,  {Lombardi} James~C. J.,   {Portegies Zwart} S.,  2010, \mn@doi
  [\mnras] {10.1111/j.1365-2966.2009.15900.x}, \href
  {https://ui.adsabs.harvard.edu/abs/2010MNRAS.402..105G} {402, 105}

\bibitem[\protect\citeauthoryear{{Gao}, {Correia}, {Eggleton}  \& {Han}}{{Gao}
  et~al.}{2018}]{2018MNRAS.479.3604G}
{Gao} Y.,  {Correia} A. C.~M.,  {Eggleton} P.~P.,   {Han} Z.,  2018, \mn@doi
  [\mnras] {10.1093/mnras/sty1558}, \href
  {https://ui.adsabs.harvard.edu/abs/2018MNRAS.479.3604G} {479, 3604}

\bibitem[\protect\citeauthoryear{{Gao}, {Toonen}, {Grishin}, {Comerford}  \&
  {Kruckow}}{{Gao} et~al.}{2020}]{2020MNRAS.491..264G}
{Gao} Y.,  {Toonen} S.,  {Grishin} E.,  {Comerford} T.,   {Kruckow} M.~U.,
  2020, \mn@doi [\mnras] {10.1093/mnras/stz3035}, \href
  {https://ui.adsabs.harvard.edu/abs/2020MNRAS.491..264G} {491, 264}

\bibitem[\protect\citeauthoryear{{Geller}, {Leigh}, {Giersz}, {Kremer}  \&
  {Rasio}}{{Geller} et~al.}{2019}]{2019ApJ...872..165G}
{Geller} A.~M.,  {Leigh} N. W.~C.,  {Giersz} M.,  {Kremer} K.,   {Rasio} F.~A.,
   2019, \mn@doi [\apj] {10.3847/1538-4357/ab0214}, \href
  {https://ui.adsabs.harvard.edu/abs/2019ApJ...872..165G} {872, 165}

\bibitem[\protect\citeauthoryear{{Giacobbo} \& {Mapelli}}{{Giacobbo} \&
  {Mapelli}}{2018}]{2018MNRAS.480.2011G}
{Giacobbo} N.,  {Mapelli} M.,  2018, \mn@doi [\mnras] {10.1093/mnras/sty1999},
  \href {https://ui.adsabs.harvard.edu/abs/2018MNRAS.480.2011G} {480, 2011}

\bibitem[\protect\citeauthoryear{{Giacobbo} \& {Mapelli}}{{Giacobbo} \&
  {Mapelli}}{2020}]{2020ApJ...891..141G}
{Giacobbo} N.,  {Mapelli} M.,  2020, \mn@doi [\apj] {10.3847/1538-4357/ab7335},
  \href {https://ui.adsabs.harvard.edu/abs/2020ApJ...891..141G} {891, 141}

\bibitem[\protect\citeauthoryear{{Giacobbo}, {Mapelli}  \& {Spera}}{{Giacobbo}
  et~al.}{2018}]{2018MNRAS.474.2959G}
{Giacobbo} N.,  {Mapelli} M.,   {Spera} M.,  2018, \mn@doi [\mnras]
  {10.1093/mnras/stx2933}, \href
  {https://ui.adsabs.harvard.edu/abs/2018MNRAS.474.2959G} {474, 2959}

\bibitem[\protect\citeauthoryear{{Giersz}, {Heggie}, {Hurley}  \&
  {Hypki}}{{Giersz} et~al.}{2013}]{2013MNRAS.431.2184G}
{Giersz} M.,  {Heggie} D.~C.,  {Hurley} J.~R.,   {Hypki} A.,  2013, \mn@doi
  [\mnras] {10.1093/mnras/stt307}, \href
  {https://ui.adsabs.harvard.edu/abs/2013MNRAS.431.2184G} {431, 2184}

\bibitem[\protect\citeauthoryear{{Glanz} \& {Perets}}{{Glanz} \&
  {Perets}}{2021}]{2021MNRAS.500.1921G}
{Glanz} H.,  {Perets} H.~B.,  2021, \mn@doi [\mnras] {10.1093/mnras/staa3242},
  \href {https://ui.adsabs.harvard.edu/abs/2021MNRAS.500.1921G} {500, 1921}

\bibitem[\protect\citeauthoryear{{Glebbeek} \& {Pols}}{{Glebbeek} \&
  {Pols}}{2008}]{2008A&A...488.1017G}
{Glebbeek} E.,  {Pols} O.~R.,  2008, \mn@doi [\aap]
  {10.1051/0004-6361:200809931}, \href
  {https://ui.adsabs.harvard.edu/abs/2008A&A...488.1017G} {488, 1017}

\bibitem[\protect\citeauthoryear{{Glebbeek}, {Pols}  \& {Hurley}}{{Glebbeek}
  et~al.}{2008}]{2008A&A...488.1007G}
{Glebbeek} E.,  {Pols} O.~R.,   {Hurley} J.~R.,  2008, \mn@doi [\aap]
  {10.1051/0004-6361:200809930}, \href
  {https://ui.adsabs.harvard.edu/abs/2008A&A...488.1007G} {488, 1007}

\bibitem[\protect\citeauthoryear{{Gragg}}{{Gragg}}{1965}]{1965SJNA....2..384G}
{Gragg} W.~B.,  1965, \mn@doi [SIAM Journal on Numerical Analysis]
  {10.1137/0702030}, \href
  {https://ui.adsabs.harvard.edu/abs/1965SJNA....2..384G} {2, 384}

\bibitem[\protect\citeauthoryear{{Grishin}, {Lai}  \& {Perets}}{{Grishin}
  et~al.}{2018a}]{2018MNRAS.474.3547G}
{Grishin} E.,  {Lai} D.,   {Perets} H.~B.,  2018a, \mn@doi [\mnras]
  {10.1093/mnras/stx3005}, \href
  {http://adsabs.harvard.edu/abs/2018MNRAS.474.3547G} {474, 3547}

\bibitem[\protect\citeauthoryear{{Grishin}, {Perets}  \& {Fragione}}{{Grishin}
  et~al.}{2018b}]{2018MNRAS.481.4907G}
{Grishin} E.,  {Perets} H.~B.,   {Fragione} G.,  2018b, \mn@doi [\mnras]
  {10.1093/mnras/sty2477}, \href
  {http://adsabs.harvard.edu/abs/2018MNRAS.481.4907G} {481, 4907}

\bibitem[\protect\citeauthoryear{{Gvaramadze}, {Gr{\"a}fener}, {Langer},
  {Maryeva}, {Kniazev}, {Moskvitin}  \& {Spiridonova}}{{Gvaramadze}
  et~al.}{2019}]{2019Natur.569..684G}
{Gvaramadze} V.~V.,  {Gr{\"a}fener} G.,  {Langer} N.,  {Maryeva} O.~V.,
  {Kniazev} A.~Y.,  {Moskvitin} A.~S.,   {Spiridonova} O.~I.,  2019, \mn@doi
  [\nat] {10.1038/s41586-019-1216-1}, \href
  {https://ui.adsabs.harvard.edu/abs/2019Natur.569..684G} {569, 684}

\bibitem[\protect\citeauthoryear{{Hadjidemetriou}}{{Hadjidemetriou}}{1963}]{1963Icar....2..440H}
{Hadjidemetriou} J.~D.,  1963, \mn@doi [\icarus]
  {10.1016/0019-1035(63)90072-1}, \href
  {http://adsabs.harvard.edu/abs/1963Icar....2..440H} {2, 440}

\bibitem[\protect\citeauthoryear{{Hadjidemetriou}}{{Hadjidemetriou}}{1969}]{1969Ap&SS...3...31H}
{Hadjidemetriou} J.~D.,  1969, \mn@doi [\apss] {10.1007/BF00649591}, \href
  {http://adsabs.harvard.edu/abs/1969Ap%26SS...3...31H} {3, 31}

\bibitem[\protect\citeauthoryear{{Hamers}}{{Hamers}}{2017a}]{2017MNRAS.466.4107H}
{Hamers} A.~S.,  2017a, \mn@doi [\mnras] {10.1093/mnras/stx035}, \href
  {https://ui.adsabs.harvard.edu/abs/2017MNRAS.466.4107H} {466, 4107}

\bibitem[\protect\citeauthoryear{{Hamers}}{{Hamers}}{2017b}]{2017ApJ...835L..24H}
{Hamers} A.~S.,  2017b, \mn@doi [\apjl] {10.3847/2041-8213/835/2/L24}, \href
  {https://ui.adsabs.harvard.edu/abs/2017ApJ...835L..24H} {835, L24}

\bibitem[\protect\citeauthoryear{{Hamers}}{{Hamers}}{2018a}]{2018MNRAS.476.4139H}
{Hamers} A.~S.,  2018a, \mn@doi [\mnras] {10.1093/mnras/sty428}, \href
  {https://ui.adsabs.harvard.edu/abs/2018MNRAS.476.4139H} {476, 4139}

\bibitem[\protect\citeauthoryear{{Hamers}}{{Hamers}}{2018b}]{2018MNRAS.478..620H}
{Hamers} A.~S.,  2018b, \mn@doi [\mnras] {10.1093/mnras/sty985}, \href
  {http://adsabs.harvard.edu/abs/2018MNRAS.478..620H} {478, 620}

\bibitem[\protect\citeauthoryear{{Hamers}}{{Hamers}}{2019}]{2019MNRAS.482.2262H}
{Hamers} A.~S.,  2019, \mn@doi [\mnras] {10.1093/mnras/sty2879}, \href
  {http://adsabs.harvard.edu/abs/2019MNRAS.482.2262H} {482, 2262}

\bibitem[\protect\citeauthoryear{{Hamers}}{{Hamers}}{2020a}]{2020MNRAS.494.5298H}
{Hamers} A.~S.,  2020a, \mn@doi [\mnras] {10.1093/mnras/staa1130}, \href
  {https://ui.adsabs.harvard.edu/abs/2020MNRAS.494.5298H} {494, 5298}

\bibitem[\protect\citeauthoryear{{Hamers}}{{Hamers}}{2020b}]{2020MNRAS.494.5492H}
{Hamers} A.~S.,  2020b, \mn@doi [\mnras] {10.1093/mnras/staa1084}, \href
  {https://ui.adsabs.harvard.edu/abs/2020MNRAS.494.5492H} {494, 5492}

\bibitem[\protect\citeauthoryear{{Hamers} \& {Dosopoulou}}{{Hamers} \&
  {Dosopoulou}}{2019}]{2019ApJ...872..119H}
{Hamers} A.~S.,  {Dosopoulou} F.,  2019, \mn@doi [\apj]
  {10.3847/1538-4357/ab001d}, \href
  {https://ui.adsabs.harvard.edu/abs/2019ApJ...872..119H} {872, 119}

\bibitem[\protect\citeauthoryear{{Hamers} \& {Lai}}{{Hamers} \&
  {Lai}}{2017}]{2017MNRAS.470.1657H}
{Hamers} A.~S.,  {Lai} D.,  2017, \mn@doi [\mnras] {10.1093/mnras/stx1319},
  \href {http://adsabs.harvard.edu/abs/2017MNRAS.470.1657H} {470, 1657}

\bibitem[\protect\citeauthoryear{{Hamers} \& {Portegies Zwart}}{{Hamers} \&
  {Portegies Zwart}}{2016a}]{2016MNRAS.459.2827H}
{Hamers} A.~S.,  {Portegies Zwart} S.~F.,  2016a, \mn@doi [\mnras]
  {10.1093/mnras/stw784}, \href
  {http://adsabs.harvard.edu/abs/2016MNRAS.459.2827H} {459, 2827}

\bibitem[\protect\citeauthoryear{{Hamers} \& {Portegies Zwart}}{{Hamers} \&
  {Portegies Zwart}}{2016b}]{2016MNRAS.462L..84H}
{Hamers} A.~S.,  {Portegies Zwart} S.~F.,  2016b, \mn@doi [\mnras]
  {10.1093/mnrasl/slw134}, \href
  {https://ui.adsabs.harvard.edu/abs/2016MNRAS.462L..84H} {462, L84}

\bibitem[\protect\citeauthoryear{{Hamers} \& {Safarzadeh}}{{Hamers} \&
  {Safarzadeh}}{2020}]{2020ApJ...898...99H}
{Hamers} A.~S.,  {Safarzadeh} M.,  2020, \mn@doi [\apj]
  {10.3847/1538-4357/ab9b27}, \href
  {https://ui.adsabs.harvard.edu/abs/2020ApJ...898...99H} {898, 99}

\bibitem[\protect\citeauthoryear{{Hamers} \& {Samsing}}{{Hamers} \&
  {Samsing}}{2019a}]{2019MNRAS.487.5630H}
{Hamers} A.~S.,  {Samsing} J.,  2019a, \mn@doi [\mnras]
  {10.1093/mnras/stz1646}, \href
  {https://ui.adsabs.harvard.edu/abs/2019MNRAS.487.5630H} {487, 5630}

\bibitem[\protect\citeauthoryear{{Hamers} \& {Samsing}}{{Hamers} \&
  {Samsing}}{2019b}]{2019MNRAS.488.5192H}
{Hamers} A.~S.,  {Samsing} J.,  2019b, \mn@doi [\mnras]
  {10.1093/mnras/stz2029}, \href
  {https://ui.adsabs.harvard.edu/abs/2019MNRAS.488.5192H} {488, 5192}

\bibitem[\protect\citeauthoryear{{Hamers} \& {Thompson}}{{Hamers} \&
  {Thompson}}{2019a}]{2019ApJ...882...24H}
{Hamers} A.~S.,  {Thompson} T.~A.,  2019a, \mn@doi [\apj]
  {10.3847/1538-4357/ab321f}, \href
  {https://ui.adsabs.harvard.edu/abs/2019ApJ...882...24H} {882, 24}

\bibitem[\protect\citeauthoryear{{Hamers} \& {Thompson}}{{Hamers} \&
  {Thompson}}{2019b}]{2019ApJ...883...23H}
{Hamers} A.~S.,  {Thompson} T.~A.,  2019b, \mn@doi [\apj]
  {10.3847/1538-4357/ab3b06}, \href
  {https://ui.adsabs.harvard.edu/abs/2019ApJ...883...23H} {883, 23}

\bibitem[\protect\citeauthoryear{{Hamers} \& {Tremaine}}{{Hamers} \&
  {Tremaine}}{2017}]{2017AJ....154..272H}
{Hamers} A.~S.,  {Tremaine} S.,  2017, \mn@doi [\aj]
  {10.3847/1538-3881/aa9926}, \href
  {https://ui.adsabs.harvard.edu/abs/2017AJ....154..272H} {154, 272}

\bibitem[\protect\citeauthoryear{{Hamers}, {Pols}, {Claeys}  \&
  {Nelemans}}{{Hamers} et~al.}{2013}]{2013MNRAS.430.2262H}
{Hamers} A.~S.,  {Pols} O.~R.,  {Claeys} J.~S.~W.,   {Nelemans} G.,  2013,
  \mn@doi [\mnras] {10.1093/mnras/stt046}, \href
  {https://ui.adsabs.harvard.edu/abs/2013MNRAS.430.2262H} {430, 2262}

\bibitem[\protect\citeauthoryear{{Hamers}, {Perets}, {Antonini}  \& {Portegies
  Zwart}}{{Hamers} et~al.}{2015}]{2015MNRAS.449.4221H}
{Hamers} A.~S.,  {Perets} H.~B.,  {Antonini} F.,   {Portegies Zwart} S.~F.,
  2015, \mn@doi [\mnras] {10.1093/mnras/stv452}, \href
  {http://adsabs.harvard.edu/abs/2015MNRAS.449.4221H} {449, 4221}

\bibitem[\protect\citeauthoryear{{Hamers}, {Perets}  \& {Portegies
  Zwart}}{{Hamers} et~al.}{2016}]{2016MNRAS.455.3180H}
{Hamers} A.~S.,  {Perets} H.~B.,   {Portegies Zwart} S.~F.,  2016, \mn@doi
  [\mnras] {10.1093/mnras/stv2447}, \href
  {http://adsabs.harvard.edu/abs/2016MNRAS.455.3180H} {455, 3180}

\bibitem[\protect\citeauthoryear{{Hamers}, {Bar-Or}, {Petrovich}  \&
  {Antonini}}{{Hamers} et~al.}{2018}]{2018ApJ...865....2H}
{Hamers} A.~S.,  {Bar-Or} B.,  {Petrovich} C.,   {Antonini} F.,  2018, \mn@doi
  [\apj] {10.3847/1538-4357/aadae2}, \href
  {http://adsabs.harvard.edu/abs/2018ApJ...865....2H} {865, 2}

\bibitem[\protect\citeauthoryear{{Harrington}}{{Harrington}}{1968}]{1968AJ.....73..190H}
{Harrington} R.~S.,  1968, \mn@doi [\aj] {10.1086/110614}, \href
  {http://adsabs.harvard.edu/abs/1968AJ.....73..190H} {73, 190}

\bibitem[\protect\citeauthoryear{{He} \& {Petrovich}}{{He} \&
  {Petrovich}}{2018}]{2018MNRAS.474...20H}
{He} M.~Y.,  {Petrovich} C.,  2018, \mn@doi [\mnras] {10.1093/mnras/stx2718},
  \href {https://ui.adsabs.harvard.edu/abs/2018MNRAS.474...20H} {474, 20}

\bibitem[\protect\citeauthoryear{{Heggie}}{{Heggie}}{1975}]{1975MNRAS.173..729H}
{Heggie} D.~C.,  1975, \mn@doi [\mnras] {10.1093/mnras/173.3.729}, \href
  {http://adsabs.harvard.edu/abs/1975MNRAS.173..729H} {173, 729}

\bibitem[\protect\citeauthoryear{{Heggie} \& {Rasio}}{{Heggie} \&
  {Rasio}}{1996}]{1996MNRAS.282.1064H}
{Heggie} D.~C.,  {Rasio} F.~A.,  1996, \mn@doi [\mnras]
  {10.1093/mnras/282.3.1064}, \href
  {https://ui.adsabs.harvard.edu/abs/1996MNRAS.282.1064H} {282, 1064}

\bibitem[\protect\citeauthoryear{{Heintz}}{{Heintz}}{1988}]{1988PASP..100..834H}
{Heintz} W.~D.,  1988, \mn@doi [\pasp] {10.1086/132244}, \href
  {https://ui.adsabs.harvard.edu/abs/1988PASP..100..834H} {100, 834}

\bibitem[\protect\citeauthoryear{{Hellstr{\"o}m} \& {Mikkola}}{{Hellstr{\"o}m}
  \& {Mikkola}}{2010}]{2010CeMDA.106..143H}
{Hellstr{\"o}m} C.,  {Mikkola} S.,  2010, \mn@doi [Celestial Mechanics and
  Dynamical Astronomy] {10.1007/s10569-009-9248-8}, \href
  {https://ui.adsabs.harvard.edu/abs/2010CeMDA.106..143H} {106, 143}

\bibitem[\protect\citeauthoryear{{Henon}}{{Henon}}{1972}]{1972A&A....19..488H}
{Henon} M.,  1972, \aap, \href
  {https://ui.adsabs.harvard.edu/abs/1972A&A....19..488H} {19, 488}

\bibitem[\protect\citeauthoryear{{Hils}, {Bender}  \& {Webbink}}{{Hils}
  et~al.}{1990}]{1990ApJ...360...75H}
{Hils} D.,  {Bender} P.~L.,   {Webbink} R.~F.,  1990, \mn@doi [\apj]
  {10.1086/169098}, \href
  {https://ui.adsabs.harvard.edu/abs/1990ApJ...360...75H} {360, 75}

\bibitem[\protect\citeauthoryear{{Hoang}, {Naoz}, {Kocsis}, {Rasio}  \&
  {Dosopoulou}}{{Hoang} et~al.}{2018}]{2018ApJ...856..140H}
{Hoang} B.-M.,  {Naoz} S.,  {Kocsis} B.,  {Rasio} F.~A.,   {Dosopoulou} F.,
  2018, \mn@doi [\apj] {10.3847/1538-4357/aaafce}, \href
  {http://adsabs.harvard.edu/abs/2018ApJ...856..140H} {856, 140}

\bibitem[\protect\citeauthoryear{{Hobbs}, {Lorimer}, {Lyne}  \&
  {Kramer}}{{Hobbs} et~al.}{2005}]{2005MNRAS.360..974H}
{Hobbs} G.,  {Lorimer} D.~R.,  {Lyne} A.~G.,   {Kramer} M.,  2005, \mn@doi
  [\mnras] {10.1111/j.1365-2966.2005.09087.x}, \href
  {https://ui.adsabs.harvard.edu/abs/2005MNRAS.360..974H} {360, 974}

\bibitem[\protect\citeauthoryear{{Hoyle} \& {Lyttleton}}{{Hoyle} \&
  {Lyttleton}}{1939}]{1939PCPS...35..405H}
{Hoyle} F.,  {Lyttleton} R.~A.,  1939, \mn@doi [Proceedings of the Cambridge
  Philosophical Society] {10.1017/S0305004100021150}, \href
  {https://ui.adsabs.harvard.edu/abs/1939PCPS...35..405H} {35, 405}

\bibitem[\protect\citeauthoryear{{Huang}}{{Huang}}{1956}]{1956AJ.....61...49H}
{Huang} S.~S.,  1956, \mn@doi [\aj] {10.1086/107290}, \href
  {http://adsabs.harvard.edu/abs/1956AJ.....61...49H} {61, 49}

\bibitem[\protect\citeauthoryear{{Humphreys} \& {Davidson}}{{Humphreys} \&
  {Davidson}}{1999}]{1999ASPC..179..216H}
{Humphreys} R.~M.,  {Davidson} K.,  1999, in {Morse} J.~A.,  {Humphreys} R.~M.,
    {Damineli} A.,  eds,  Astronomical Society of the Pacific Conference Series
  Vol. 179, Eta Carinae at The Millennium. p.~216

\bibitem[\protect\citeauthoryear{{Hurley}, {Pols}  \& {Tout}}{{Hurley}
  et~al.}{2000}]{2000MNRAS.315..543H}
{Hurley} J.~R.,  {Pols} O.~R.,   {Tout} C.~A.,  2000, \mn@doi [\mnras]
  {10.1046/j.1365-8711.2000.03426.x}, \href
  {http://adsabs.harvard.edu/abs/2000MNRAS.315..543H} {315, 543}

\bibitem[\protect\citeauthoryear{{Hurley}, {Tout}  \& {Pols}}{{Hurley}
  et~al.}{2002}]{2002MNRAS.329..897H}
{Hurley} J.~R.,  {Tout} C.~A.,   {Pols} O.~R.,  2002, \mn@doi [\mnras]
  {10.1046/j.1365-8711.2002.05038.x}, \href
  {http://adsabs.harvard.edu/abs/2002MNRAS.329..897H} {329, 897}

\bibitem[\protect\citeauthoryear{{Hut}}{{Hut}}{1980}]{1980A&A....92..167H}
{Hut} P.,  1980, \aap, \href
  {http://cdsads.u-strasbg.fr/abs/1980A%26A....92..167H} {92, 167}

\bibitem[\protect\citeauthoryear{{Hut}}{{Hut}}{1981}]{1981A&A....99..126H}
{Hut} P.,  1981, \aap, \href
  {http://cdsads.u-strasbg.fr/abs/1981A%26A....99..126H} {99, 126}

\bibitem[\protect\citeauthoryear{{Hypki} \& {Giersz}}{{Hypki} \&
  {Giersz}}{2013}]{2013MNRAS.429.1221H}
{Hypki} A.,  {Giersz} M.,  2013, \mn@doi [\mnras] {10.1093/mnras/sts415}, \href
  {https://ui.adsabs.harvard.edu/abs/2013MNRAS.429.1221H} {429, 1221}

\bibitem[\protect\citeauthoryear{{Iben} \& {Livio}}{{Iben} \&
  {Livio}}{1993}]{1993PASP..105.1373I}
{Iben} Icko J.,  {Livio} M.,  1993, \mn@doi [\pasp] {10.1086/133321}, \href
  {https://ui.adsabs.harvard.edu/abs/1993PASP..105.1373I} {105, 1373}

\bibitem[\protect\citeauthoryear{{Innes}}{{Innes}}{1917}]{1917CiUO...40..331I}
{Innes} R.~T.~A.,  1917, Circular of the Union Observatory Johannesburg, \href
  {https://ui.adsabs.harvard.edu/abs/1917CiUO...40..331I} {40, 331}

\bibitem[\protect\citeauthoryear{{Ito} \& {Ohtsuka}}{{Ito} \&
  {Ohtsuka}}{2019}]{2019MEEP....7....1I}
{Ito} T.,  {Ohtsuka} K.,  2019, \mn@doi [Monographs on Environment, Earth and
  Planets] {10.5047/meep.2019.00701.0001}, \href
  {https://ui.adsabs.harvard.edu/abs/2019MEEP....7....1I} {7, 1}

\bibitem[\protect\citeauthoryear{{Ivanova} et~al.,}{{Ivanova}
  et~al.}{2013}]{2013A&ARv..21...59I}
{Ivanova} N.,  et~al., 2013, \mn@doi [\aapr] {10.1007/s00159-013-0059-2}, \href
  {https://ui.adsabs.harvard.edu/abs/2013A&ARv..21...59I} {21, 59}

\bibitem[\protect\citeauthoryear{{Izzard}, {Tout}, {Karakas}  \&
  {Pols}}{{Izzard} et~al.}{2004}]{2004MNRAS.350..407I}
{Izzard} R.~G.,  {Tout} C.~A.,  {Karakas} A.~I.,   {Pols} O.~R.,  2004, \mn@doi
  [\mnras] {10.1111/j.1365-2966.2004.07446.x}, \href
  {https://ui.adsabs.harvard.edu/abs/2004MNRAS.350..407I} {350, 407}

\bibitem[\protect\citeauthoryear{{Izzard}, {Dray}, {Karakas}, {Lugaro}  \&
  {Tout}}{{Izzard} et~al.}{2006}]{2006A&A...460..565I}
{Izzard} R.~G.,  {Dray} L.~M.,  {Karakas} A.~I.,  {Lugaro} M.,   {Tout} C.~A.,
  2006, \mn@doi [\aap] {10.1051/0004-6361:20066129}, \href
  {https://ui.adsabs.harvard.edu/abs/2006A&A...460..565I} {460, 565}

\bibitem[\protect\citeauthoryear{{Izzard}, {Glebbeek}, {Stancliffe}  \&
  {Pols}}{{Izzard} et~al.}{2009}]{2009A&A...508.1359I}
{Izzard} R.~G.,  {Glebbeek} E.,  {Stancliffe} R.~J.,   {Pols} O.~R.,  2009,
  \mn@doi [\aap] {10.1051/0004-6361/200912827}, \href
  {https://ui.adsabs.harvard.edu/abs/2009A&A...508.1359I} {508, 1359}

\bibitem[\protect\citeauthoryear{{Janka}}{{Janka}}{2012}]{2012ARNPS..62..407J}
{Janka} H.-T.,  2012, \mn@doi [Annual Review of Nuclear and Particle Science]
  {10.1146/annurev-nucl-102711-094901}, \href
  {https://ui.adsabs.harvard.edu/abs/2012ARNPS..62..407J} {62, 407}

\bibitem[\protect\citeauthoryear{{Jordan}, {Perets}, {Fisher}  \& {van
  Rossum}}{{Jordan} et~al.}{2012}]{2012ApJ...761L..23J}
{Jordan} George~C. I.,  {Perets} H.~B.,  {Fisher} R.~T.,   {van Rossum} D.~R.,
  2012, \mn@doi [\apjl] {10.1088/2041-8205/761/2/L23}, \href
  {https://ui.adsabs.harvard.edu/abs/2012ApJ...761L..23J} {761, L23}

\bibitem[\protect\citeauthoryear{{Joshi}, {Rasio}  \& {Portegies
  Zwart}}{{Joshi} et~al.}{2000}]{2000ApJ...540..969J}
{Joshi} K.~J.,  {Rasio} F.~A.,   {Portegies Zwart} S.,  2000, \mn@doi [\apj]
  {10.1086/309350}, \href
  {https://ui.adsabs.harvard.edu/abs/2000ApJ...540..969J} {540, 969}

\bibitem[\protect\citeauthoryear{{Joshi}, {Nave}  \& {Rasio}}{{Joshi}
  et~al.}{2001}]{2001ApJ...550..691J}
{Joshi} K.~J.,  {Nave} C.~P.,   {Rasio} F.~A.,  2001, \mn@doi [\apj]
  {10.1086/319771}, \href
  {https://ui.adsabs.harvard.edu/abs/2001ApJ...550..691J} {550, 691}

\bibitem[\protect\citeauthoryear{{Kato} \& {Hachisu}}{{Kato} \&
  {Hachisu}}{2004}]{2004ApJ...613L.129K}
{Kato} M.,  {Hachisu} I.,  2004, \mn@doi [\apjl] {10.1086/425249}, \href
  {https://ui.adsabs.harvard.edu/abs/2004ApJ...613L.129K} {613, L129}

\bibitem[\protect\citeauthoryear{{Kato}, {Hachisu}  \& {Saio}}{{Kato}
  et~al.}{2017}]{2017gacv.workE..56K}
{Kato} M.,  {Hachisu} I.,   {Saio} H.,  2017, in The Golden Age of Cataclysmic
  Variables and Related Objects IV. p.~56 (\mn@eprint {arXiv} {1711.01529})

\bibitem[\protect\citeauthoryear{{Katz}, {Dong}  \& {Malhotra}}{{Katz}
  et~al.}{2011}]{2011PhRvL.107r1101K}
{Katz} B.,  {Dong} S.,   {Malhotra} R.,  2011, \mn@doi [Physical Review
  Letters] {10.1103/PhysRevLett.107.181101}, \href
  {http://adsabs.harvard.edu/abs/2011PhRvL.107r1101K} {107, 181101}

\bibitem[\protect\citeauthoryear{{Kidder}}{{Kidder}}{1995}]{1995PhRvD..52..821K}
{Kidder} L.~E.,  1995, \mn@doi [\prd] {10.1103/PhysRevD.52.821}, \href
  {https://ui.adsabs.harvard.edu/abs/1995PhRvD..52..821K} {52, 821}

\bibitem[\protect\citeauthoryear{{Kiseleva}, {Eggleton}  \&
  {Mikkola}}{{Kiseleva} et~al.}{1998}]{1998MNRAS.300..292K}
{Kiseleva} L.~G.,  {Eggleton} P.~P.,   {Mikkola} S.,  1998, \mn@doi [\mnras]
  {10.1046/j.1365-8711.1998.01903.x}, \href
  {http://cdsads.u-strasbg.fr/abs/1998MNRAS.300..292K} {300, 292}

\bibitem[\protect\citeauthoryear{{Kostov}, {McCullough}, {Hinse}, {Tsvetanov},
  {H{\'e}brard}, {D{\'\i}az}, {Deleuil}  \& {Valenti}}{{Kostov}
  et~al.}{2013}]{2013ApJ...770...52K}
{Kostov} V.~B.,  {McCullough} P.~R.,  {Hinse} T.~C.,  {Tsvetanov} Z.~I.,
  {H{\'e}brard} G.,  {D{\'\i}az} R.~F.,  {Deleuil} M.,   {Valenti} J.~A.,
  2013, \mn@doi [\apj] {10.1088/0004-637X/770/1/52}, \href
  {https://ui.adsabs.harvard.edu/abs/2013ApJ...770...52K} {770, 52}

\bibitem[\protect\citeauthoryear{{Kostov} et~al.,}{{Kostov}
  et~al.}{2016}]{2016ApJ...827...86K}
{Kostov} V.~B.,  et~al., 2016, \mn@doi [\apj] {10.3847/0004-637X/827/1/86},
  \href {https://ui.adsabs.harvard.edu/abs/2016ApJ...827...86K} {827, 86}

\bibitem[\protect\citeauthoryear{{Kozai}}{{Kozai}}{1962}]{1962AJ.....67..591K}
{Kozai} Y.,  1962, \mn@doi [\aj] {10.1086/108790}, \href
  {http://cdsads.u-strasbg.fr/abs/1962AJ.....67..591K} {67, 591}

\bibitem[\protect\citeauthoryear{{Kroupa}, {Tout}  \& {Gilmore}}{{Kroupa}
  et~al.}{1993}]{1993MNRAS.262..545K}
{Kroupa} P.,  {Tout} C.~A.,   {Gilmore} G.,  1993, \mn@doi [\mnras]
  {10.1093/mnras/262.3.545}, \href
  {http://adsabs.harvard.edu/abs/1993MNRAS.262..545K} {262, 545}

\bibitem[\protect\citeauthoryear{{Kruckow}, {Tauris}, {Langer}, {Kramer}  \&
  {Izzard}}{{Kruckow} et~al.}{2018}]{2018MNRAS.481.1908K}
{Kruckow} M.~U.,  {Tauris} T.~M.,  {Langer} N.,  {Kramer} M.,   {Izzard} R.~G.,
   2018, \mn@doi [\mnras] {10.1093/mnras/sty2190}, \href
  {https://ui.adsabs.harvard.edu/abs/2018MNRAS.481.1908K} {481, 1908}

\bibitem[\protect\citeauthoryear{{Kruszewski}}{{Kruszewski}}{1964}]{1964AcA....14..241K}
{Kruszewski} A.,  1964, \actaa, \href
  {http://adsabs.harvard.edu/abs/1964AcA....14..241K} {14, 241}

\bibitem[\protect\citeauthoryear{{Lajoie} \& {Sills}}{{Lajoie} \&
  {Sills}}{2011a}]{2011ApJ...726...66L}
{Lajoie} C.-P.,  {Sills} A.,  2011a, \mn@doi [\apj]
  {10.1088/0004-637X/726/2/66}, \href
  {http://adsabs.harvard.edu/abs/2011ApJ...726...66L} {726, 66}

\bibitem[\protect\citeauthoryear{{Lajoie} \& {Sills}}{{Lajoie} \&
  {Sills}}{2011b}]{2011ApJ...726...67L}
{Lajoie} C.-P.,  {Sills} A.,  2011b, \mn@doi [\apj]
  {10.1088/0004-637X/726/2/67}, \href
  {http://adsabs.harvard.edu/abs/2011ApJ...726...67L} {726, 67}

\bibitem[\protect\citeauthoryear{{Lang}}{{Lang}}{1992}]{1992adps.book.....L}
{Lang} K.~R.,  1992, {Astrophysical Data I. Planets and Stars.}

\bibitem[\protect\citeauthoryear{{Lei}}{{Lei}}{2019}]{2019MNRAS.490.4756L}
{Lei} H.,  2019, \mn@doi [\mnras] {10.1093/mnras/stz2917}, \href
  {https://ui.adsabs.harvard.edu/abs/2019MNRAS.490.4756L} {490, 4756}

\bibitem[\protect\citeauthoryear{{Lei}, {Circi}  \& {Ortore}}{{Lei}
  et~al.}{2018}]{2018MNRAS.481.4602L}
{Lei} H.,  {Circi} C.,   {Ortore} E.,  2018, \mn@doi [\mnras]
  {10.1093/mnras/sty2619}, \href
  {http://adsabs.harvard.edu/abs/2018MNRAS.481.4602L} {481, 4602}

\bibitem[\protect\citeauthoryear{{Leigh}, {Toonen}, {Portegies Zwart}  \&
  {Perna}}{{Leigh} et~al.}{2020}]{2020MNRAS.496.1819L}
{Leigh} N. W.~C.,  {Toonen} S.,  {Portegies Zwart} S.~F.,   {Perna} R.,  2020,
  \mn@doi [\mnras] {10.1093/mnras/staa1670}, \href
  {https://ui.adsabs.harvard.edu/abs/2020MNRAS.496.1819L} {496, 1819}

\bibitem[\protect\citeauthoryear{{Li}, {Naoz}, {Holman}  \& {Loeb}}{{Li}
  et~al.}{2014}]{2014ApJ...791...86L}
{Li} G.,  {Naoz} S.,  {Holman} M.,   {Loeb} A.,  2014, \mn@doi [\apj]
  {10.1088/0004-637X/791/2/86}, \href
  {http://adsabs.harvard.edu/abs/2014ApJ...791...86L} {791, 86}

\bibitem[\protect\citeauthoryear{{Lidov}}{{Lidov}}{1962}]{1962P&SS....9..719L}
{Lidov} M.~L.,  1962, \mn@doi [\planss] {10.1016/0032-0633(62)90129-0}, \href
  {http://cdsads.u-strasbg.fr/abs/1962P%26SS....9..719L} {9, 719}

\bibitem[\protect\citeauthoryear{{Lim} \& {Rodriguez}}{{Lim} \&
  {Rodriguez}}{2020}]{2020PhRvD.102f4033L}
{Lim} H.,  {Rodriguez} C.~L.,  2020, \mn@doi [\prd]
  {10.1103/PhysRevD.102.064033}, \href
  {https://ui.adsabs.harvard.edu/abs/2020PhRvD.102f4033L} {102, 064033}

\bibitem[\protect\citeauthoryear{{Lipunov} \& {Postnov}}{{Lipunov} \&
  {Postnov}}{1987}]{1987SvA....31..228L}
{Lipunov} V.~M.,  {Postnov} K.~A.,  1987, \sovast, \href
  {https://ui.adsabs.harvard.edu/abs/1987SvA....31..228L} {31, 228}

\bibitem[\protect\citeauthoryear{{Lithwick} \& {Naoz}}{{Lithwick} \&
  {Naoz}}{2011}]{2011ApJ...742...94L}
{Lithwick} Y.,  {Naoz} S.,  2011, \mn@doi [\apj] {10.1088/0004-637X/742/2/94},
  \href {http://adsabs.harvard.edu/abs/2011ApJ...742...94L} {742, 94}

\bibitem[\protect\citeauthoryear{{Liu} \& {Lai}}{{Liu} \&
  {Lai}}{2017}]{2017ApJ...846L..11L}
{Liu} B.,  {Lai} D.,  2017, \mn@doi [\apjl] {10.3847/2041-8213/aa8727}, \href
  {http://adsabs.harvard.edu/abs/2017ApJ...846L..11L} {846, L11}

\bibitem[\protect\citeauthoryear{{Liu} \& {Lai}}{{Liu} \&
  {Lai}}{2018}]{2018ApJ...863...68L}
{Liu} B.,  {Lai} D.,  2018, \mn@doi [\apj] {10.3847/1538-4357/aad09f}, \href
  {https://ui.adsabs.harvard.edu/abs/2018ApJ...863...68L} {863, 68}

\bibitem[\protect\citeauthoryear{{Liu} \& {Lai}}{{Liu} \&
  {Lai}}{2019}]{2019MNRAS.483.4060L}
{Liu} B.,  {Lai} D.,  2019, \mn@doi [\mnras] {10.1093/mnras/sty3432}, \href
  {https://ui.adsabs.harvard.edu/abs/2019MNRAS.483.4060L} {483, 4060}

\bibitem[\protect\citeauthoryear{{Livio} \& {Soker}}{{Livio} \&
  {Soker}}{1988}]{1988ApJ...329..764L}
{Livio} M.,  {Soker} N.,  1988, \mn@doi [\apj] {10.1086/166419}, \href
  {https://ui.adsabs.harvard.edu/abs/1988ApJ...329..764L} {329, 764}

\bibitem[\protect\citeauthoryear{{Lombardi}, {Rasio}  \& {Shapiro}}{{Lombardi}
  et~al.}{1995}]{1995ApJ...445L.117L}
{Lombardi} James~C. J.,  {Rasio} F.~A.,   {Shapiro} S.~L.,  1995, \mn@doi
  [\apjl] {10.1086/187903}, \href
  {https://ui.adsabs.harvard.edu/abs/1995ApJ...445L.117L} {445, L117}

\bibitem[\protect\citeauthoryear{{Lombardi}, {Rasio}  \& {Shapiro}}{{Lombardi}
  et~al.}{1996}]{1996ApJ...468..797L}
{Lombardi} James~C. J.,  {Rasio} F.~A.,   {Shapiro} S.~L.,  1996, \mn@doi
  [\apj] {10.1086/177736}, \href
  {https://ui.adsabs.harvard.edu/abs/1996ApJ...468..797L} {468, 797}

\bibitem[\protect\citeauthoryear{{Longland}, {Lor{\'e}n-Aguilar}, {Jos{\'e}},
  {Garc{\'\i}a-Berro}, {Althaus}  \& {Isern}}{{Longland}
  et~al.}{2011}]{2011ApJ...737L..34L}
{Longland} R.,  {Lor{\'e}n-Aguilar} P.,  {Jos{\'e}} J.,  {Garc{\'\i}a-Berro}
  E.,  {Althaus} L.~G.,   {Isern} J.,  2011, \mn@doi [\apjl]
  {10.1088/2041-8205/737/2/L34}, \href
  {https://ui.adsabs.harvard.edu/abs/2011ApJ...737L..34L} {737, L34}

\bibitem[\protect\citeauthoryear{{Lousto}, {Campanelli}, {Zlochower}  \&
  {Nakano}}{{Lousto} et~al.}{2010}]{2010CQGra..27k4006L}
{Lousto} C.~O.,  {Campanelli} M.,  {Zlochower} Y.,   {Nakano} H.,  2010,
  \mn@doi [Classical and Quantum Gravity] {10.1088/0264-9381/27/11/114006},
  \href {https://ui.adsabs.harvard.edu/abs/2010CQGra..27k4006L} {27, 114006}

\bibitem[\protect\citeauthoryear{{Lubow} \& {Ogilvie}}{{Lubow} \&
  {Ogilvie}}{2017}]{2017MNRAS.469.4292L}
{Lubow} S.~H.,  {Ogilvie} G.~I.,  2017, \mn@doi [\mnras]
  {10.1093/mnras/stx990}, \href
  {https://ui.adsabs.harvard.edu/abs/2017MNRAS.469.4292L} {469, 4292}

\bibitem[\protect\citeauthoryear{{Luo}, {Katz}  \& {Dong}}{{Luo}
  et~al.}{2016}]{2016MNRAS.458.3060L}
{Luo} L.,  {Katz} B.,   {Dong} S.,  2016, \mn@doi [\mnras]
  {10.1093/mnras/stw475}, \href
  {http://cdsads.u-strasbg.fr/abs/2016MNRAS.458.3060L} {458, 3060}

\bibitem[\protect\citeauthoryear{{Mamajek}, {Kenworthy}, {Hinz}  \&
  {Meyer}}{{Mamajek} et~al.}{2010}]{2010AJ....139..919M}
{Mamajek} E.~E.,  {Kenworthy} M.~A.,  {Hinz} P.~M.,   {Meyer} M.~R.,  2010,
  \mn@doi [\aj] {10.1088/0004-6256/139/3/919}, \href
  {https://ui.adsabs.harvard.edu/abs/2010AJ....139..919M} {139, 919}

\bibitem[\protect\citeauthoryear{{Mandel}}{{Mandel}}{2016}]{2016MNRAS.456..578M}
{Mandel} I.,  2016, \mn@doi [\mnras] {10.1093/mnras/stv2733}, \href
  {https://ui.adsabs.harvard.edu/abs/2016MNRAS.456..578M} {456, 578}

\bibitem[\protect\citeauthoryear{{Mandel} \& {M{\"u}ller}}{{Mandel} \&
  {M{\"u}ller}}{2020}]{2020MNRAS.499.3214M}
{Mandel} I.,  {M{\"u}ller} B.,  2020, \mn@doi [\mnras]
  {10.1093/mnras/staa3043}, \href
  {https://ui.adsabs.harvard.edu/abs/2020MNRAS.499.3214M} {499, 3214}

\bibitem[\protect\citeauthoryear{{Mapelli}}{{Mapelli}}{2016}]{2016MNRAS.459.3432M}
{Mapelli} M.,  2016, \mn@doi [\mnras] {10.1093/mnras/stw869}, \href
  {https://ui.adsabs.harvard.edu/abs/2016MNRAS.459.3432M} {459, 3432}

\bibitem[\protect\citeauthoryear{{Mardling} \& {Aarseth}}{{Mardling} \&
  {Aarseth}}{2001}]{2001MNRAS.321..398M}
{Mardling} R.~A.,  {Aarseth} S.~J.,  2001, \mn@doi [\mnras]
  {10.1046/j.1365-8711.2001.03974.x}, \href
  {http://adsabs.harvard.edu/abs/2001MNRAS.321..398M} {321, 398}

\bibitem[\protect\citeauthoryear{{Martin} \& {Franchini}}{{Martin} \&
  {Franchini}}{2019}]{2019MNRAS.489.1797M}
{Martin} R.~G.,  {Franchini} A.,  2019, \mn@doi [\mnras]
  {10.1093/mnras/stz2250}, \href
  {https://ui.adsabs.harvard.edu/abs/2019MNRAS.489.1797M} {489, 1797}

\bibitem[\protect\citeauthoryear{{Martin}, {Nixon}, {Lubow}, {Armitage},
  {Price}, {Do{\u g}an}  \& {King}}{{Martin}
  et~al.}{2014}]{2014ApJ...792L..33M}
{Martin} R.~G.,  {Nixon} C.,  {Lubow} S.~H.,  {Armitage} P.~J.,  {Price} D.~J.,
   {Do{\u g}an} S.,   {King} A.,  2014, \mn@doi [\apjl]
  {10.1088/2041-8205/792/2/L33}, \href
  {http://adsabs.harvard.edu/abs/2014ApJ...792L..33M} {792, L33}

\bibitem[\protect\citeauthoryear{{Martin}, {Mazeh}  \& {Fabrycky}}{{Martin}
  et~al.}{2015}]{2015MNRAS.453.3554M}
{Martin} D.~V.,  {Mazeh} T.,   {Fabrycky} D.~C.,  2015, \mn@doi [\mnras]
  {10.1093/mnras/stv1870}, \href
  {https://ui.adsabs.harvard.edu/abs/2015MNRAS.453.3554M} {453, 3554}

\bibitem[\protect\citeauthoryear{{Martinez} et~al.,}{{Martinez}
  et~al.}{2020}]{2020ApJ...903...67M}
{Martinez} M. A.~S.,  et~al., 2020, \mn@doi [\apj] {10.3847/1538-4357/abba25},
  \href {https://ui.adsabs.harvard.edu/abs/2020ApJ...903...67M} {903, 67}

\bibitem[\protect\citeauthoryear{{Matese} \& {Whitmire}}{{Matese} \&
  {Whitmire}}{1983}]{1983ApJ...266..776M}
{Matese} J.~J.,  {Whitmire} D.~P.,  1983, \mn@doi [\apj] {10.1086/160825},
  \href {http://adsabs.harvard.edu/abs/1983ApJ...266..776M} {266, 776}

\bibitem[\protect\citeauthoryear{{Matese} \& {Whitmire}}{{Matese} \&
  {Whitmire}}{1984}]{1984ApJ...282..522M}
{Matese} J.~J.,  {Whitmire} D.~P.,  1984, \mn@doi [\apj] {10.1086/162230},
  \href {http://adsabs.harvard.edu/abs/1984ApJ...282..522M} {282, 522}

\bibitem[\protect\citeauthoryear{{Mazeh} \& {Shaham}}{{Mazeh} \&
  {Shaham}}{1979}]{1979A&A....77..145M}
{Mazeh} T.,  {Shaham} J.,  1979, \aap, \href
  {http://cdsads.u-strasbg.fr/abs/1979A%26A....77..145M} {77, 145}

\bibitem[\protect\citeauthoryear{{Meltzer}}{{Meltzer}}{1957}]{1957ApJ...125..359M}
{Meltzer} A.~S.,  1957, \mn@doi [\apj] {10.1086/146313}, \href
  {https://ui.adsabs.harvard.edu/abs/1957ApJ...125..359M} {125, 359}

\bibitem[\protect\citeauthoryear{{Mennekens}, {Vanbeveren}, {De Greve}  \& {De
  Donder}}{{Mennekens} et~al.}{2010}]{2010A&A...515A..89M}
{Mennekens} N.,  {Vanbeveren} D.,  {De Greve} J.~P.,   {De Donder} E.,  2010,
  \mn@doi [\aap] {10.1051/0004-6361/201014115}, \href
  {https://ui.adsabs.harvard.edu/abs/2010A&A...515A..89M} {515, A89}

\bibitem[\protect\citeauthoryear{{Michaely} \& {Perets}}{{Michaely} \&
  {Perets}}{2014}]{2014ApJ...794..122M}
{Michaely} E.,  {Perets} H.~B.,  2014, \mn@doi [\apj]
  {10.1088/0004-637X/794/2/122}, \href
  {https://ui.adsabs.harvard.edu/abs/2014ApJ...794..122M} {794, 122}

\bibitem[\protect\citeauthoryear{{Michaely} \& {Perets}}{{Michaely} \&
  {Perets}}{2019}]{2019MNRAS.484.4711M}
{Michaely} E.,  {Perets} H.~B.,  2019, \mn@doi [\mnras] {10.1093/mnras/stz352},
  \href {https://ui.adsabs.harvard.edu/abs/2019MNRAS.484.4711M} {484, 4711}

\bibitem[\protect\citeauthoryear{{Mikkola} \& {Merritt}}{{Mikkola} \&
  {Merritt}}{2006}]{2006MNRAS.372..219M}
{Mikkola} S.,  {Merritt} D.,  2006, \mn@doi [\mnras]
  {10.1111/j.1365-2966.2006.10854.x}, \href
  {https://ui.adsabs.harvard.edu/abs/2006MNRAS.372..219M} {372, 219}

\bibitem[\protect\citeauthoryear{{Mikkola} \& {Merritt}}{{Mikkola} \&
  {Merritt}}{2008}]{2008AJ....135.2398M}
{Mikkola} S.,  {Merritt} D.,  2008, \mn@doi [\aj]
  {10.1088/0004-6256/135/6/2398}, \href
  {https://ui.adsabs.harvard.edu/abs/2008AJ....135.2398M} {135, 2398}

\bibitem[\protect\citeauthoryear{{Mikkola} \& {Tanikawa}}{{Mikkola} \&
  {Tanikawa}}{1999}]{1999MNRAS.310..745M}
{Mikkola} S.,  {Tanikawa} K.,  1999, \mn@doi [\mnras]
  {10.1046/j.1365-8711.1999.02982.x}, \href
  {https://ui.adsabs.harvard.edu/abs/1999MNRAS.310..745M} {310, 745}

\bibitem[\protect\citeauthoryear{{Moe} \& {Di Stefano}}{{Moe} \& {Di
  Stefano}}{2017}]{2017ApJS..230...15M}
{Moe} M.,  {Di Stefano} R.,  2017, \mn@doi [\apjs] {10.3847/1538-4365/aa6fb6},
  \href {http://adsabs.harvard.edu/abs/2017ApJS..230...15M} {230, 15}

\bibitem[\protect\citeauthoryear{{Moe} \& {Kratter}}{{Moe} \&
  {Kratter}}{2018}]{2018ApJ...854...44M}
{Moe} M.,  {Kratter} K.~M.,  2018, \mn@doi [\apj] {10.3847/1538-4357/aaa6d2},
  \href {https://ui.adsabs.harvard.edu/abs/2018ApJ...854...44M} {854, 44}

\bibitem[\protect\citeauthoryear{{Mora} \& {Will}}{{Mora} \&
  {Will}}{2004}]{2004PhRvD..69j4021M}
{Mora} T.,  {Will} C.~M.,  2004, \mn@doi [\prd] {10.1103/PhysRevD.69.104021},
  \href {https://ui.adsabs.harvard.edu/abs/2004PhRvD..69j4021M} {69, 104021}

\bibitem[\protect\citeauthoryear{{Mu{\~n}oz} \& {Lai}}{{Mu{\~n}oz} \&
  {Lai}}{2015}]{2015PNAS..112.9264M}
{Mu{\~n}oz} D.~J.,  {Lai} D.,  2015, \mn@doi [Proceedings of the National
  Academy of Science] {10.1073/pnas.1505671112}, \href
  {https://ui.adsabs.harvard.edu/abs/2015PNAS..112.9264M} {112, 9264}

\bibitem[\protect\citeauthoryear{{Mugrauer}, {Neuh{\"a}user}  \&
  {Mazeh}}{{Mugrauer} et~al.}{2007}]{2007A&A...469..755M}
{Mugrauer} M.,  {Neuh{\"a}user} R.,   {Mazeh} T.,  2007, \mn@doi [\aap]
  {10.1051/0004-6361:20065883}, \href
  {https://ui.adsabs.harvard.edu/abs/2007A&A...469..755M} {469, 755}

\bibitem[\protect\citeauthoryear{{Naoz}}{{Naoz}}{2016}]{2016ARA&A..54..441N}
{Naoz} S.,  2016, \mn@doi [\araa] {10.1146/annurev-astro-081915-023315}, \href
  {http://adsabs.harvard.edu/abs/2016ARA%26A..54..441N} {54, 441}

\bibitem[\protect\citeauthoryear{{Naoz} \& {Fabrycky}}{{Naoz} \&
  {Fabrycky}}{2014}]{2014ApJ...793..137N}
{Naoz} S.,  {Fabrycky} D.~C.,  2014, \mn@doi [\apj]
  {10.1088/0004-637X/793/2/137}, \href
  {http://adsabs.harvard.edu/abs/2014ApJ...793..137N} {793, 137}

\bibitem[\protect\citeauthoryear{{Naoz}, {Farr}  \& {Rasio}}{{Naoz}
  et~al.}{2012}]{2012ApJ...754L..36N}
{Naoz} S.,  {Farr} W.~M.,   {Rasio} F.~A.,  2012, \mn@doi [\apjl]
  {10.1088/2041-8205/754/2/L36}, \href
  {http://cdsads.u-strasbg.fr/abs/2012ApJ...754L..36N} {754, L36}

\bibitem[\protect\citeauthoryear{{Naoz}, {Farr}, {Lithwick}, {Rasio}  \&
  {Teyssandier}}{{Naoz} et~al.}{2013a}]{2013MNRAS.431.2155N}
{Naoz} S.,  {Farr} W.~M.,  {Lithwick} Y.,  {Rasio} F.~A.,   {Teyssandier} J.,
  2013a, \mn@doi [\mnras] {10.1093/mnras/stt302}, \href
  {http://adsabs.harvard.edu/abs/2013MNRAS.431.2155N} {431, 2155}

\bibitem[\protect\citeauthoryear{{Naoz}, {Kocsis}, {Loeb}  \& {Yunes}}{{Naoz}
  et~al.}{2013b}]{2013ApJ...773..187N}
{Naoz} S.,  {Kocsis} B.,  {Loeb} A.,   {Yunes} N.,  2013b, \mn@doi [\apj]
  {10.1088/0004-637X/773/2/187}, \href
  {http://adsabs.harvard.edu/abs/2013ApJ...773..187N} {773, 187}

\bibitem[\protect\citeauthoryear{{Neijssel} et~al.,}{{Neijssel}
  et~al.}{2019}]{2019MNRAS.490.3740N}
{Neijssel} C.~J.,  et~al., 2019, \mn@doi [\mnras] {10.1093/mnras/stz2840},
  \href {https://ui.adsabs.harvard.edu/abs/2019MNRAS.490.3740N} {490, 3740}

\bibitem[\protect\citeauthoryear{{Nelemans}, {Yungelson}, {Portegies Zwart}  \&
  {Verbunt}}{{Nelemans} et~al.}{2001a}]{2001A&A...365..491N}
{Nelemans} G.,  {Yungelson} L.~R.,  {Portegies Zwart} S.~F.,   {Verbunt} F.,
  2001a, \mn@doi [\aap] {10.1051/0004-6361:20000147}, \href
  {https://ui.adsabs.harvard.edu/abs/2001A&A...365..491N} {365, 491}

\bibitem[\protect\citeauthoryear{{Nelemans}, {Portegies Zwart}, {Verbunt}  \&
  {Yungelson}}{{Nelemans} et~al.}{2001b}]{2001A&A...368..939N}
{Nelemans} G.,  {Portegies Zwart} S.~F.,  {Verbunt} F.,   {Yungelson} L.~R.,
  2001b, \mn@doi [\aap] {10.1051/0004-6361:20010049}, \href
  {https://ui.adsabs.harvard.edu/abs/2001A&A...368..939N} {368, 939}

\bibitem[\protect\citeauthoryear{{Nie}, {Wood}  \& {Nicholls}}{{Nie}
  et~al.}{2017}]{2017ApJ...835..209N}
{Nie} J.~D.,  {Wood} P.~R.,   {Nicholls} C.~P.,  2017, \mn@doi [\apj]
  {10.3847/1538-4357/835/2/209}, \href
  {https://ui.adsabs.harvard.edu/abs/2017ApJ...835..209N} {835, 209}

\bibitem[\protect\citeauthoryear{{Nomoto} \& {Kondo}}{{Nomoto} \&
  {Kondo}}{1991}]{1991ApJ...367L..19N}
{Nomoto} K.,  {Kondo} Y.,  1991, \mn@doi [\apjl] {10.1086/185922}, \href
  {https://ui.adsabs.harvard.edu/abs/1991ApJ...367L..19N} {367, L19}

\bibitem[\protect\citeauthoryear{{Ogilvie}}{{Ogilvie}}{2014}]{2014ARA&A..52..171O}
{Ogilvie} G.~I.,  2014, \mn@doi [\araa] {10.1146/annurev-astro-081913-035941},
  \href {https://ui.adsabs.harvard.edu/abs/2014ARA&A..52..171O} {52, 171}

\bibitem[\protect\citeauthoryear{{Ott}}{{Ott}}{1993}]{1993cds..book.....O}
{Ott} E.,  1993, {Chaos in dynamical systems}

\bibitem[\protect\citeauthoryear{{Paczynski}}{{Paczynski}}{1976}]{1976IAUS...73...75P}
{Paczynski} B.,  1976, in {Eggleton} P.,  {Mitton} S.,   {Whelan} J.,  eds,
  {Structure and Evolution of Close Binary Systems} Vol. 73, Structure and
  Evolution of Close Binary Systems. p.~75

\bibitem[\protect\citeauthoryear{{Paczy{\'n}ski} \&
  {Sienkiewicz}}{{Paczy{\'n}ski} \& {Sienkiewicz}}{1972}]{1972AcA....22...73P}
{Paczy{\'n}ski} B.,  {Sienkiewicz} R.,  1972, \actaa, \href
  {http://adsabs.harvard.edu/abs/1972AcA....22...73P} {22, 73}

\bibitem[\protect\citeauthoryear{{Pejcha}, {Antognini}, {Shappee}  \&
  {Thompson}}{{Pejcha} et~al.}{2013}]{2013MNRAS.435..943P}
{Pejcha} O.,  {Antognini} J.~M.,  {Shappee} B.~J.,   {Thompson} T.~A.,  2013,
  \mn@doi [\mnras] {10.1093/mnras/stt1281}, \href
  {http://adsabs.harvard.edu/abs/2013MNRAS.435..943P} {435, 943}

\bibitem[\protect\citeauthoryear{{Pelupessy}, {van Elteren}, {de Vries},
  {McMillan}, {Drost}  \& {Portegies Zwart}}{{Pelupessy}
  et~al.}{2013}]{2013A&A...557A..84P}
{Pelupessy} F.~I.,  {van Elteren} A.,  {de Vries} N.,  {McMillan} S.~L.~W.,
  {Drost} N.,   {Portegies Zwart} S.~F.,  2013, \mn@doi [\aap]
  {10.1051/0004-6361/201321252}, \href
  {http://adsabs.harvard.edu/abs/2013A%26A...557A..84P} {557, A84}

\bibitem[\protect\citeauthoryear{{Perets} \& {Fabrycky}}{{Perets} \&
  {Fabrycky}}{2009}]{2009ApJ...697.1048P}
{Perets} H.~B.,  {Fabrycky} D.~C.,  2009, \mn@doi [\apj]
  {10.1088/0004-637X/697/2/1048}, \href
  {https://ui.adsabs.harvard.edu/abs/2009ApJ...697.1048P} {697, 1048}

\bibitem[\protect\citeauthoryear{{Perets} \& {Kratter}}{{Perets} \&
  {Kratter}}{2012}]{2012ApJ...760...99P}
{Perets} H.~B.,  {Kratter} K.~M.,  2012, \mn@doi [\apj]
  {10.1088/0004-637X/760/2/99}, \href
  {https://ui.adsabs.harvard.edu/abs/2012ApJ...760...99P} {760, 99}

\bibitem[\protect\citeauthoryear{{Peters}}{{Peters}}{1964}]{1964PhRv..136.1224P}
{Peters} P.~C.,  1964, \mn@doi [Physical Review] {10.1103/PhysRev.136.B1224},
  \href {https://ui.adsabs.harvard.edu/abs/1964PhRv..136.1224P} {136, 1224}

\bibitem[\protect\citeauthoryear{{Petrovich}}{{Petrovich}}{2015a}]{2015ApJ...799...27P}
{Petrovich} C.,  2015a, \mn@doi [\apj] {10.1088/0004-637X/799/1/27}, \href
  {http://cdsads.u-strasbg.fr/abs/2015ApJ...799...27P} {799, 27}

\bibitem[\protect\citeauthoryear{{Petrovich}}{{Petrovich}}{2015b}]{2015ApJ...808..120P}
{Petrovich} C.,  2015b, \mn@doi [\apj] {10.1088/0004-637X/808/2/120}, \href
  {https://ui.adsabs.harvard.edu/abs/2015ApJ...808..120P} {808, 120}

\bibitem[\protect\citeauthoryear{{Petrovich} \& {Mu{\~n}oz}}{{Petrovich} \&
  {Mu{\~n}oz}}{2017}]{2017ApJ...834..116P}
{Petrovich} C.,  {Mu{\~n}oz} D.~J.,  2017, \mn@doi [\apj]
  {10.3847/1538-4357/834/2/116}, \href
  {https://ui.adsabs.harvard.edu/abs/2017ApJ...834..116P} {834, 116}

\bibitem[\protect\citeauthoryear{{Petrovich} \& {Tremaine}}{{Petrovich} \&
  {Tremaine}}{2016}]{2016ApJ...829..132P}
{Petrovich} C.,  {Tremaine} S.,  2016, \mn@doi [\apj]
  {10.3847/0004-637X/829/2/132}, \href
  {http://adsabs.harvard.edu/abs/2016ApJ...829..132P} {829, 132}

\bibitem[\protect\citeauthoryear{{Piotrowski}}{{Piotrowski}}{1964}]{1964AcA....14..251P}
{Piotrowski} S.~L.,  1964, \actaa, \href
  {http://adsabs.harvard.edu/abs/1964AcA....14..251P} {14, 251}

\bibitem[\protect\citeauthoryear{{Pols}, {Schr{\"o}der}, {Hurley}, {Tout}  \&
  {Eggleton}}{{Pols} et~al.}{1998}]{1998MNRAS.298..525P}
{Pols} O.~R.,  {Schr{\"o}der} K.-P.,  {Hurley} J.~R.,  {Tout} C.~A.,
  {Eggleton} P.~P.,  1998, \mn@doi [\mnras] {10.1046/j.1365-8711.1998.01658.x},
  \href {https://ui.adsabs.harvard.edu/abs/1998MNRAS.298..525P} {298, 525}

\bibitem[\protect\citeauthoryear{{Portegies Zwart} \& {Verbunt}}{{Portegies
  Zwart} \& {Verbunt}}{1996}]{1996A&A...309..179P}
{Portegies Zwart} S.~F.,  {Verbunt} F.,  1996, \aap, \href
  {https://ui.adsabs.harvard.edu/abs/1996A&A...309..179P} {309, 179}

\bibitem[\protect\citeauthoryear{{Portegies Zwart} \& {van den
  Heuvel}}{{Portegies Zwart} \& {van den Heuvel}}{2016}]{2016MNRAS.456.3401P}
{Portegies Zwart} S.~F.,  {van den Heuvel} E.~P.~J.,  2016, \mn@doi [\mnras]
  {10.1093/mnras/stv2787}, \href
  {https://ui.adsabs.harvard.edu/abs/2016MNRAS.456.3401P} {456, 3401}

\bibitem[\protect\citeauthoryear{{Portegies Zwart}, {Makino}, {McMillan}  \&
  {Hut}}{{Portegies Zwart} et~al.}{1999}]{1999A&A...348..117P}
{Portegies Zwart} S.~F.,  {Makino} J.,  {McMillan} S.~L.~W.,   {Hut} P.,  1999,
  \aap, \href {https://ui.adsabs.harvard.edu/abs/1999A&A...348..117P} {348,
  117}

\bibitem[\protect\citeauthoryear{{Portegies Zwart}, {Belleman}  \&
  {Geldof}}{{Portegies Zwart} et~al.}{2007}]{2007NewA...12..641P}
{Portegies Zwart} S.~F.,  {Belleman} R.~G.,   {Geldof} P.~M.,  2007, \mn@doi
  [\na] {10.1016/j.newast.2007.05.004}, \href
  {https://ui.adsabs.harvard.edu/abs/2007NewA...12..641P} {12, 641}

\bibitem[\protect\citeauthoryear{{Portegies Zwart} et~al.,}{{Portegies Zwart}
  et~al.}{2009}]{2009NewA...14..369P}
{Portegies Zwart} S.,  et~al., 2009, \mn@doi [\na]
  {10.1016/j.newast.2008.10.006}, \href
  {https://ui.adsabs.harvard.edu/abs/2009NewA...14..369P} {14, 369}

\bibitem[\protect\citeauthoryear{{Preto} \& {Tremaine}}{{Preto} \&
  {Tremaine}}{1999}]{1999AJ....118.2532P}
{Preto} M.,  {Tremaine} S.,  1999, \mn@doi [\aj] {10.1086/301102}, \href
  {https://ui.adsabs.harvard.edu/abs/1999AJ....118.2532P} {118, 2532}

\bibitem[\protect\citeauthoryear{{Pustylnik}}{{Pustylnik}}{1998}]{1998A&AT...15..357P}
{Pustylnik} I.,  1998, \mn@doi [Astronomical and Astrophysical Transactions]
  {10.1080/10556799808201791}, \href
  {https://ui.adsabs.harvard.edu/abs/1998A&AT...15..357P} {15, 357}

\bibitem[\protect\citeauthoryear{{Raghavan} et~al.,}{{Raghavan}
  et~al.}{2010}]{2010ApJS..190....1R}
{Raghavan} D.,  et~al., 2010, \mn@doi [\apjs] {10.1088/0067-0049/190/1/1},
  \href {http://adsabs.harvard.edu/abs/2010ApJS..190....1R} {190, 1}

\bibitem[\protect\citeauthoryear{{Randall} \& {Xianyu}}{{Randall} \&
  {Xianyu}}{2018a}]{2018ApJ...853...93R}
{Randall} L.,  {Xianyu} Z.-Z.,  2018a, \mn@doi [\apj]
  {10.3847/1538-4357/aaa1a2}, \href
  {https://ui.adsabs.harvard.edu/abs/2018ApJ...853...93R} {853, 93}

\bibitem[\protect\citeauthoryear{{Randall} \& {Xianyu}}{{Randall} \&
  {Xianyu}}{2018b}]{2018ApJ...864..134R}
{Randall} L.,  {Xianyu} Z.-Z.,  2018b, \mn@doi [\apj]
  {10.3847/1538-4357/aad7fe}, \href
  {http://adsabs.harvard.edu/abs/2018ApJ...864..134R} {864, 134}

\bibitem[\protect\citeauthoryear{{Rantala}, {Pihajoki}, {Johansson}, {Naab},
  {Lah{\'e}n}  \& {Sawala}}{{Rantala} et~al.}{2017}]{2017ApJ...840...53R}
{Rantala} A.,  {Pihajoki} P.,  {Johansson} P.~H.,  {Naab} T.,  {Lah{\'e}n} N.,
   {Sawala} T.,  2017, \mn@doi [\apj] {10.3847/1538-4357/aa6d65}, \href
  {https://ui.adsabs.harvard.edu/abs/2017ApJ...840...53R} {840, 53}

\bibitem[\protect\citeauthoryear{{Rantala}, {Pihajoki}, {Mannerkoski},
  {Johansson}  \& {Naab}}{{Rantala} et~al.}{2020}]{2020MNRAS.492.4131R}
{Rantala} A.,  {Pihajoki} P.,  {Mannerkoski} M.,  {Johansson} P.~H.,   {Naab}
  T.,  2020, \mn@doi [\mnras] {10.1093/mnras/staa084}, \href
  {https://ui.adsabs.harvard.edu/abs/2020MNRAS.492.4131R} {492, 4131}

\bibitem[\protect\citeauthoryear{{Rasio}, {Tout}, {Lubow}  \& {Livio}}{{Rasio}
  et~al.}{1996}]{1996ApJ...470.1187R}
{Rasio} F.~A.,  {Tout} C.~A.,  {Lubow} S.~H.,   {Livio} M.,  1996, \mn@doi
  [\apj] {10.1086/177941}, \href
  {https://ui.adsabs.harvard.edu/abs/1996ApJ...470.1187R} {470, 1187}

\bibitem[\protect\citeauthoryear{{Reg{\"o}s}, {Bailey}  \&
  {Mardling}}{{Reg{\"o}s} et~al.}{2005}]{2005MNRAS.358..544R}
{Reg{\"o}s} E.,  {Bailey} V.~C.,   {Mardling} R.,  2005, \mn@doi [\mnras]
  {10.1111/j.1365-2966.2005.08813.x}, \href
  {http://adsabs.harvard.edu/abs/2005MNRAS.358..544R} {358, 544}

\bibitem[\protect\citeauthoryear{{Repetto} \& {Nelemans}}{{Repetto} \&
  {Nelemans}}{2015}]{2015MNRAS.453.3341R}
{Repetto} S.,  {Nelemans} G.,  2015, \mn@doi [\mnras] {10.1093/mnras/stv1753},
  \href {https://ui.adsabs.harvard.edu/abs/2015MNRAS.453.3341R} {453, 3341}

\bibitem[\protect\citeauthoryear{{Ritter}, {Politano}, {Livio}  \&
  {Webbink}}{{Ritter} et~al.}{1991}]{1991ApJ...376..177R}
{Ritter} H.,  {Politano} M.,  {Livio} M.,   {Webbink} R.~F.,  1991, \mn@doi
  [\apj] {10.1086/170265}, \href
  {https://ui.adsabs.harvard.edu/abs/1991ApJ...376..177R} {376, 177}

\bibitem[\protect\citeauthoryear{{Rizzuto} et~al.,}{{Rizzuto}
  et~al.}{2021}]{2021MNRAS.501.5257R}
{Rizzuto} F.~P.,  et~al., 2021, \mn@doi [\mnras] {10.1093/mnras/staa3634},
  \href {https://ui.adsabs.harvard.edu/abs/2021MNRAS.501.5257R} {501, 5257}

\bibitem[\protect\citeauthoryear{{Rodriguez} \& {Antonini}}{{Rodriguez} \&
  {Antonini}}{2018}]{2018ApJ...863....7R}
{Rodriguez} C.~L.,  {Antonini} F.,  2018, \mn@doi [\apj]
  {10.3847/1538-4357/aacea4}, \href
  {https://ui.adsabs.harvard.edu/abs/2018ApJ...863....7R} {863, 7}

\bibitem[\protect\citeauthoryear{{Safarzadeh}, {Hamers}, {Loeb}  \&
  {Berger}}{{Safarzadeh} et~al.}{2020}]{2020ApJ...888L...3S}
{Safarzadeh} M.,  {Hamers} A.~S.,  {Loeb} A.,   {Berger} E.,  2020, \mn@doi
  [\apjl] {10.3847/2041-8213/ab5dc8}, \href
  {https://ui.adsabs.harvard.edu/abs/2020ApJ...888L...3S} {888, L3}

\bibitem[\protect\citeauthoryear{{Salpeter}}{{Salpeter}}{1955}]{1955ApJ...121..161S}
{Salpeter} E.~E.,  1955, \mn@doi [\apj] {10.1086/145971}, \href
  {https://ui.adsabs.harvard.edu/abs/1955ApJ...121..161S} {121, 161}

\bibitem[\protect\citeauthoryear{{Schwab}, {Quataert}  \& {Kasen}}{{Schwab}
  et~al.}{2016}]{2016MNRAS.463.3461S}
{Schwab} J.,  {Quataert} E.,   {Kasen} D.,  2016, \mn@doi [\mnras]
  {10.1093/mnras/stw2249}, \href
  {https://ui.adsabs.harvard.edu/abs/2016MNRAS.463.3461S} {463, 3461}

\bibitem[\protect\citeauthoryear{{Schwamb} et~al.,}{{Schwamb}
  et~al.}{2013}]{2013ApJ...768..127S}
{Schwamb} M.~E.,  et~al., 2013, \mn@doi [\apj] {10.1088/0004-637X/768/2/127},
  \href {https://ui.adsabs.harvard.edu/abs/2013ApJ...768..127S} {768, 127}

\bibitem[\protect\citeauthoryear{{Sepinsky}, {Willems}  \&
  {Kalogera}}{{Sepinsky} et~al.}{2007a}]{2007ApJ...660.1624S}
{Sepinsky} J.~F.,  {Willems} B.,   {Kalogera} V.,  2007a, \mn@doi [\apj]
  {10.1086/513736}, \href {http://adsabs.harvard.edu/abs/2007ApJ...660.1624S}
  {660, 1624}

\bibitem[\protect\citeauthoryear{{Sepinsky}, {Willems}, {Kalogera}  \&
  {Rasio}}{{Sepinsky} et~al.}{2007b}]{2007ApJ...667.1170S}
{Sepinsky} J.~F.,  {Willems} B.,  {Kalogera} V.,   {Rasio} F.~A.,  2007b,
  \mn@doi [\apj] {10.1086/520911}, \href
  {http://adsabs.harvard.edu/abs/2007ApJ...667.1170S} {667, 1170}

\bibitem[\protect\citeauthoryear{{Sepinsky}, {Willems}, {Kalogera}  \&
  {Rasio}}{{Sepinsky} et~al.}{2009}]{2009ApJ...702.1387S}
{Sepinsky} J.~F.,  {Willems} B.,  {Kalogera} V.,   {Rasio} F.~A.,  2009,
  \mn@doi [\apj] {10.1088/0004-637X/702/2/1387}, \href
  {http://adsabs.harvard.edu/abs/2009ApJ...702.1387S} {702, 1387}

\bibitem[\protect\citeauthoryear{{Sepinsky}, {Willems}, {Kalogera}  \&
  {Rasio}}{{Sepinsky} et~al.}{2010}]{2010ApJ...724..546S}
{Sepinsky} J.~F.,  {Willems} B.,  {Kalogera} V.,   {Rasio} F.~A.,  2010,
  \mn@doi [\apj] {10.1088/0004-637X/724/1/546}, \href
  {http://adsabs.harvard.edu/abs/2010ApJ...724..546S} {724, 546}

\bibitem[\protect\citeauthoryear{{Shappee} \& {Thompson}}{{Shappee} \&
  {Thompson}}{2013}]{2013ApJ...766...64S}
{Shappee} B.~J.,  {Thompson} T.~A.,  2013, \mn@doi [\apj]
  {10.1088/0004-637X/766/1/64}, \href
  {https://ui.adsabs.harvard.edu/abs/2013ApJ...766...64S} {766, 64}

\bibitem[\protect\citeauthoryear{{Shevchenko}}{{Shevchenko}}{2017}]{2017ASSL..441.....S}
{Shevchenko} I.~I.,  2017, {The Lidov-Kozai Effect - Applications in Exoplanet
  Research and Dynamical Astronomy}.
 {Part of the Astrophysics and Space Science Library book series (ASSL, volume
  441)} Vol. 441, \mn@doi{10.1007/978-3-319-43522-0, }

\bibitem[\protect\citeauthoryear{{Sills} \& {Bailyn}}{{Sills} \&
  {Bailyn}}{1999}]{1999ApJ...513..428S}
{Sills} A.,  {Bailyn} C.~D.,  1999, \mn@doi [\apj] {10.1086/306840}, \href
  {https://ui.adsabs.harvard.edu/abs/1999ApJ...513..428S} {513, 428}

\bibitem[\protect\citeauthoryear{{Sills}, {Lombardi}, {Bailyn}, {Demarque},
  {Rasio}  \& {Shapiro}}{{Sills} et~al.}{1997}]{1997ApJ...487..290S}
{Sills} A.,  {Lombardi} James~C. J.,  {Bailyn} C.~D.,  {Demarque} P.,  {Rasio}
  F.~A.,   {Shapiro} S.~L.,  1997, \mn@doi [\apj] {10.1086/304588}, \href
  {https://ui.adsabs.harvard.edu/abs/1997ApJ...487..290S} {487, 290}

\bibitem[\protect\citeauthoryear{{Sills}, {Faber}, {Lombardi}, {Rasio}  \&
  {Warren}}{{Sills} et~al.}{2001}]{2001ApJ...548..323S}
{Sills} A.,  {Faber} J.~A.,  {Lombardi} James~C. J.,  {Rasio} F.~A.,   {Warren}
  A.~R.,  2001, \mn@doi [\apj] {10.1086/318689}, \href
  {https://ui.adsabs.harvard.edu/abs/2001ApJ...548..323S} {548, 323}

\bibitem[\protect\citeauthoryear{{Sills}, {Adams}, {Davies}  \& {Bate}}{{Sills}
  et~al.}{2002}]{2002MNRAS.332...49S}
{Sills} A.,  {Adams} T.,  {Davies} M.~B.,   {Bate} M.~R.,  2002, \mn@doi
  [\mnras] {10.1046/j.1365-8711.2002.05266.x}, \href
  {https://ui.adsabs.harvard.edu/abs/2002MNRAS.332...49S} {332, 49}

\bibitem[\protect\citeauthoryear{{Silsbee} \& {Tremaine}}{{Silsbee} \&
  {Tremaine}}{2017}]{2017ApJ...836...39S}
{Silsbee} K.,  {Tremaine} S.,  2017, \mn@doi [\apj] {10.3847/1538-4357/aa5729},
  \href {http://adsabs.harvard.edu/abs/2017ApJ...836...39S} {836, 39}

\bibitem[\protect\citeauthoryear{{Soberman}, {Phinney}  \& {van den
  Heuvel}}{{Soberman} et~al.}{1997}]{1997A&A...327..620S}
{Soberman} G.~E.,  {Phinney} E.~S.,   {van den Heuvel} E.~P.~J.,  1997, \aap,
  \href {https://ui.adsabs.harvard.edu/abs/1997A&A...327..620S} {327, 620}

\bibitem[\protect\citeauthoryear{{Spera} \& {Mapelli}}{{Spera} \&
  {Mapelli}}{2017}]{2017MNRAS.470.4739S}
{Spera} M.,  {Mapelli} M.,  2017, \mn@doi [\mnras] {10.1093/mnras/stx1576},
  \href {https://ui.adsabs.harvard.edu/abs/2017MNRAS.470.4739S} {470, 4739}

\bibitem[\protect\citeauthoryear{{Spera}, {Mapelli}  \& {Bressan}}{{Spera}
  et~al.}{2015}]{2015MNRAS.451.4086S}
{Spera} M.,  {Mapelli} M.,   {Bressan} A.,  2015, \mn@doi [\mnras]
  {10.1093/mnras/stv1161}, \href
  {https://ui.adsabs.harvard.edu/abs/2015MNRAS.451.4086S} {451, 4086}

\bibitem[\protect\citeauthoryear{{Spera}, {Mapelli}, {Giacobbo}, {Trani},
  {Bressan}  \& {Costa}}{{Spera} et~al.}{2019}]{2019MNRAS.485..889S}
{Spera} M.,  {Mapelli} M.,  {Giacobbo} N.,  {Trani} A.~A.,  {Bressan} A.,
  {Costa} G.,  2019, \mn@doi [\mnras] {10.1093/mnras/stz359}, \href
  {https://ui.adsabs.harvard.edu/abs/2019MNRAS.485..889S} {485, 889}

\bibitem[\protect\citeauthoryear{{Spurzem}}{{Spurzem}}{1999}]{1999JCoAM.109..407S}
{Spurzem} R.,  1999, Journal of Computational and Applied Mathematics, \href
  {https://ui.adsabs.harvard.edu/abs/1999JCoAM.109..407S} {109, 407}

\bibitem[\protect\citeauthoryear{{Spurzem}, {Berentzen}, {Berczik}, {Merritt},
  {Amaro-Seoane}, {Harfst}  \& {Gualand ris}}{{Spurzem}
  et~al.}{2008}]{2008LNP...760..377S}
{Spurzem} R.,  {Berentzen} I.,  {Berczik} P.,  {Merritt} D.,  {Amaro-Seoane}
  P.,  {Harfst} S.,   {Gualand ris} A.,  2008, {Parallelization, Special
  Hardware and Post-Newtonian Dynamics in Direct N - Body Simulations}.
p.~377, \mn@doi{10.1007/978-1-4020-8431-7_15}

\bibitem[\protect\citeauthoryear{{Spurzem}, {Giersz}, {Heggie}  \&
  {Lin}}{{Spurzem} et~al.}{2009}]{2009ApJ...697..458S}
{Spurzem} R.,  {Giersz} M.,  {Heggie} D.~C.,   {Lin} D.~N.~C.,  2009, \mn@doi
  [\apj] {10.1088/0004-637X/697/1/458}, \href
  {https://ui.adsabs.harvard.edu/abs/2009ApJ...697..458S} {697, 458}

\bibitem[\protect\citeauthoryear{{Stanway} \& {Eldridge}}{{Stanway} \&
  {Eldridge}}{2018}]{2018MNRAS.479...75S}
{Stanway} E.~R.,  {Eldridge} J.~J.,  2018, \mn@doi [\mnras]
  {10.1093/mnras/sty1353}, \href
  {https://ui.adsabs.harvard.edu/abs/2018MNRAS.479...75S} {479, 75}

\bibitem[\protect\citeauthoryear{{Stephan}, {Naoz}, {Ghez}, {Witzel},
  {Sitarski}, {Do}  \& {Kocsis}}{{Stephan} et~al.}{2016}]{2016MNRAS.460.3494S}
{Stephan} A.~P.,  {Naoz} S.,  {Ghez} A.~M.,  {Witzel} G.,  {Sitarski} B.~N.,
  {Do} T.,   {Kocsis} B.,  2016, \mn@doi [\mnras] {10.1093/mnras/stw1220},
  \href {http://adsabs.harvard.edu/abs/2016MNRAS.460.3494S} {460, 3494}

\bibitem[\protect\citeauthoryear{{Stephan} et~al.,}{{Stephan}
  et~al.}{2019}]{2019ApJ...878...58S}
{Stephan} A.~P.,  et~al., 2019, \mn@doi [\apj] {10.3847/1538-4357/ab1e4d},
  \href {https://ui.adsabs.harvard.edu/abs/2019ApJ...878...58S} {878, 58}

\bibitem[\protect\citeauthoryear{{Stevenson}, {Vigna-G{\'o}mez}, {Mandel},
  {Barrett}, {Neijssel}, {Perkins}  \& {de Mink}}{{Stevenson}
  et~al.}{2017}]{2017NatCo...814906S}
{Stevenson} S.,  {Vigna-G{\'o}mez} A.,  {Mandel} I.,  {Barrett} J.~W.,
  {Neijssel} C.~J.,  {Perkins} D.,   {de Mink} S.~E.,  2017, \mn@doi [Nature
  Communications] {10.1038/ncomms14906}, \href
  {https://ui.adsabs.harvard.edu/abs/2017NatCo...814906S} {8, 14906}

\bibitem[\protect\citeauthoryear{{Stevenson}, {Sampson}, {Powell},
  {Vigna-G{\'o}mez}, {Neijssel}, {Sz{\'e}csi}  \& {Mandel}}{{Stevenson}
  et~al.}{2019}]{2019ApJ...882..121S}
{Stevenson} S.,  {Sampson} M.,  {Powell} J.,  {Vigna-G{\'o}mez} A.,  {Neijssel}
  C.~J.,  {Sz{\'e}csi} D.,   {Mandel} I.,  2019, \mn@doi [\apj]
  {10.3847/1538-4357/ab3981}, \href
  {https://ui.adsabs.harvard.edu/abs/2019ApJ...882..121S} {882, 121}

\bibitem[\protect\citeauthoryear{{Stryker}}{{Stryker}}{1993}]{1993PASP..105.1081S}
{Stryker} L.~L.,  1993, \mn@doi [\pasp] {10.1086/133286}, \href
  {https://ui.adsabs.harvard.edu/abs/1993PASP..105.1081S} {105, 1081}

\bibitem[\protect\citeauthoryear{{Sun}, {Maund}, {Hirai}, {Crowther}  \&
  {Podsiadlowski}}{{Sun} et~al.}{2020}]{2020MNRAS.491.6000S}
{Sun} N.-C.,  {Maund} J.~R.,  {Hirai} R.,  {Crowther} P.~A.,   {Podsiadlowski}
  P.,  2020, \mn@doi [\mnras] {10.1093/mnras/stz3431}, \href
  {https://ui.adsabs.harvard.edu/abs/2020MNRAS.491.6000S} {491, 6000}

\bibitem[\protect\citeauthoryear{{Thompson}}{{Thompson}}{2011}]{2011ApJ...741...82T}
{Thompson} T.~A.,  2011, \mn@doi [\apj] {10.1088/0004-637X/741/2/82}, \href
  {http://adsabs.harvard.edu/abs/2011ApJ...741...82T} {741, 82}

\bibitem[\protect\citeauthoryear{{Thorne} \& {Zytkow}}{{Thorne} \&
  {Zytkow}}{1977}]{1977ApJ...212..832T}
{Thorne} K.~S.,  {Zytkow} A.~N.,  1977, \mn@doi [\apj] {10.1086/155109}, \href
  {https://ui.adsabs.harvard.edu/abs/1977ApJ...212..832T} {212, 832}

\bibitem[\protect\citeauthoryear{{Tokovinin}}{{Tokovinin}}{1997}]{1997A&AS..124...75T}
{Tokovinin} A.~A.,  1997, \mn@doi [\aaps] {10.1051/aas:1997181}, \href
  {https://ui.adsabs.harvard.edu/abs/1997A&AS..124...75T} {124, 75}

\bibitem[\protect\citeauthoryear{{Tokovinin}}{{Tokovinin}}{2014a}]{2014AJ....147...87T}
{Tokovinin} A.,  2014a, \mn@doi [\aj] {10.1088/0004-6256/147/4/87}, \href
  {http://adsabs.harvard.edu/abs/2014AJ....147...87T} {147, 87}

\bibitem[\protect\citeauthoryear{{Tokovinin}}{{Tokovinin}}{2014b}]{2014AJ....147...86T}
{Tokovinin} A.,  2014b, \mn@doi [\aj] {10.1088/0004-6256/147/4/86}, \href
  {https://ui.adsabs.harvard.edu/abs/2014AJ....147...86T} {147, 86}

\bibitem[\protect\citeauthoryear{{Tokovinin}}{{Tokovinin}}{2018}]{2018ApJS..235....6T}
{Tokovinin} A.,  2018, \mn@doi [\apjs] {10.3847/1538-4365/aaa1a5}, \href
  {https://ui.adsabs.harvard.edu/abs/2018ApJS..235....6T} {235, 6}

\bibitem[\protect\citeauthoryear{{Toonen} \& {Nelemans}}{{Toonen} \&
  {Nelemans}}{2013}]{2013A&A...557A..87T}
{Toonen} S.,  {Nelemans} G.,  2013, \mn@doi [\aap]
  {10.1051/0004-6361/201321753}, \href
  {https://ui.adsabs.harvard.edu/abs/2013A&A...557A..87T} {557, A87}

\bibitem[\protect\citeauthoryear{{Toonen}, {Nelemans}  \& {Portegies
  Zwart}}{{Toonen} et~al.}{2012}]{2012A&A...546A..70T}
{Toonen} S.,  {Nelemans} G.,   {Portegies Zwart} S.,  2012, \mn@doi [\aap]
  {10.1051/0004-6361/201218966}, \href
  {https://ui.adsabs.harvard.edu/abs/2012A&A...546A..70T} {546, A70}

\bibitem[\protect\citeauthoryear{{Toonen}, {Claeys}, {Mennekens}  \&
  {Ruiter}}{{Toonen} et~al.}{2014}]{2014A&A...562A..14T}
{Toonen} S.,  {Claeys} J.~S.~W.,  {Mennekens} N.,   {Ruiter} A.~J.,  2014,
  \mn@doi [\aap] {10.1051/0004-6361/201321576}, \href
  {https://ui.adsabs.harvard.edu/abs/2014A&A...562A..14T} {562, A14}

\bibitem[\protect\citeauthoryear{{Toonen}, {Hamers}  \& {Portegies
  Zwart}}{{Toonen} et~al.}{2016}]{2016ComAC...3....6T}
{Toonen} S.,  {Hamers} A.,   {Portegies Zwart} S.,  2016, \mn@doi
  [Computational Astrophysics and Cosmology] {10.1186/s40668-016-0019-0}, \href
  {http://adsabs.harvard.edu/abs/2016ComAC...3....6T} {3, 6}

\bibitem[\protect\citeauthoryear{{Toonen}, {Perets}  \& {Hamers}}{{Toonen}
  et~al.}{2018}]{2018A&A...610A..22T}
{Toonen} S.,  {Perets} H.~B.,   {Hamers} A.~S.,  2018, \mn@doi [\aap]
  {10.1051/0004-6361/201731874}, \href
  {http://adsabs.harvard.edu/abs/2018A%26A...610A..22T} {610, A22}

\bibitem[\protect\citeauthoryear{{Toonen}, {Portegies Zwart}, {Hamers}  \&
  {Band opadhyay}}{{Toonen} et~al.}{2020}]{2020A&A...640A..16T}
{Toonen} S.,  {Portegies Zwart} S.,  {Hamers} A.~S.,   {Band opadhyay} D.,
  2020, \mn@doi [\aap] {10.1051/0004-6361/201936835}, \href
  {https://ui.adsabs.harvard.edu/abs/2020A&A...640A..16T} {640, A16}

\bibitem[\protect\citeauthoryear{{Tout}, {Aarseth}, {Pols}  \&
  {Eggleton}}{{Tout} et~al.}{1997}]{1997MNRAS.291..732T}
{Tout} C.~A.,  {Aarseth} S.~J.,  {Pols} O.~R.,   {Eggleton} P.~P.,  1997,
  \mn@doi [\mnras] {10.1093/mnras/291.4.732}, \href
  {https://ui.adsabs.harvard.edu/abs/1997MNRAS.291..732T} {291, 732}

\bibitem[\protect\citeauthoryear{{Ulrich} \& {Burger}}{{Ulrich} \&
  {Burger}}{1976}]{1976ApJ...206..509U}
{Ulrich} R.~K.,  {Burger} H.~L.,  1976, \mn@doi [\apj] {10.1086/154406}, \href
  {https://ui.adsabs.harvard.edu/abs/1976ApJ...206..509U} {206, 509}

\bibitem[\protect\citeauthoryear{{Umbreit}, {Fregeau}, {Chatterjee}  \&
  {Rasio}}{{Umbreit} et~al.}{2012}]{2012ApJ...750...31U}
{Umbreit} S.,  {Fregeau} J.~M.,  {Chatterjee} S.,   {Rasio} F.~A.,  2012,
  \mn@doi [\apj] {10.1088/0004-637X/750/1/31}, \href
  {https://ui.adsabs.harvard.edu/abs/2012ApJ...750...31U} {750, 31}

\bibitem[\protect\citeauthoryear{{Vanbeveren}, {Mennekens}  \& {De
  Greve}}{{Vanbeveren} et~al.}{2012}]{2012A&A...543A...4V}
{Vanbeveren} D.,  {Mennekens} N.,   {De Greve} J.~P.,  2012, \mn@doi [\aap]
  {10.1051/0004-6361/201118081}, \href
  {https://ui.adsabs.harvard.edu/abs/2012A&A...543A...4V} {543, A4}

\bibitem[\protect\citeauthoryear{{Vennes}, {Nemeth}, {Kawka}, {Thorstensen},
  {Khalack}, {Ferrario}  \& {Alper}}{{Vennes}
  et~al.}{2017}]{2017Sci...357..680V}
{Vennes} S.,  {Nemeth} P.,  {Kawka} A.,  {Thorstensen} J.~R.,  {Khalack} V.,
  {Ferrario} L.,   {Alper} E.~H.,  2017, \mn@doi [Science]
  {10.1126/science.aam8378}, \href
  {https://ui.adsabs.harvard.edu/abs/2017Sci...357..680V} {357, 680}

\bibitem[\protect\citeauthoryear{{Veras} \& {Tout}}{{Veras} \&
  {Tout}}{2012}]{2012MNRAS.422.1648V}
{Veras} D.,  {Tout} C.~A.,  2012, \mn@doi [\mnras]
  {10.1111/j.1365-2966.2012.20741.x}, \href
  {http://adsabs.harvard.edu/abs/2012MNRAS.422.1648V} {422, 1648}

\bibitem[\protect\citeauthoryear{{Veras}, {Wyatt}, {Mustill}, {Bonsor}  \&
  {Eldridge}}{{Veras} et~al.}{2011}]{2011MNRAS.417.2104V}
{Veras} D.,  {Wyatt} M.~C.,  {Mustill} A.~J.,  {Bonsor} A.,   {Eldridge} J.~J.,
   2011, \mn@doi [\mnras] {10.1111/j.1365-2966.2011.19393.x}, \href
  {http://adsabs.harvard.edu/abs/2011MNRAS.417.2104V} {417, 2104}

\bibitem[\protect\citeauthoryear{{Veras}, {Hadjidemetriou}  \& {Tout}}{{Veras}
  et~al.}{2013}]{2013MNRAS.435.2416V}
{Veras} D.,  {Hadjidemetriou} J.~D.,   {Tout} C.~A.,  2013, \mn@doi [\mnras]
  {10.1093/mnras/stt1451}, \href
  {http://adsabs.harvard.edu/abs/2013MNRAS.435.2416V} {435, 2416}

\bibitem[\protect\citeauthoryear{{Veras}, {Evans}, {Wyatt}  \& {Tout}}{{Veras}
  et~al.}{2014}]{2014MNRAS.437.1127V}
{Veras} D.,  {Evans} N.~W.,  {Wyatt} M.~C.,   {Tout} C.~A.,  2014, \mn@doi
  [\mnras] {10.1093/mnras/stt1905}, \href
  {http://adsabs.harvard.edu/abs/2014MNRAS.437.1127V} {437, 1127}

\bibitem[\protect\citeauthoryear{{Vick} \& {Lai}}{{Vick} \&
  {Lai}}{2020}]{2020MNRAS.496.3767V}
{Vick} M.,  {Lai} D.,  2020, \mn@doi [\mnras] {10.1093/mnras/staa1784}, \href
  {https://ui.adsabs.harvard.edu/abs/2020MNRAS.496.3767V} {496, 3767}

\bibitem[\protect\citeauthoryear{{Vigna-G{\'o}mez} et~al.,}{{Vigna-G{\'o}mez}
  et~al.}{2018}]{2018MNRAS.481.4009V}
{Vigna-G{\'o}mez} A.,  et~al., 2018, \mn@doi [\mnras] {10.1093/mnras/sty2463},
  \href {https://ui.adsabs.harvard.edu/abs/2018MNRAS.481.4009V} {481, 4009}

\bibitem[\protect\citeauthoryear{{Vila}}{{Vila}}{1977}]{1977ApJ...213..464V}
{Vila} S.~C.,  1977, \mn@doi [\apj] {10.1086/155177}, \href
  {https://ui.adsabs.harvard.edu/abs/1977ApJ...213..464V} {213, 464}

\bibitem[\protect\citeauthoryear{{Vokrouhlick{\'y}}}{{Vokrouhlick{\'y}}}{2016}]{2016MNRAS.461.3964V}
{Vokrouhlick{\'y}} D.,  2016, \mn@doi [\mnras] {10.1093/mnras/stw1596}, \href
  {https://ui.adsabs.harvard.edu/abs/2016MNRAS.461.3964V} {461, 3964}

\bibitem[\protect\citeauthoryear{{Wang}, {Spurzem}, {Aarseth}, {Nitadori},
  {Berczik}, {Kouwenhoven}  \& {Naab}}{{Wang}
  et~al.}{2015}]{2015MNRAS.450.4070W}
{Wang} L.,  {Spurzem} R.,  {Aarseth} S.,  {Nitadori} K.,  {Berczik} P.,
  {Kouwenhoven} M.~B.~N.,   {Naab} T.,  2015, \mn@doi [\mnras]
  {10.1093/mnras/stv817}, \href
  {https://ui.adsabs.harvard.edu/abs/2015MNRAS.450.4070W} {450, 4070}

\bibitem[\protect\citeauthoryear{{Webbink}}{{Webbink}}{1984}]{1984ApJ...277..355W}
{Webbink} R.~F.,  1984, \mn@doi [\apj] {10.1086/161701}, \href
  {https://ui.adsabs.harvard.edu/abs/1984ApJ...277..355W} {277, 355}

\bibitem[\protect\citeauthoryear{{Webbink} \& {Han}}{{Webbink} \&
  {Han}}{1998}]{1998AIPC..456...61W}
{Webbink} R.~F.,  {Han} Z.,  1998, in {Folkner} W.~M.,  ed.,  American
  Institute of Physics Conference Series Vol. 456, Laser Interferometer Space
  Antenna, Second International LISA Symposium on the Detection and Observation
  of Gravitational Waves in Space. pp 61--67, \mn@doi{10.1063/1.57428}

\bibitem[\protect\citeauthoryear{{Weinberg}}{{Weinberg}}{1972}]{1972gcpa.book.....W}
{Weinberg} S.,  1972, {Gravitation and Cosmology: Principles and Applications
  of the General Theory of Relativity}

\bibitem[\protect\citeauthoryear{{Weppner}, {McKelvey}, {Thielen}  \&
  {Zielinski}}{{Weppner} et~al.}{2015}]{2015MNRAS.452.1375W}
{Weppner} S.~P.,  {McKelvey} J.~P.,  {Thielen} K.~D.,   {Zielinski} A.~K.,
  2015, \mn@doi [\mnras] {10.1093/mnras/stv1397}, \href
  {http://adsabs.harvard.edu/abs/2015MNRAS.452.1375W} {452, 1375}

\bibitem[\protect\citeauthoryear{{Whyte} \& {Eggleton}}{{Whyte} \&
  {Eggleton}}{1985}]{1985MNRAS.214..357W}
{Whyte} C.~A.,  {Eggleton} P.~P.,  1985, \mn@doi [\mnras]
  {10.1093/mnras/214.3.357}, \href
  {https://ui.adsabs.harvard.edu/abs/1985MNRAS.214..357W} {214, 357}

\bibitem[\protect\citeauthoryear{{Will}}{{Will}}{2006}]{2006LRR.....9....3W}
{Will} C.~M.,  2006, \mn@doi [Living Reviews in Relativity]
  {10.12942/lrr-2006-3}, \href
  {https://ui.adsabs.harvard.edu/abs/2006LRR.....9....3W} {9, 3}

\bibitem[\protect\citeauthoryear{{Will}}{{Will}}{2014}]{2014CQGra..31x4001W}
{Will} C.~M.,  2014, \mn@doi [Classical and Quantum Gravity]
  {10.1088/0264-9381/31/24/244001}, \href
  {https://ui.adsabs.harvard.edu/abs/2014CQGra..31x4001W} {31, 244001}

\bibitem[\protect\citeauthoryear{{Will}}{{Will}}{2017}]{2017PhRvD..96b3017W}
{Will} C.~M.,  2017, \mn@doi [\prd] {10.1103/PhysRevD.96.023017}, \href
  {https://ui.adsabs.harvard.edu/abs/2017PhRvD..96b3017W} {96, 023017}

\bibitem[\protect\citeauthoryear{{Winn} \& {Fabrycky}}{{Winn} \&
  {Fabrycky}}{2015}]{2015ARA&A..53..409W}
{Winn} J.~N.,  {Fabrycky} D.~C.,  2015, \mn@doi [\araa]
  {10.1146/annurev-astro-082214-122246}, \href
  {https://ui.adsabs.harvard.edu/abs/2015ARA&A..53..409W} {53, 409}

\bibitem[\protect\citeauthoryear{{Wu} \& {Murray}}{{Wu} \&
  {Murray}}{2003}]{2003ApJ...589..605W}
{Wu} Y.,  {Murray} N.,  2003, \mn@doi [\apj] {10.1086/374598}, \href
  {http://cdsads.u-strasbg.fr/abs/2003ApJ...589..605W} {589, 605}

\bibitem[\protect\citeauthoryear{{Zahn}}{{Zahn}}{1975}]{1975A&A....41..329Z}
{Zahn} J.~P.,  1975, \aap, \href
  {https://ui.adsabs.harvard.edu/abs/1975A&A....41..329Z} {41, 329}

\bibitem[\protect\citeauthoryear{{Zahn}}{{Zahn}}{1977}]{1977A&A....57..383Z}
{Zahn} J.-P.,  1977, \aap, \href
  {http://adsabs.harvard.edu/abs/1977A%26A....57..383Z} {57, 383}

\bibitem[\protect\citeauthoryear{{Zanazzi} \& {Lai}}{{Zanazzi} \&
  {Lai}}{2017}]{2017MNRAS.467.1957Z}
{Zanazzi} J.~J.,  {Lai} D.,  2017, \mn@doi [\mnras] {10.1093/mnras/stx208},
  \href {https://ui.adsabs.harvard.edu/abs/2017MNRAS.467.1957Z} {467, 1957}

\bibitem[\protect\citeauthoryear{{Zanazzi} \& {Lai}}{{Zanazzi} \&
  {Lai}}{2018}]{2018MNRAS.477.5207Z}
{Zanazzi} J.~J.,  {Lai} D.,  2018, \mn@doi [\mnras] {10.1093/mnras/sty951},
  \href {https://ui.adsabs.harvard.edu/abs/2018MNRAS.477.5207Z} {477, 5207}

\bibitem[\protect\citeauthoryear{{de Kool}}{{de
  Kool}}{1990}]{1990ApJ...358..189D}
{de Kool} M.,  1990, \mn@doi [\apj] {10.1086/168974}, \href
  {https://ui.adsabs.harvard.edu/abs/1990ApJ...358..189D} {358, 189}

\bibitem[\protect\citeauthoryear{{de Mink}, {Sana}, {Langer}, {Izzard}  \&
  {Schneider}}{{de Mink} et~al.}{2014}]{2014ApJ...782....7D}
{de Mink} S.~E.,  {Sana} H.,  {Langer} N.,  {Izzard} R.~G.,   {Schneider}
  F.~R.~N.,  2014, \mn@doi [\apj] {10.1088/0004-637X/782/1/7}, \href
  {https://ui.adsabs.harvard.edu/abs/2014ApJ...782....7D} {782, 7}

\bibitem[\protect\citeauthoryear{{de Vries}, {Portegies Zwart}  \&
  {Figueira}}{{de Vries} et~al.}{2014}]{2014MNRAS.438.1909D}
{de Vries} N.,  {Portegies Zwart} S.,   {Figueira} J.,  2014, \mn@doi [\mnras]
  {10.1093/mnras/stt1688}, \href
  {https://ui.adsabs.harvard.edu/abs/2014MNRAS.438.1909D} {438, 1909}

\bibitem[\protect\citeauthoryear{{van Haaften}, {Nelemans}, {Voss}, {Toonen},
  {Portegies Zwart}, {Yungelson}  \& {van der Sluys}}{{van Haaften}
  et~al.}{2013}]{2013A&A...552A..69V}
{van Haaften} L.~M.,  {Nelemans} G.,  {Voss} R.,  {Toonen} S.,  {Portegies
  Zwart} S.~F.,  {Yungelson} L.~R.,   {van der Sluys} M.~V.,  2013, \mn@doi
  [\aap] {10.1051/0004-6361/201220552}, \href
  {https://ui.adsabs.harvard.edu/abs/2013A&A...552A..69V} {552, A69}

\bibitem[\protect\citeauthoryear{{van Paradijs}, {van den Heuvel},
  {Kouveliotou}, {Fishman}, {Finger}  \& {Lewin}}{{van Paradijs}
  et~al.}{1997}]{1997A&A...317L...9V}
{van Paradijs} J.,  {van den Heuvel} E.~P.~J.,  {Kouveliotou} C.,  {Fishman}
  G.~J.,  {Finger} M.~H.,   {Lewin} W.~H.~G.,  1997, \aap, \href
  {https://ui.adsabs.harvard.edu/abs/1997A&A...317L...9V} {317, L9}

\bibitem[\protect\citeauthoryear{{van den Berk}, {Portegies Zwart}  \&
  {McMillan}}{{van den Berk} et~al.}{2007}]{2007MNRAS.379..111V}
{van den Berk} J.,  {Portegies Zwart} S.~F.,   {McMillan} S.~L.~W.,  2007,
  \mn@doi [\mnras] {10.1111/j.1365-2966.2007.11913.x}, \href
  {https://ui.adsabs.harvard.edu/abs/2007MNRAS.379..111V} {379, 111}

\bibitem[\protect\citeauthoryear{{van den Heuvel}}{{van den
  Heuvel}}{1976}]{1976IAUS...73...35V}
{van den Heuvel} E.~P.~J.,  1976, in {Eggleton} P.,  {Mitton} S.,   {Whelan}
  J.,  eds,  {Structure and Evolution of Close Binary Systems} Vol. 73,
  Structure and Evolution of Close Binary Systems. p.~35

\bibitem[\protect\citeauthoryear{{van der Helm}, {Portegies Zwart}  \&
  {Pols}}{{van der Helm} et~al.}{2016}]{2016MNRAS.455..462V}
{van der Helm} E.,  {Portegies Zwart} S.,   {Pols} O.,  2016, \mn@doi [\mnras]
  {10.1093/mnras/stv2318}, \href
  {http://adsabs.harvard.edu/abs/2016MNRAS.455..462V} {455, 462}

\bibitem[\protect\citeauthoryear{{von Zeipel}}{{von
  Zeipel}}{1910}]{1910AN....183..345V}
{von Zeipel} H.,  1910, \mn@doi [Astronomische Nachrichten]
  {10.1002/asna.19091832202}, \href
  {https://ui.adsabs.harvard.edu/abs/1910AN....183..345V} {183, 345}

\makeatother
\end{thebibliography}


\label{lastpage}
\end{document}